\documentclass[preprint,authoryear,12pt]{elsarticle}
\usepackage{amssymb}
\usepackage{color}
\usepackage{amsmath}
\usepackage{subfigure}
\usepackage{graphicx}
\usepackage{graphics}
\usepackage{booktabs}
\usepackage{threeparttable}
\usepackage{appendix}
\usepackage{geometry}
\usepackage[displaymath]{lineno}
\usepackage{setspace}
\usepackage{hyperref}
\usepackage{soul}
\usepackage{bm}
\usepackage[table]{xcolor}
\definecolor{light-gray}{gray}{0.95}    

\usepackage{mathpazo}

\journal{Transportation Research Part A, 2021}

\begin{document}
\begin{spacing}{1.3}

\begin{frontmatter}

\title{Portraying ride-hailing mobility using multi-day trip order data: \\A case study of Beijing, China}


\author{Zhengbing He}

\address{Beijing Key Laboratory of Traffic Engineering, Beijing University of Technology, Beijing, China\\ \vspace{0.1in} Email: he.zb@hotmail.com}

\begin{abstract}
As a newly-emerging travel mode in the era of mobile internet, ride-hailing that connects passengers with private-car drivers via an online platform has been very popular all over the world. Although it attracts much attention in both practice and theory, the understanding of ride-hailing is still very limited largely because of the lack of related data. For the first time, this paper introduces ride-hailing drivers' multi-day trip order data and portrays ride-hailing mobility in Beijing, China, from the regional and driver's perspectives. The analyses from the regional perspective help understand the spatiotemporal flowing of the ride-hailing demand, and those from the driver's perspective characterize the ride-hailing drivers' preferences in providing ride-hailing services. A series of findings are obtained, such as the observation of the spatiotemporal rhythm of a city in using ride-hailing services and two categories of ride-hailing drivers in terms of the correlation between the activity space and working time. Those findings contribute to the understanding of ride-hailing activities, the prediction of ride-hailing demand, the modeling of ride-hailing drivers' preferences, and the management of ride-hailing services. 

\vspace{0.1in}

\end{abstract}


\end{frontmatter}




\newpage

\section{Introduction}

In the era of mobile internet, ride-hailing that allows a passenger to hail a private car or a taxi for traveling through a mobile application is a newly-emerging travel mode and it is attracting much attention and many users all over the world \citep{Furuhata2013,Jin2018,Tirachini2019}. 
For the ride-hailing, the transportation network company (TNC) provides a mobile internet-based platform, and a passenger matches his/her origin and destination with the driver (of a private vehicle or a taxi) who would like to give a ride for the purpose of earning money or saving travel cost \citep{Wang2019c}.

Ever since its emergence, a variety of studies have been carried out to understand the ride-hailing services provided by fixed- or part-time drivers as well as the usage of passengers. Three kinds of data sources, namely, questionnaire survey data, self-collected operation data, TNC-released operation data, are the widely-used data source of the existing studies, largely determining the output of the studies.

Questionnaire survey is one of the most important research tools in the existing ride-hailing studies, which unveils the characterization of ride-hailing, in particular from the microscopic perspective that is related to personal selection.
For example, \cite{Anderson2014} made ethnographic interviews and identified three types of driving strategies for providing ride-hailing services, which were incidental, part- and full-time driving. 
However, as mentioned by the authors, the ratio of the three types was unknown limited by the lack of data reflecting the overall population.
\cite{Rayle2016} compared their intercept survey results with taxi trip data in San Francisco, United States and showed that taxis and ride-hailing were different in user characteristics, wait times, etc.
It was found that at least half of ride-hailing trips replaced travel modes such as public transits and private cars, other than taxis. 
\cite{Tang2019} designed a questionnaire for frequent ride-hailing users and conducted an app-based survey through the platform of DiDi Chuxing.
A total of 9762 survey responses were obtained and travelers' behavior changes impacted by ride-hailing were investigated.
Taking Santiago de Chile as an example, \cite{Tirachini2019a} examined the characterization of people's selection on ride-hailing services and its effects on travel behavior.
Through a household travel survey conducted in Toronto, Canada, \cite{Young2019} answered the questions regarding ride-hailing usages, such as who, when, and why people use the ride-hailing.
From the perspective of ride-hailing users, \cite{Alemi2018,Alemi2019} collected 1975 samples through an online survey and unveiled the factors that affected the adoption and the usage frequency of ride-hailing services in California, United States. 
More recently, \cite{Tirachini2020} found that ride-hailing services usually increased vehicle kilometers traveled by conducting an online survey data-based Monte Carlo simulation.
\textcolor{black}{
\cite{Vij2020} surveyed 3,985 Australians on their attitudes and opinions towards on-demand transportation services and found that the services have the potential to increase public transportation usage although current market is still limited. }

Although the questionnaire survey could reveal many latent details such as the intension of selection, the cost of sending questionnaires is high and respondents' answers may not be completely consistent with their daily behaviors.
To avoid the shortcoming, researchers developed various interesting ways to collect ride-hailing data and investigated ride-hailing activities using more observed behavior data. 
For example, \cite{Cooper2018} repeatedly sent synthetic requests to the ride-hailing platform through computer programs at 200 locations across all San Francisco.
Responses of the ride-hailing vehicles nearby were recorded and then employed to estimate the spatial-temporal characteristics of the ride-hailing services in San Francisco.
Also using that data, \cite{Erhardt2019} conducted a before-and-after assessment and found that ride-hailing services were the biggest contributor to the growing traffic congestion in San Francisco.
\textcolor{black}{
One of the authors of \cite{Henao2019a,Henao2019b} drove a ride-hailing vehicle himself to collect trip data and passengers' feedback in Denver, United States.
Using the data, they not only found that the ride-hailing resulted in 83\% more vehicle kilometers traveled than that when no ride-hailing existed \citep{Henao2019a}, but also investigated the actual earning of a ride-hailing driver when considering those factors such as time spent without passengers, driver residential location \citep{Henao2019b}. }
%
\cite{Qian2020} developed a web crawler on the Uber mobile platform to collect ride-hailing data in New York, United States.
A variety of aspects of ride-hailing services were then characterized such as the market share and the distributions of the origin and destination.

TNCs are not willing to share their data with researchers or the public \citep{Costain2012,Li2019,Henao2019a,Henao2019b},
resulting in the fact that little is known about the aggregated characteristics of ride-hailing-related urban mobility,
even though almost ten years have passed since ride-hailing first appeared. 
Until recently, some TNCs conditionally released a part of their data and changed the status quo to some extent.
Leveraging those big-sample or even overall-population data, ride-hailing behavior and its related human mobility are understood more comprehensively.
Based on the ride-hailing trip data provided by a TNC in Austin, United States, \cite{Yu2019a,Yu2019b} found strong relationship between ride-hailing demand and built environment through the geographically weighted Poisson regression and the structural equation model. 
\textcolor{black}{
\cite{Sun2019} proposed a two-level growth model and investigated the spatiotemporal evolution of ride-hailing markets under the new restriction policy of Shanghai, China by using randomly sampled 20,000 ride-hailing and 33,500 taxi orders data of Shanghai.}
\cite{Zhang2020} identified the distribution of regions with high travel intensity and explored the correlation between travel intensity and points of interest by using 209,423 ride-hailing order records in a day in Chengdu, China.
From the labor (driver) side, 
\cite{Dong2018} analyzed 6,471 ride-sharing\footnote{The ride-sharing in \cite{Dong2018} refers to DiDi Hitch.
It is a carpooling service provided by DiDi Chuxing, in which a driver shares a trip with other travelers who have similar origin and/or destinations \citep{Li2019}.
} drivers' activities in a month and identified two kinds of ride-sharing drivers, i.e., daily home-work commuting providers and no-constant-origin-destination providers, in which the daily home-work commuting providers accounted for only a small part of total drivers.
Moreover, it was found that ride-sharing drivers intended to make long distance trips compared with taxi drivers.
\cite{Chen2017c} found that ride-hailing drivers could benefit significantly from the flexibility, which is deemed as one of the attractions of ride-hailing, by analyzing hourly earning data for Uber drivers.
Moreover, \cite{Hall2018} explored ride-hailing drivers' preference and showed that ride-hailing drivers tended to work substantially fewer hours compared with taxi drivers.

From the questionnaire survey data to the ride-hailing operation data, our knowledge regarding ride-hailing is gradually deeper and wider.
However, the data employed for analysis is still very limited, which slows the steps of understanding the special travel mode in the spotlight and its impact on our society. 
To further enrich the knowledge, the paper analyzes multi-day ride-hailing driver activity data in an entire city.
Such data contains not only the spatiotemporal dynamics of ride-hailing demand in a city but also the characterization of ride-hailing driver's behavior of provision-of-service.
The uniqueness of the data makes the paper, to the best of our knowledge, the first that examines ride-hailing from a multi-day perspective using TNC-provided ride-hailing trip data.
More specifically, this paper portrays ride-hailing activities from the perspectives of regional mobility and drivers' multi-day behaviors, respectively.
Many details, such as the temporal varying and spatial flowing of ride-hailing trips, and the spatiotemporal characterization of ride-hailing drivers' behaviors, are explicitly investigated. 
Those findings, which are obtained by directly analyzing the data, contribute to the understanding of ride-hailing activities, the prediction of ride-hailing demand, the modeling of ride-hailing drivers' preferences, and the management of ride-hailing services.

The rest of the paper is organized as follows. 
Section \ref{sec:perspective} introduces the notations, the multi-day trip order data used here, and the perspectives of analyses in the paper, i.e., regional and ride-hailing driver's perspectives.
In Section \ref{sec:Macro}, the flowing dynamics of ride-hailing demand is understood from a regional perspective, and in Section \ref{sec:Micro} the ride-hailing driver's preferences to the newly-emerging job are characterized from a driver's perspective.
At last, Section \ref{sec:Conclusion} concludes the study with discussions.

\section{Analytical framework}\label{sec:perspective}

\subsection{Notations}

The notations regarding ride-hailing trips are defined as follows.
Let $i=1, 2, ..., I$ denote drivers, where $I$ is the total number of drivers.
Let $TR_{ij}= \{O_{ij}, D_{ij}\}$ be driver $i$'s trip $j=1, 2, ..., J$, where $J$ is the total number of driver $i$'s trips and $O_{ij}$ and $D_{ij}$ are the information regarding the origin and destination points of a trip, respectively.
To save space, denote by $\Xi$ either $O$ or $D$.
Then, the information regarding the origin or destination point of a trip is $\Xi_{ij}= \{t^\Xi_{ij},\ lon^\Xi_{ij},\ lat^\Xi_{ij}\}$,
where $t^\Xi_{ij}$ is the time when the trip started (i.e., a passenger was picked up) or ended (i.e., a passenger was dropped off);
$lon^\Xi_{ij}$ and $lat^\Xi_{ij}$ are the longitude and latitude of the location where a passenger was picked up or dropped off, respectively.
Regarding a trip, we can calculate its duration and displacement, denoted by $T_{ij}$ and $L_{ij}$, respectively, as follows.
\begin{equation}\label{equ:TL}
	\begin{cases}
		T_{ij}= t^D_{ij} - t^O_{ij}\\
		L_{ij}= \text{dis}\left((lon^D_{ij}, lat^D_{ij}), (lon^O_{ij}, lat^O_{ij})\right)
	\end{cases}
\end{equation}
where $\text{dis}(\cdot)$ is the function of calculating the distance between two points on the earth surface. 
\textcolor{black}{
Note that, different from vehicle kilometers traveled, the displacement only reflects the straight-line distance between the origin and destination points of a trip, while it is widely used in studying human mobility \citep{Liang2012,He2020b}.}

\subsection{Data description}

\begin{figure}[!htbp]
    \centering
    \includegraphics[width=4in]{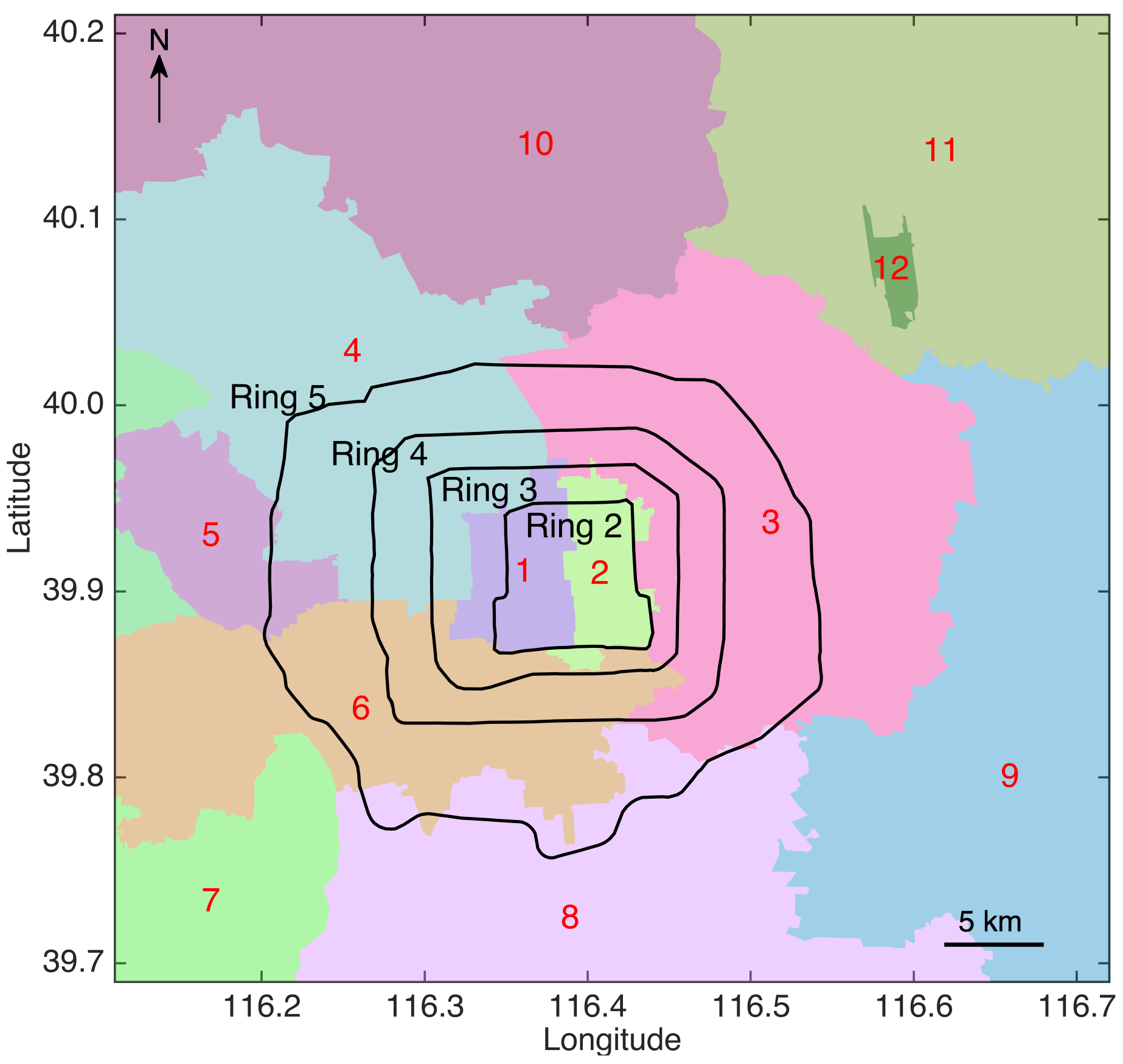}
    \caption[Beijing and its districts]{Study area of Beijing. 
    Those numbers indicate the districts of Beijing:
    1-{\it Dongcheng District};
    2-{\it Xicheng District};
    3-{\it Chaoyang District};
    4-{\it Haidian District};
    5-{\it Shijingshan District};
    6-{\it Fengtai District};
    7-{\it Fangshan District};
    8-{\it Daxing District};
    9-{\it Tongzhou District};
    10-{\it Changping District};
    11-{\it Shunyi District}.
    Among the districts, Districts 1-6 are the central districts of Beijing.
    In addition, 12 indicates {\it Beijing Capital International Airport}.}
    \label{fig:BeijingMap}
\end{figure}

Beijing, the capital of China, is one of the largest cities in the world.
By 2018, the total population of Beijing was approximately 21.5 million.
The central area is enclosed by four urban freeways, namely, Rings 2$\sim$5 (Figure \ref{fig:BeijingMap}), and commonly, Ring 5 is treated as a separate line of the central and suburban areas.
The lengths of Rings 2$\sim$5 are 32.7 km, 48.3 km, 65.3 km, 98.6 km, and 188.0 km, respectively, and the areas inside Rings 2$\sim$5 are 62 km$^2$, 159 km$^2$, 302 km$^2$ and 667 km$^2$, respectively.

The dataset that we used here is provided by one of the biggest TNCs in China and it contains all one-week trip order information of 138,138 drivers with their private vehicles, i.e., no taxis here. 
Those drivers made a total of 5,289,045 ride-hailing trips (orders) during a week in August, 2018.
We finally obtain 5,041,455 (95.3\%) valid trips after removing the invalid trips whose $T_{ij}\leq 0$.

\subsection{Perspectives of analyses}

A ride-hailing system, which involves users, drivers, and an online platform, usually operates as follows.
A user who is looking for a ride-hailing service first sends a request to the platform.
The platform immediately matches the request with the drivers who are close to the origin of the demand, and then assigns the request to one of the qualified drivers according to some mechanism of selection.
Based on the data we obtained and our daily experience, the standby ride-hailing vehicles are usually sufficient in Beijing, China, where ride-hailing has been very popular \citep{Nie2017}.
Therefore, a user can usually get served immediately after sending a request out.

Looking at the ride-hailing mobility from a regional perspective, a travel demand could be met by one of the nearby drivers.
Therefore, analyzing the ride-hailing data at a regional level, i.e., the inter-region transfer of passengers, could uncover the dynamics of travel demand.


\textcolor{black}{From a ride-hailing driver's perspective, after a travel request is sent, one of the nearby ride-hailing drivers will receive the request that is sent by the TNC. 
Although ride-hailing drivers are unable to directly determine what kind of requests are assigned, they could make limited selections by pre-defining the range where they prefer to provide services or by canceling the assignment \citep{Yang2018b}.}
The decision making is determined by their preferences to ride-hailing, such as the time when they would like to provide services and the place where they would like to go.
Although the TNCs have some mechanism of selection and assignment of transactions, the free-market essence of ride-hailing determines that the drivers have a certain freedom of selection. 
If the temporal and spatial activity pattern exhibited by a driver appears in a single day, we may say that it depends on both the demand (passenger) and supply (driver) sides.
However, if the activity pattern of a driver is repeated for a few days, we can reasonably deem that the recurrent pattern is mainly determined by the preferences of the driver who provides the service. 
Therefore, analyzing ride-hailing driver's multi-day trip data could better characterize ride-hailing drivers and help understand the labor market at the supply side.

\section{Regional perspective: spatiotemporal flowing of ride-hailing trips}\label{sec:Macro}

\subsection{Temporal varying of ride-hailing trips}

\begin{figure}[!htbp]
    \centering
    \subfigure[Trip numbers in days]{
    \includegraphics[width=6in]{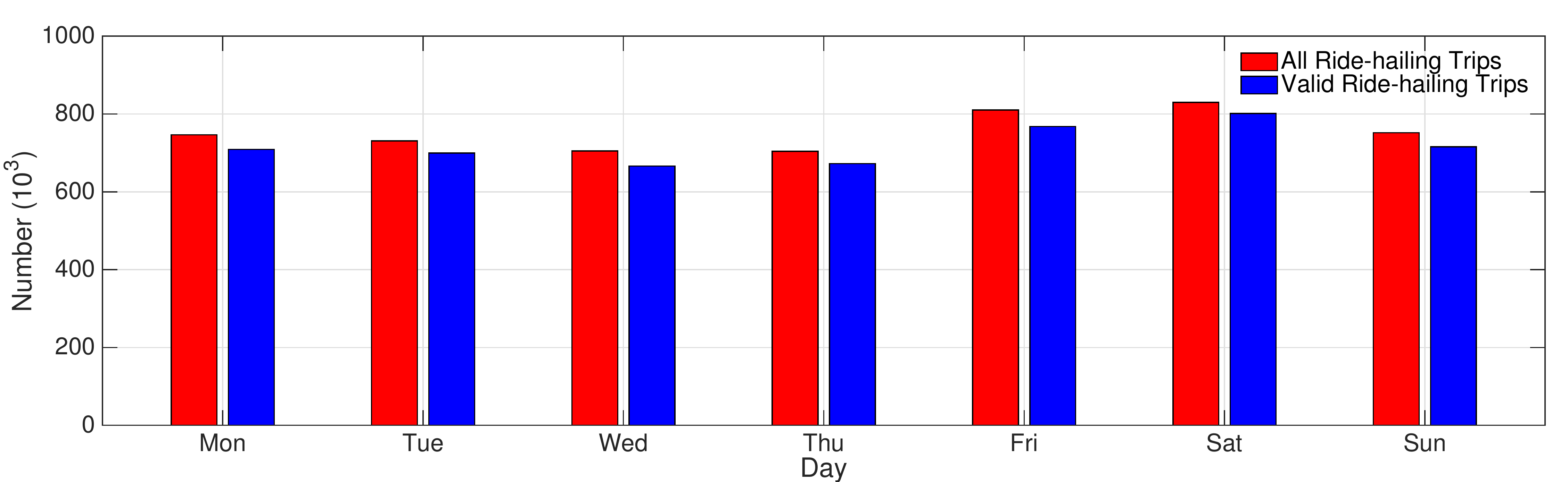}}
    \subfigure[Trip numbers in hours]{
    \includegraphics[width=6in]{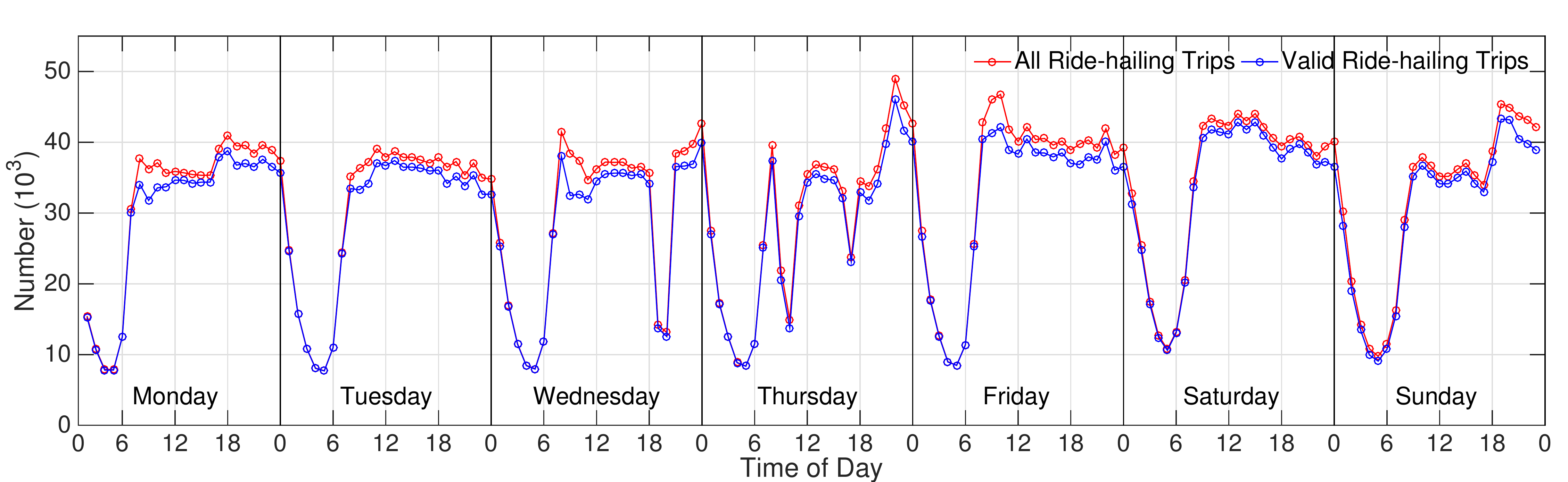}}    
    \caption[Daily and hourly changes of trip generations in the studied dataset]
    {Daily and hourly changes of trip generations (represented by $t^O_{ij}$) in the studied dataset.}
    \label{fig:Z_TimeSeries_TripNumber}
\end{figure}

We are interested in the temporal varying pattern of the generations of ride-hailing trips.
Therefore, we present the daily and hourly changes of trip numbers in the studied dataset in Figure \ref{fig:Z_TimeSeries_TripNumber}
and we have the following observations.

\begin{itemize}
\setlength{\itemsep}{0pt}
\setlength{\parsep}{0pt}
\setlength{\parskip}{0pt}

	\item[{[1]}] The total ride-hailing demands on Friday and Saturday are slightly larger than those on the other days (Figure \ref{fig:Z_TimeSeries_TripNumber}(a)).

	\item[{[2]}] Once the number of ride-hailing trips reach \textcolor{black}{high values around 40,000} in the morning (Figure \ref{fig:Z_TimeSeries_TripNumber}(b)), 
	it generally stabilizes at \textcolor{black}{high values} and lasts to midnight (except for those on Wednesday and Thursday\footnote
	{We carefully checked the data and we did not find any evidence that indicates that the fluctuations are resulted from data damage. \textcolor{black}{Currently, it is still difficult to clearly explain the ``abnormal" pattern on Wednesday and Thursday partially due to the lack of other source data for cross validation. We report it here and expect an ongoing work for better interpretation.}}).
	It turns out that 
	(i) there is no clear off-peak period at noon and 
	(ii) the trip number drops until midnight instead of late evening. 
	The observations are different from the pattern of distinct morning-and-evening peaks exhibited by daily traffic and taxi demand \citep{He2017,He2019b}.
	
\end{itemize}

\subsection{Spatial flowing of ride-hailing trips}

We attempt to unveil how ride-hailing trips spatially flow in a city.
To the end, we take the area presented in Figure \ref{fig:BeijingMap} to be the study area, which is 52 km long from the west (longitude=116.11) to the east (longitude=116.72) and 58 km long from the south (latitude=39.69) to the north (latitude=40.21).
\textcolor{black}{The total numbers of the origin and destination points in the area are 4,763,115 and 4,743,235, respectively, which are approximately 94\% of the total contained in the dataset that we use here.}

We divide the \textcolor{black}{above} area using \textcolor{black}{same-size} square grids with a side length of $L$\textcolor{black}{; $L$ is equal to $L_\text{lon}$ and $L_\text{lat}$ in longitude and latitude, respectively.
Therefore, }
for an area of $XL\times YL$ \textcolor{black}{(i.e., $XL_\text{lon}\times YL_\text{lat}$),} we have $X\times Y$ grids in total.
Here, we set $L=1$ km and thus $X=52$ and $Y=58$.

Then, we map those origin or destination points (i.e., $\Xi_{ij}$) into the grids by using basic arithmetic operations as follows.
\begin{equation}\label{equ:Mapping}
	\begin{cases}
		 x = \left\lceil  \displaystyle\frac{lon^\Xi_{ij} - A_\text{left}}{L_\text{lon}} \right\rceil \in [1, X] \vspace{3mm} \\
		 y  =  \left\lceil \displaystyle\frac{lat^\Xi_{ij}- A_\text{bottom}}{L_\text{lat}} \right\rceil \in [1, Y]  \\
	\end{cases}
\end{equation}
where $(x,y)$ indicates the grid that an origin or destination point belongs to;
$A_\text{left}$ and $A_\text{bottom}$ are the left and bottom edges of the selected area;
$\lceil \cdot \rceil$ is an operation that rounds a number up to an integer.
As a consequence, we obtain the number of the origin or destination points in grid $(x,y)$ within time interval $k$ of day $g$; 
denoted by $N^\Xi_{x,y}(g,k)$.
In this study, we set $k\in K=\{0, 1, 2, ..., 23\}$, i.e., 24 hours of a day. 
In addition, we define the grids satisfying $N^\Xi_{x,y}(g,k)>N^*$ as {\it dense origin or destination grids}, where $N^*$ is a threshold.

\begin{figure}[!htbp]
    \centering
    \subfigure[\scriptsize{Morning \textcolor{black}{(8:00)}: Origin}]{
    \includegraphics[width=2.2in]{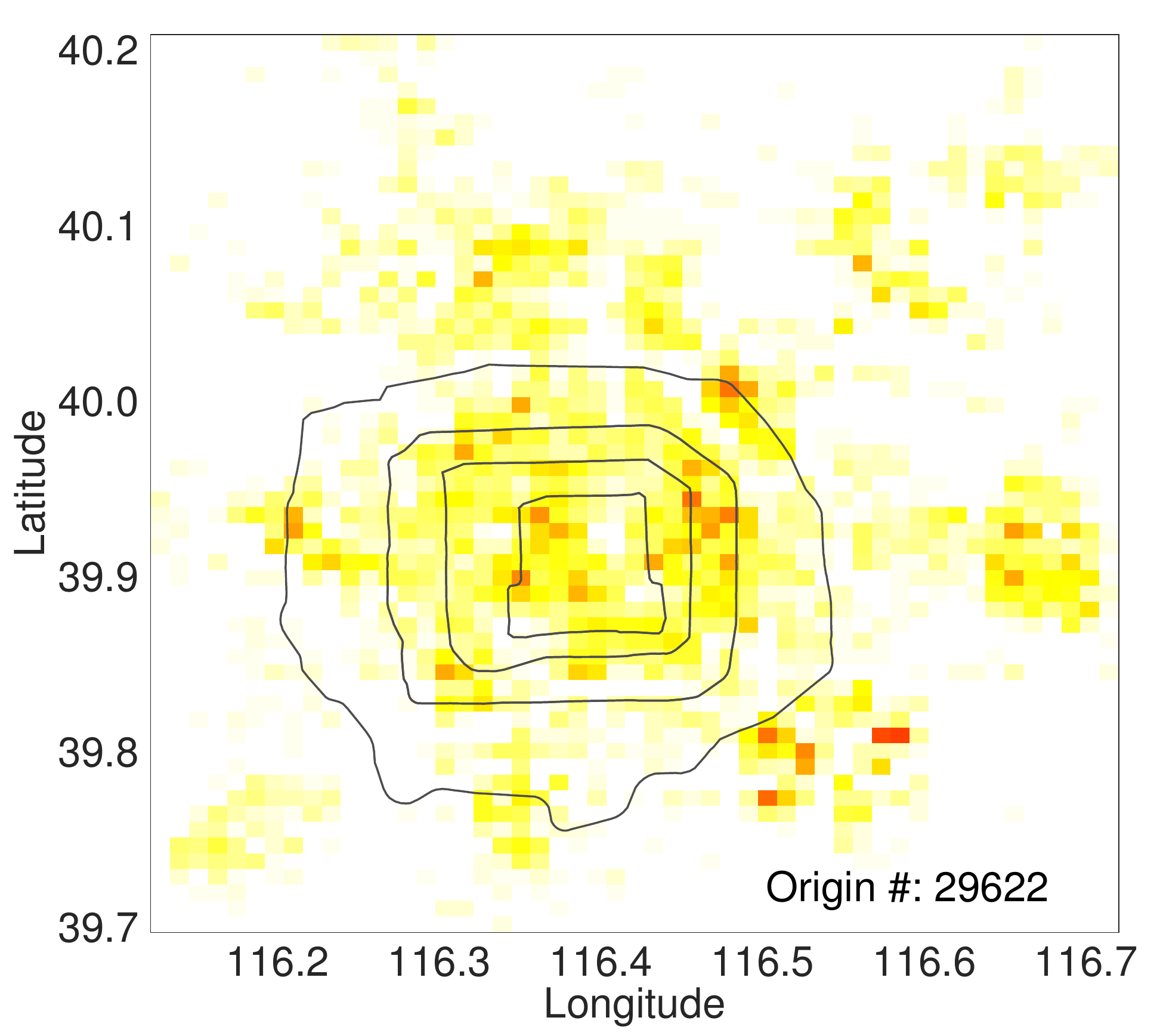}}
    \subfigure[\scriptsize{Morning \textcolor{black}{(8:00)}: Destination}]{
    \includegraphics[width=2.2in]{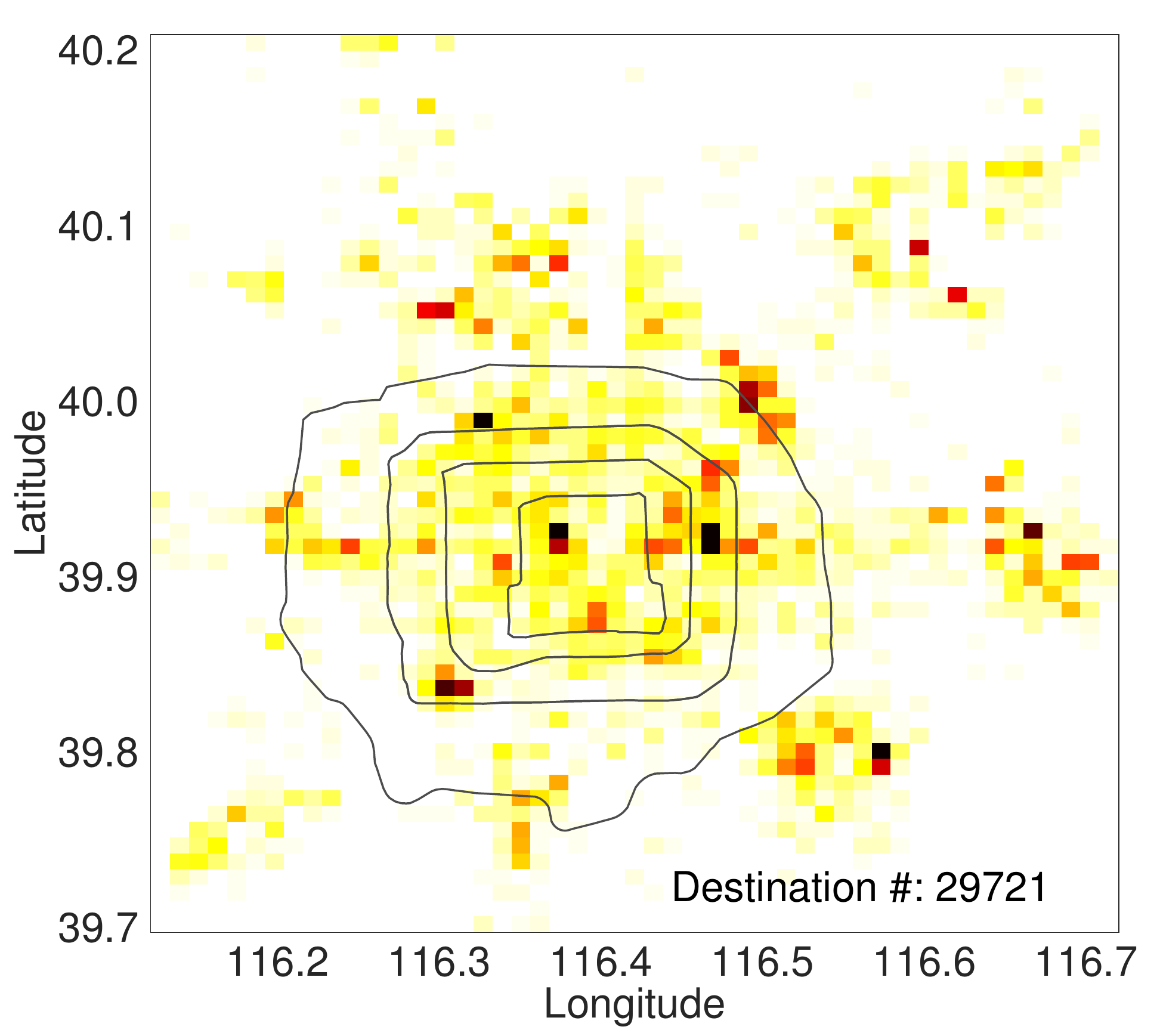}}
    \subfigure[\scriptsize{Morning \textcolor{black}{(8:00)}: Frequency comparison}]{
    \includegraphics[width=2.2in]{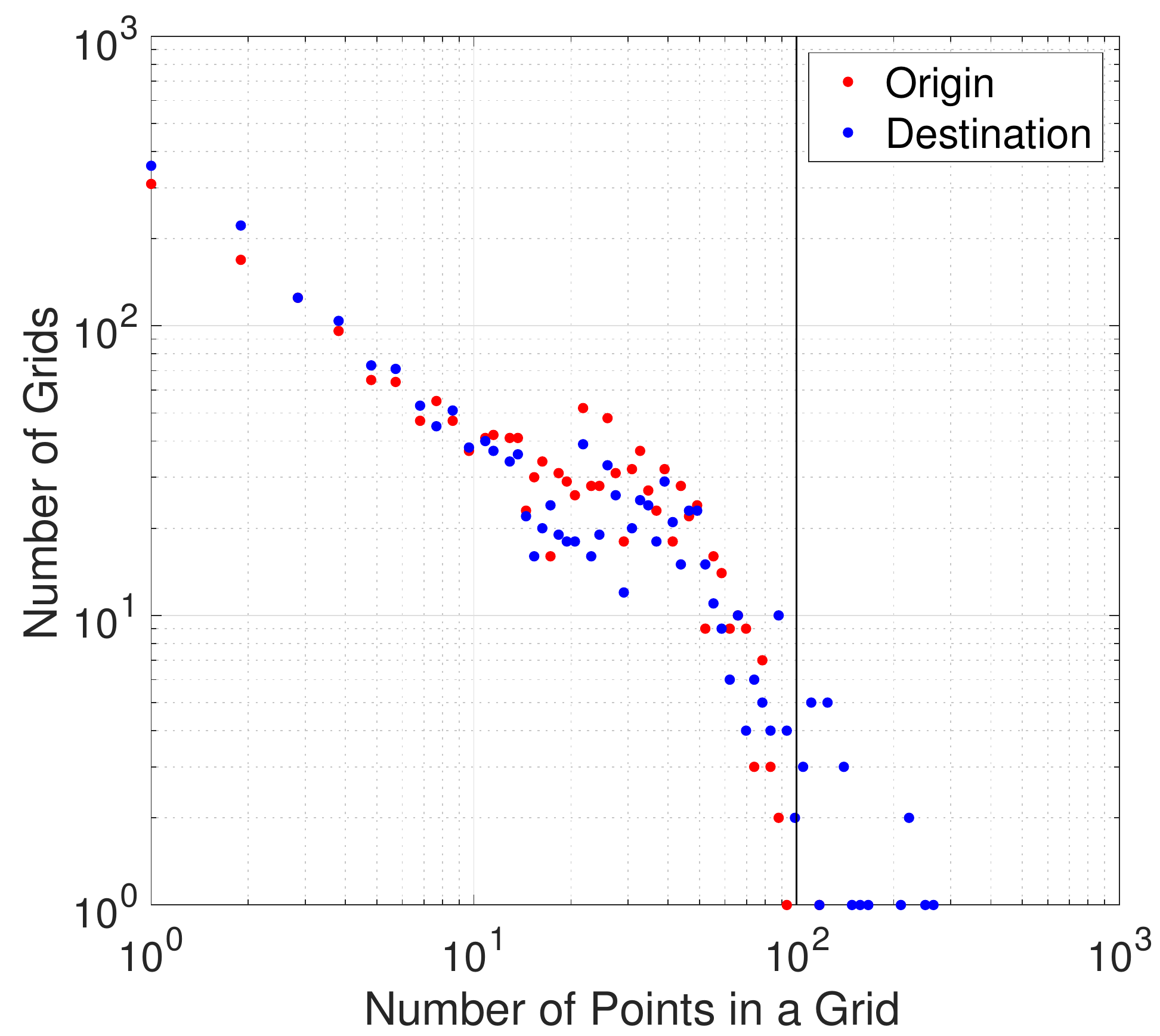}}
    \subfigure[\scriptsize{Evening \textcolor{black}{(17:00)}: Origin}]{
    \includegraphics[width=2.2in]{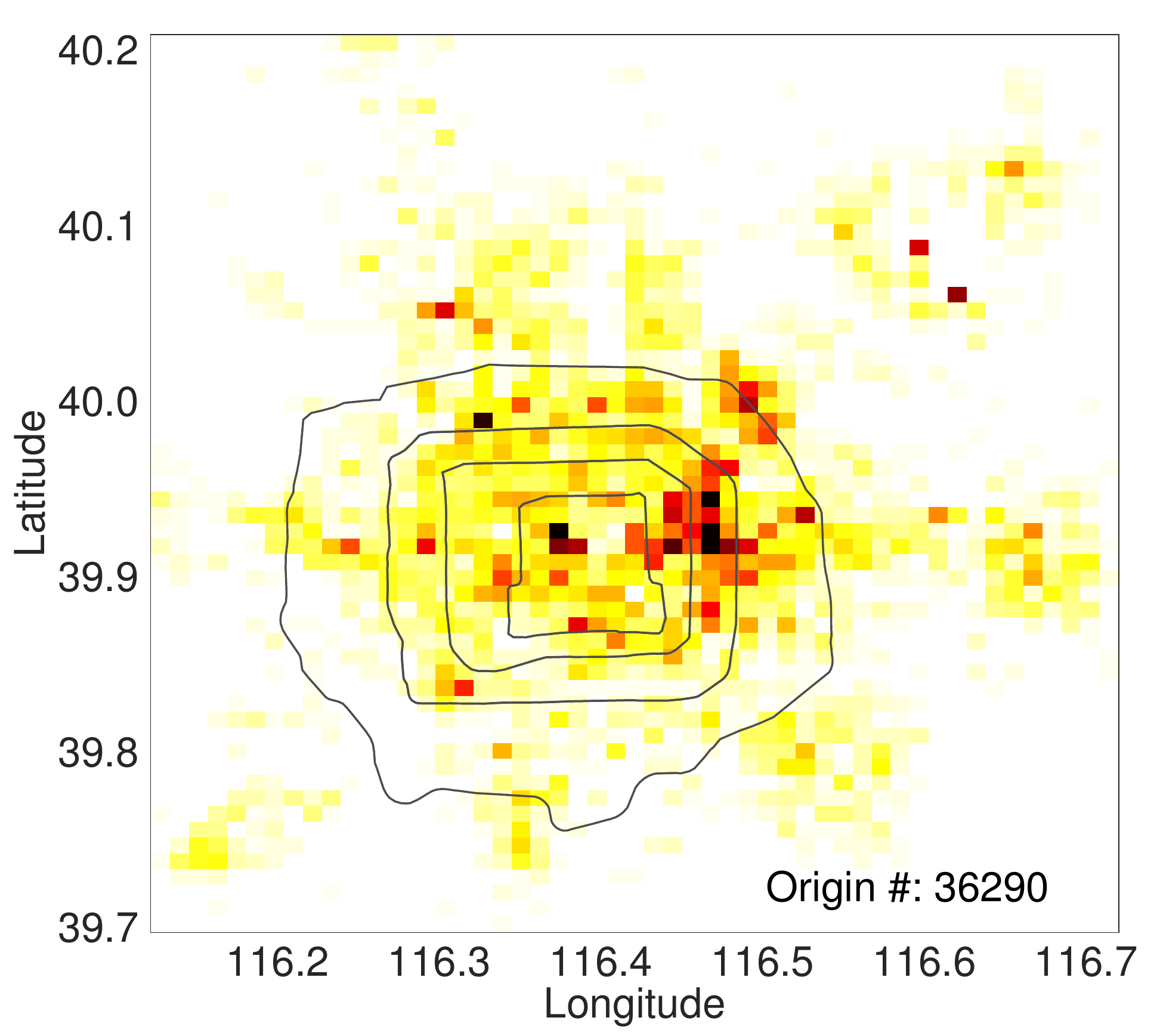}}
    \subfigure[\scriptsize{Evening \textcolor{black}{(17:00)}: Destination}]{
    \includegraphics[width=2.2in]{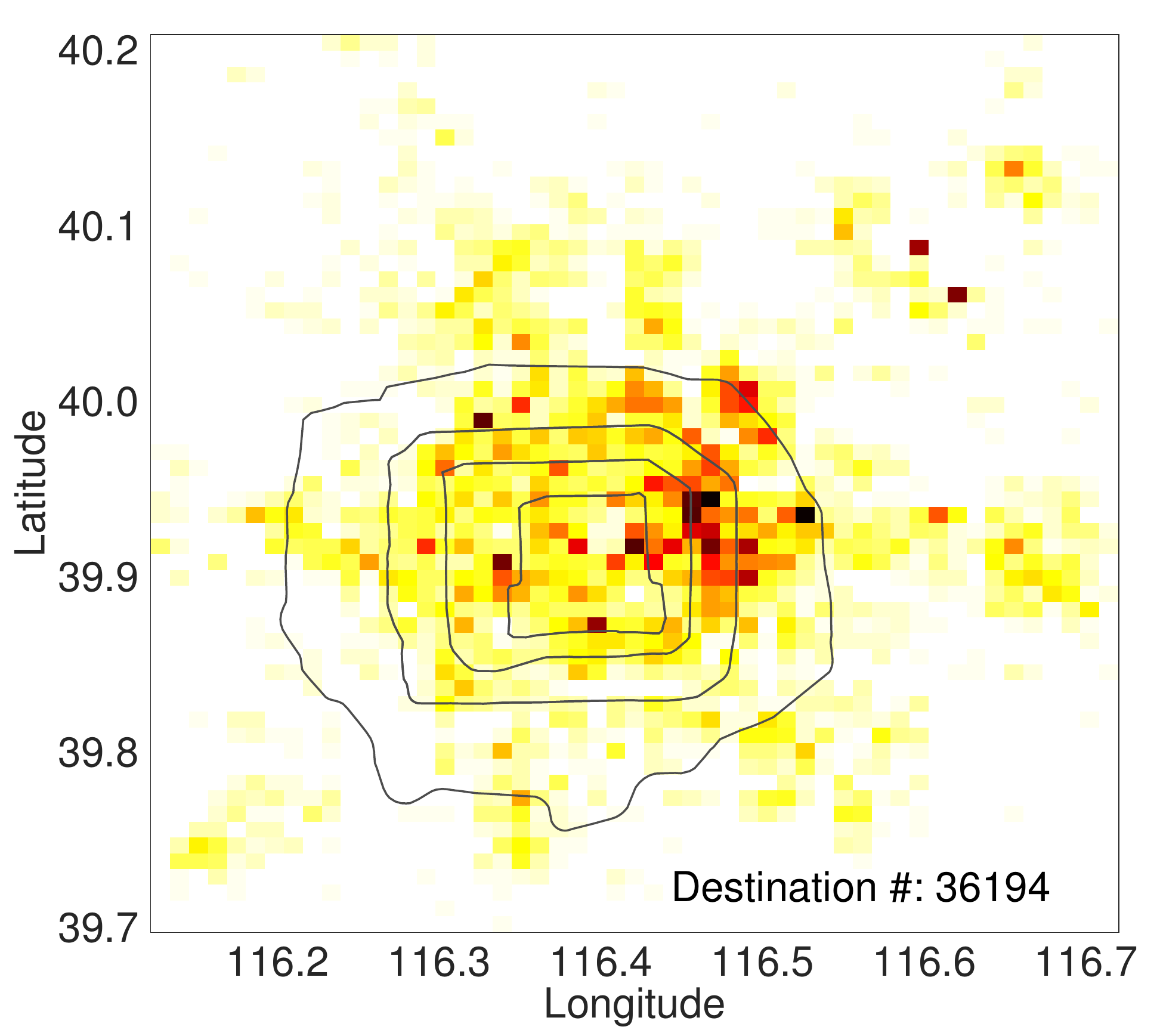}}
    \subfigure[\scriptsize{Evening \textcolor{black}{(17:00)}: Frequency comparison}]{
    \includegraphics[width=2.2in]{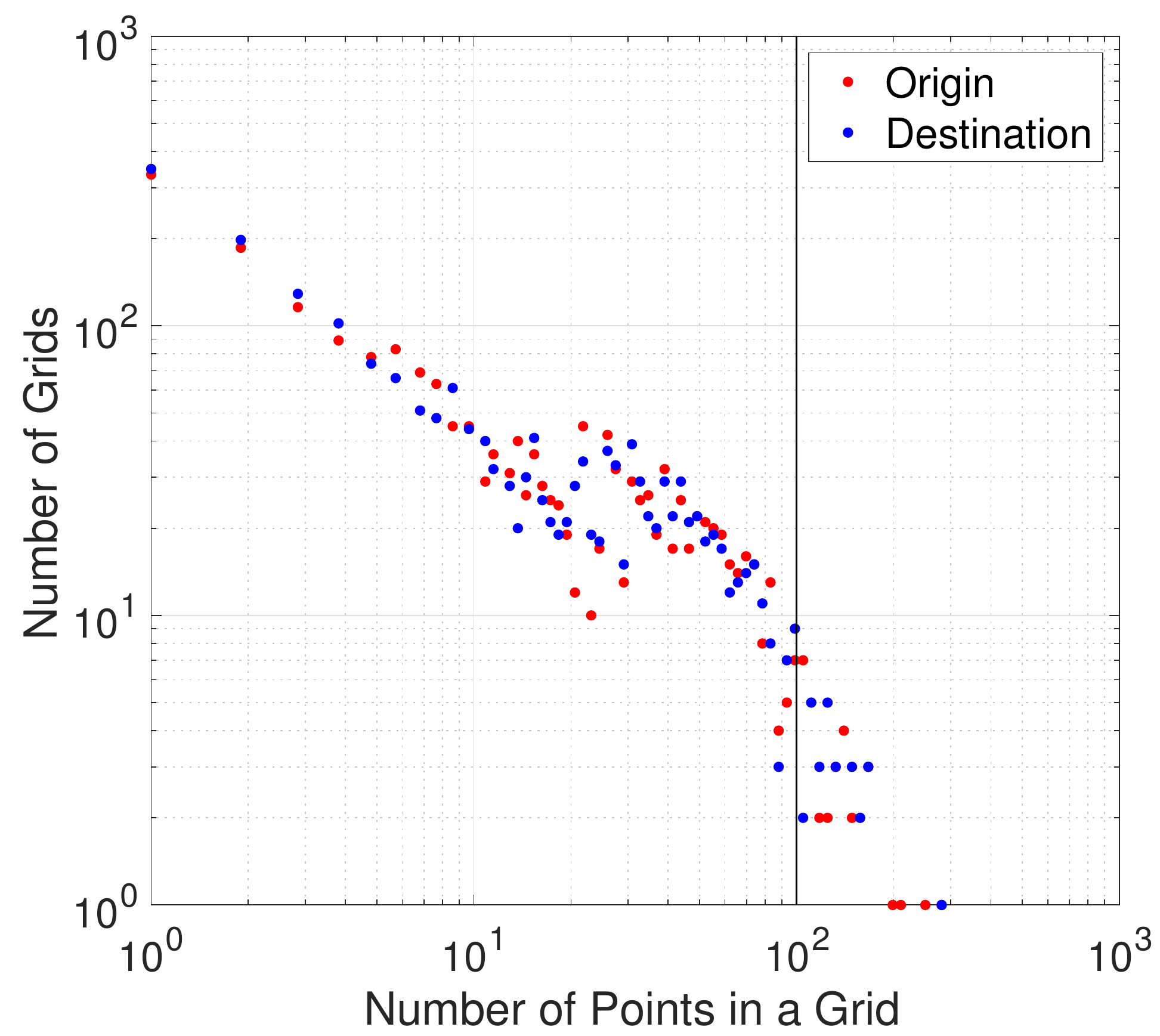}}
    \subfigure[\scriptsize{Midnight \textcolor{black}{(22:00)}: Origin}]{
    \includegraphics[width=2.2in]{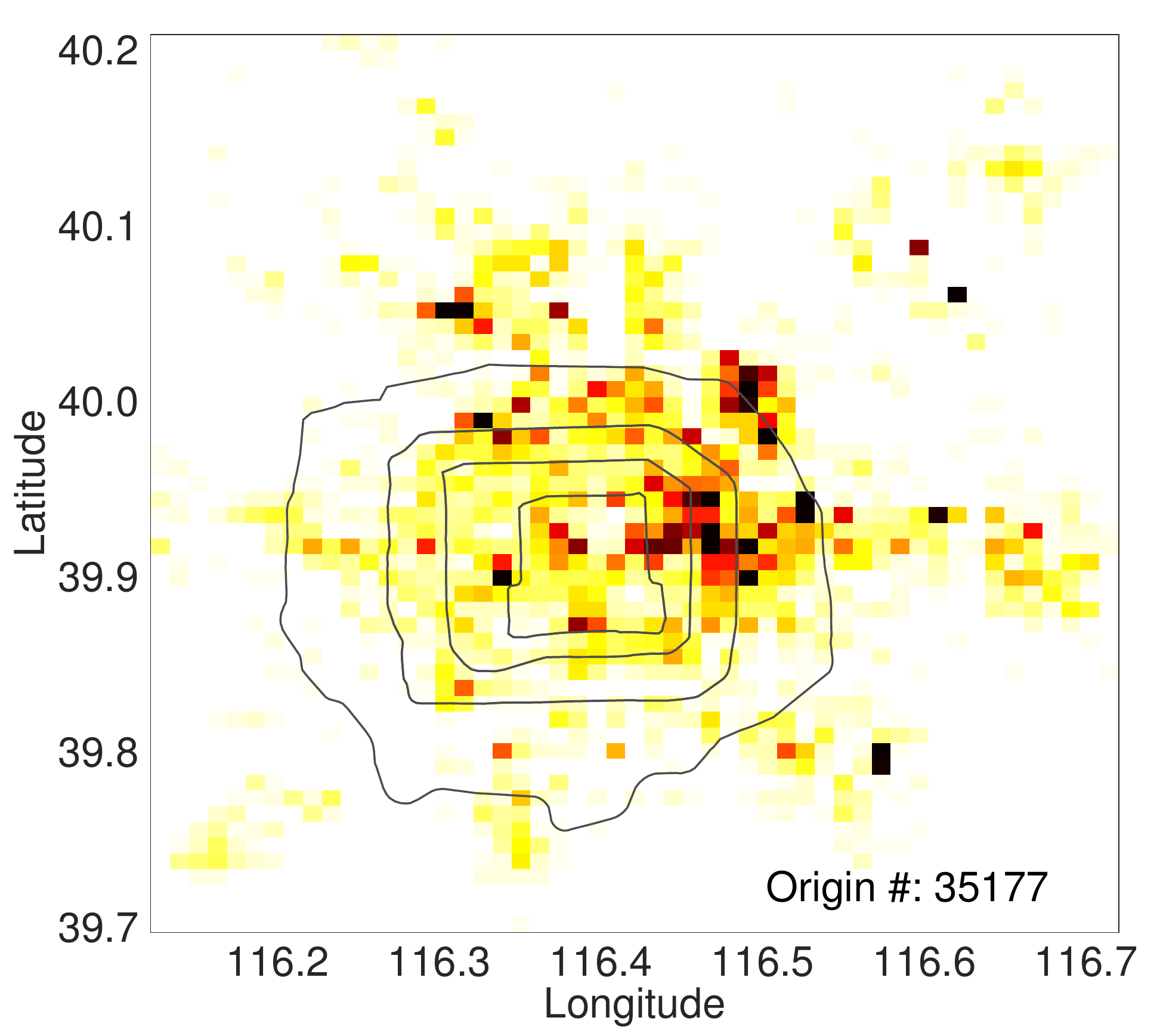}}
    \subfigure[\scriptsize{Midnight \textcolor{black}{(22:00)}: Destination}]{
    \includegraphics[width=2.2in]{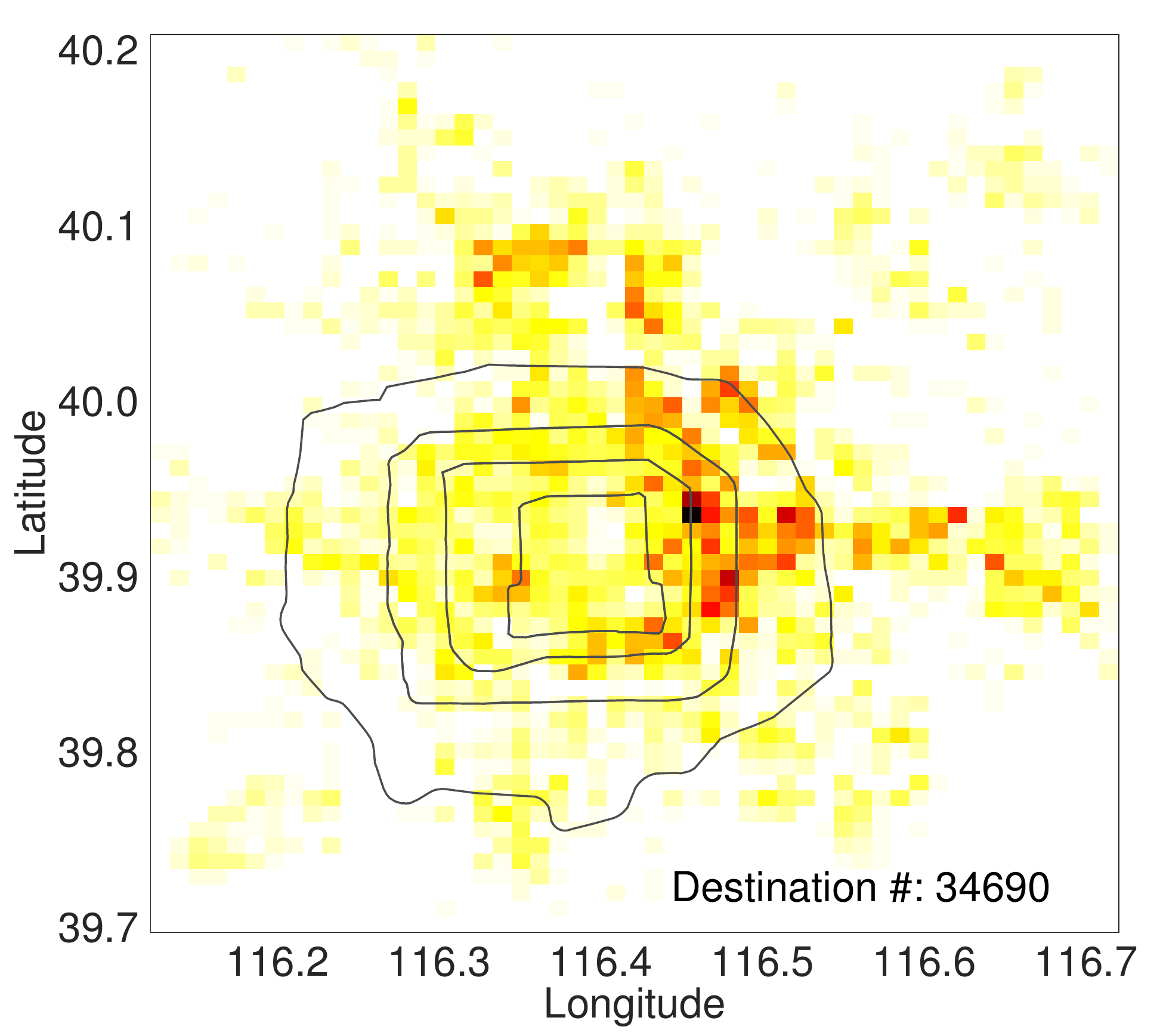}}
    \subfigure[\scriptsize{Midnight \textcolor{black}{(22:00)}: Frequency comparison}]{
    \includegraphics[width=2.2in]{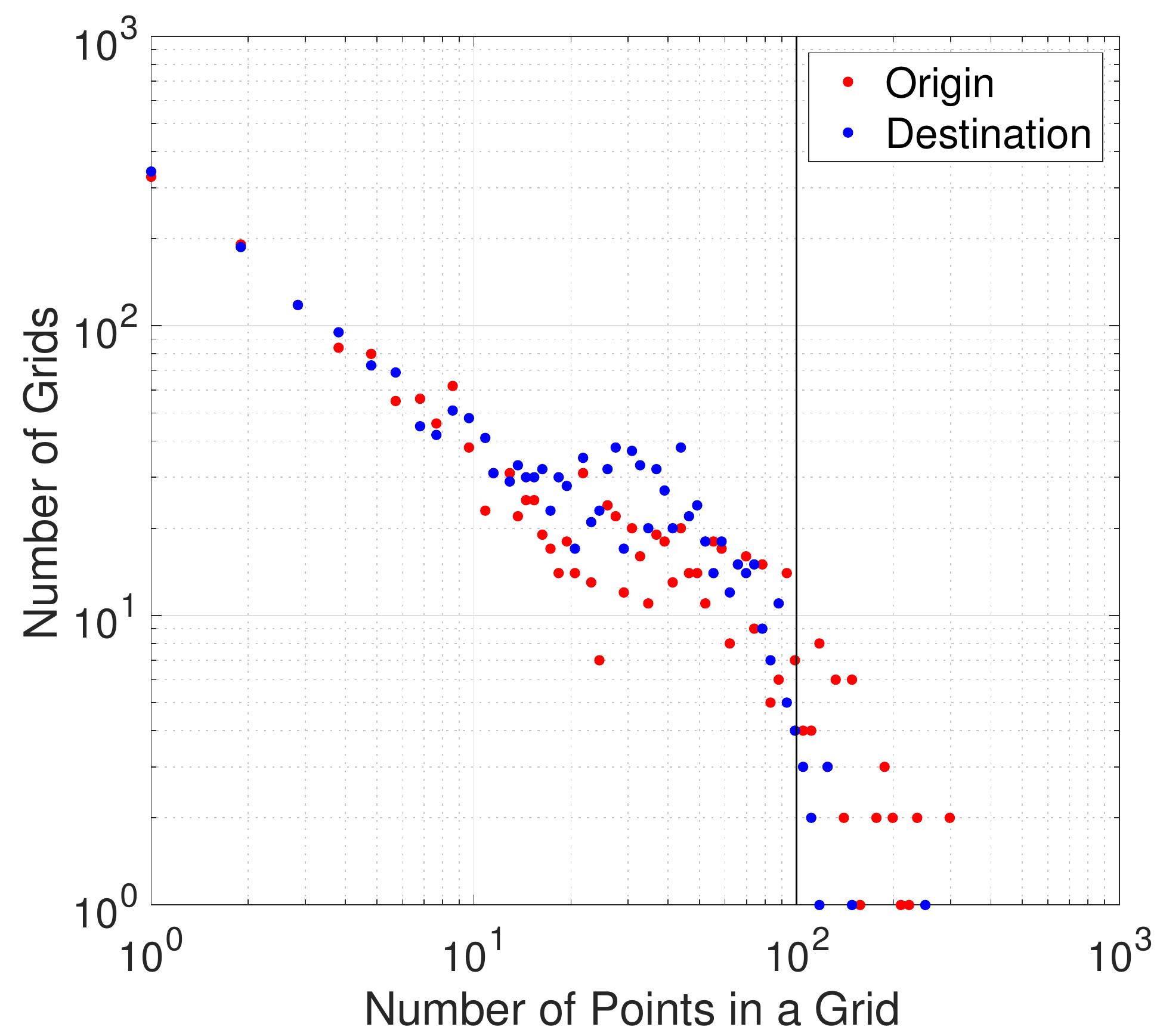}}
    \subfigure{
    \includegraphics[width=3in]{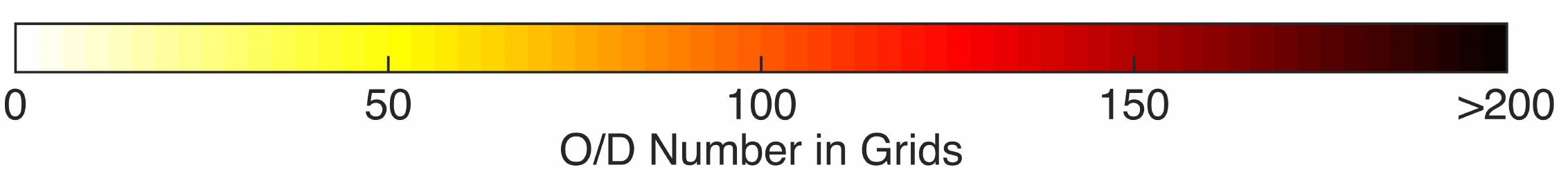}}
    \caption[Spatial distribution of the origin and destination points of ride-hailing trips]
    {Spatial distribution of the origin and destination points of ride-hailing trips (Monday, 2018).}
    \label{fig:Z_HeatmapOfAllTripOD}
\end{figure}

From the resulting $N^\Xi_{x,y}(g,k)$, we initially find three distinct time periods, i.e., {\it morning}, {\it evening} and {\it midnight}, and we separately plot them in Figure \ref{fig:Z_HeatmapOfAllTripOD}.
By comparing the heat maps of origins (Figure \ref{fig:Z_HeatmapOfAllTripOD}(a)(d)(g)) with those of destinations (Figure \ref{fig:Z_HeatmapOfAllTripOD}(b)(e)(h)), 
directional flowing of trips can be observed as follows.
In the {\it morning}, the number of dense destination grids (Figure \ref{fig:Z_HeatmapOfAllTripOD}(a)) is larger than that of dense origin grids (Figure \ref{fig:Z_HeatmapOfAllTripOD}(b)), given the condition that the total numbers of the origin and destination points are approximately equal.
Figure \ref{fig:Z_HeatmapOfAllTripOD}(c) more directly compares the frequencies of the grids with different numbers of the origin and destination points.
It shows that the number of the grids that contain more than (e.g.) 100 destination points is clearly larger than that of the grids that contain more than (e.g.) 100 origin points.
Likewise, in the {\it evening}, the numbers of dense origin and destination grids (Figure \ref{fig:Z_HeatmapOfAllTripOD}(d)(e)) are close (Figure \ref{fig:Z_HeatmapOfAllTripOD}(f)), 
while, in the {\it midnight}, there are more dense origin grids (Figure \ref{fig:Z_HeatmapOfAllTripOD}(g)), compared with dense destination grids (Figure \ref{fig:Z_HeatmapOfAllTripOD}(h)); also see Figure \ref{fig:Z_HeatmapOfAllTripOD}(i).


\begin{figure}[!htbp]
    \centering
    \includegraphics[width=6in]{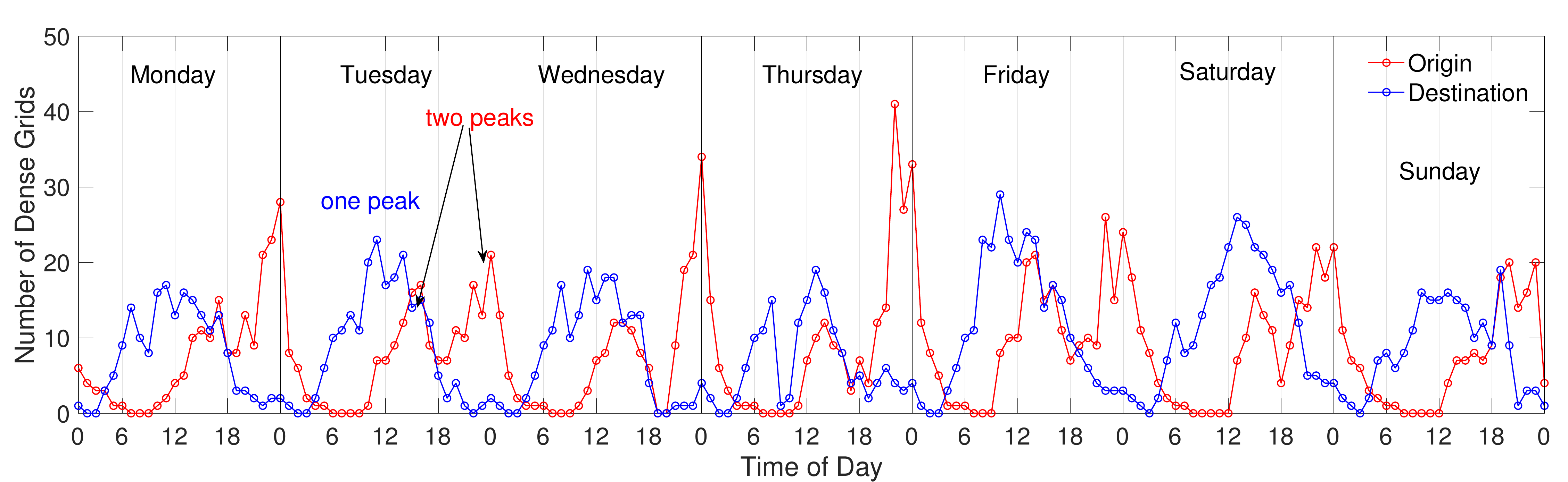}
    \caption[Flowing of aggregated trips]
    {Flowing of aggregated trips. The threshold of defining dense origin and destination grids is set to be 150 per hour (i.e., $N^*$=150).}
    \label{fig:Z_FlowingOfDenseODArea}
\end{figure}

To enhance more in the direction, we compare the time series of the numbers of dense origin and destination grids in Figure \ref{fig:Z_FlowingOfDenseODArea} after setting $N^*$=150.
The rise and fall of the time series in Figure \ref{fig:Z_FlowingOfDenseODArea} reflect the spatiotemporal rhythm of the city in using ride-hailing services, which can be interpreted as follows. 
Note that the total numbers of the origin and destination points in an hour in the studied area are approximately equal.

\begin{itemize}
\setlength{\itemsep}{0pt}
\setlength{\parsep}{0pt}
\setlength{\parskip}{0pt}	    

	\item[{[3]}] There are two peaks for the appearance of the dense origin grids during a day. 
	One appears at approximately 15:00 and the other one at approximately midnight.
	The second peak is higher than the first one. 
	The observations indicate that, during a day, the origins of trips usually spatially shrink twice. 
	One is at approximately 15:00, which may be associated with the purpose of back-home-from-work, 
	and the other at approximately midnight, probably related to back-home-from-overtime and back-home-from-entertainment.
	The second is more intensive, implying that ride-hailing services are more needed at night.

	\item[{[4]}] There is only one peak for the appearance of the dense destination regions during a day, which occurs at noon. 
	The reason cannot be directly speculated, while it is worth studying more carefully in the future. 

	\item[{[5]}] Comparing the rise and fall of the two time series, the demand for ride-hailing services scatters in the city in the morning, 
	while the destinations are more concentrative largely because of to-work activities; at night, the trend is in the opposite.
	This observation implies that, in the city, the places of residence may be more scattered than the places of work.

\end{itemize}

\subsection{Multi-day repeatability of regional origin or destination numbers}\label{sec:repeatability}

We are interested in understanding the multi-day repeatability of the regional intensity (i.e., number) of the origin and destination points, since it is closely related to the prediction of ride-hailing activities.
First, we measure the following coefficient of variation of the origin or destination points that fall into grid $(x,y)$ during time interval $k$ on all weekdays.
\begin{equation}
	\text{cv}^\Xi_{x,y}(k) = \frac{\sigma^\Xi_{x,y}(k)}{\mu^\Xi_{x,y}(k)}
\end{equation}
where $\mu^\Xi_{x,y}(k)$ and $\sigma^\Xi_{x,y}(k)$ are the mean and standard deviation of the numbers of the origin and destination points that fall into grid $(x,y)$ during time interval $k$ on all weekdays, which are written as follows. 
\begin{subequations}
\begin{align} 
	&\mu^\Xi_{x,y}(k)= \frac{1}{|G|}  \sum_{g\in G} N_{x,y}^\Xi (g,k)\\
	&\sigma^\Xi_{x,y}(k)= \left(\frac{1}{|G|}  \sum_{g\in G} (N_{x,y}^\Xi (g,k) - \mu^\Xi_{x,y}(k))^2 \right)^{\frac{1}{2}}
\end{align}
\end{subequations}
where $G=$\{Mon, Tue, Wed, Thu, Fri\}.
Only the trips occurred on weekdays are considered here because travel demands on weekdays and weekends are usually different.

It is natural to believe that the grids with very few origin and destination points have no intense real-world functions.
Therefore, we only take the grids whose $\mu^\Xi_{x,y}(k)\geqslant 10$ into account.
After the refinement, the number of the considered grids for origins is 1954 (64.8\% of the total number of grids) and 2025 (67.1\%) for destinations.
The numbers of the origin and destination points remained in those grids are 3,322,246 (99.6\%) and 3,306,820 (99.5\%), respectively, meaning that the trips are greatly remained.

\begin{figure}[!htbp]
    \centering
    \subfigure[Origin]{
    \includegraphics[width=6in]{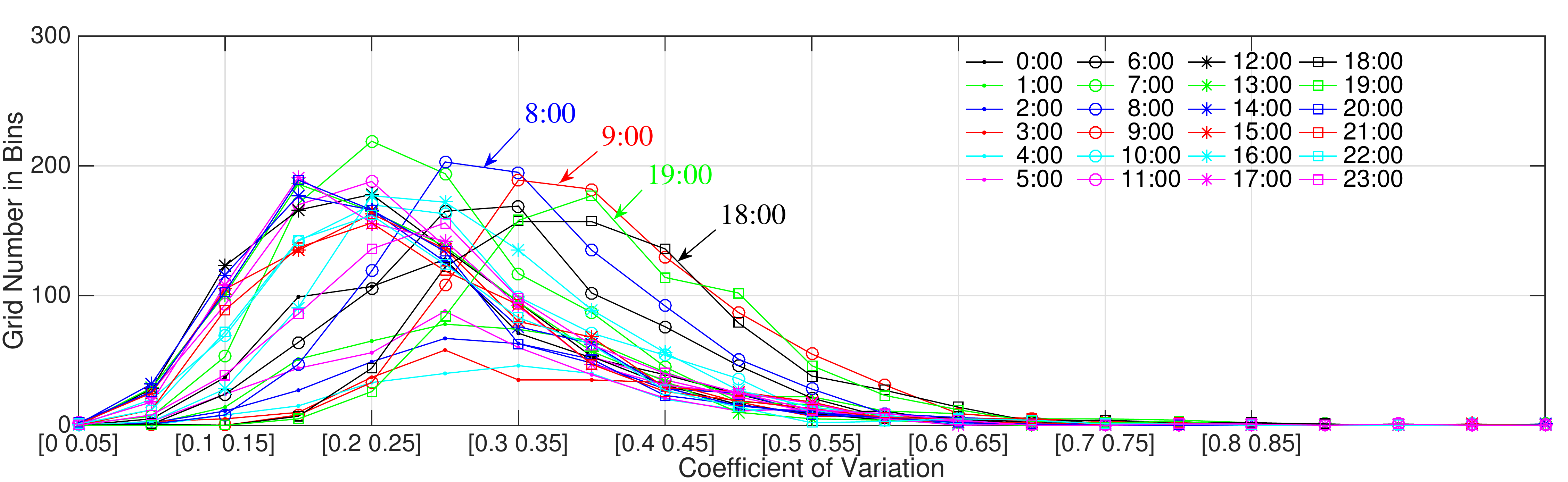}}
    \subfigure[Destination]{
    \includegraphics[width=6in]{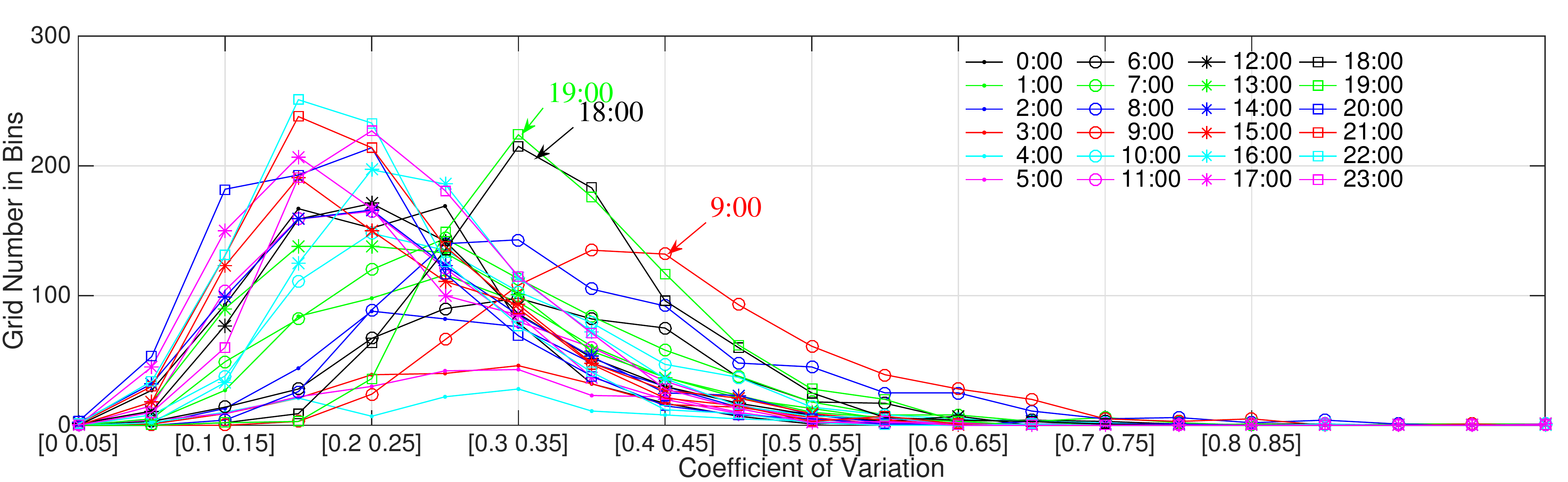}}    
    \caption{Distributions of coefficient of variation of the origin and destination of trips in grids at different times.}
    \label{fig:Z_D2DInvariantOfGridOD_Hour}
\end{figure}

Figure \ref{fig:Z_D2DInvariantOfGridOD_Hour} presents the numbers of grids within different values of $\text{cv}^\Xi_{x,y}(k)$ 
and we have the following observations.

\begin{itemize}
\setlength{\itemsep}{0pt}
\setlength{\parsep}{0pt}
\setlength{\parskip}{0pt}

	\item[{[6]}] The values of $\text{cv}^\Xi_{x,y}(k)$ at most time are between 0.15 and 0.25, 
	\textcolor{black}{indicating that the number of either the origin or destination points in a region essentially does not change dramatically at the same time of each weekday.
	Therefore, only when an advanced prediction method, 
	which is expected to forecast the future accurately, 
	could result in smaller prediction deviations, we say that the method is effective.}

	\item[{[7]}] For several time intervals (Origin: 8:00, 9:00, 18:00, 19:00; Destination: 9:00, 18:00, 19:00), 
	the values of $\text{cv}^\Xi_{x,y}(k)$ are relatively large and range between 0.3 and 0.4, 
	indicating relatively low daily repeatability of regional origin and destination numbers. 
	It implies that making predictions by utilizing multi-day repeatability is more difficult for those time intervals.
\end{itemize}

\begin{figure}[!htbp]
    \centering
    \includegraphics[width=3.1in]{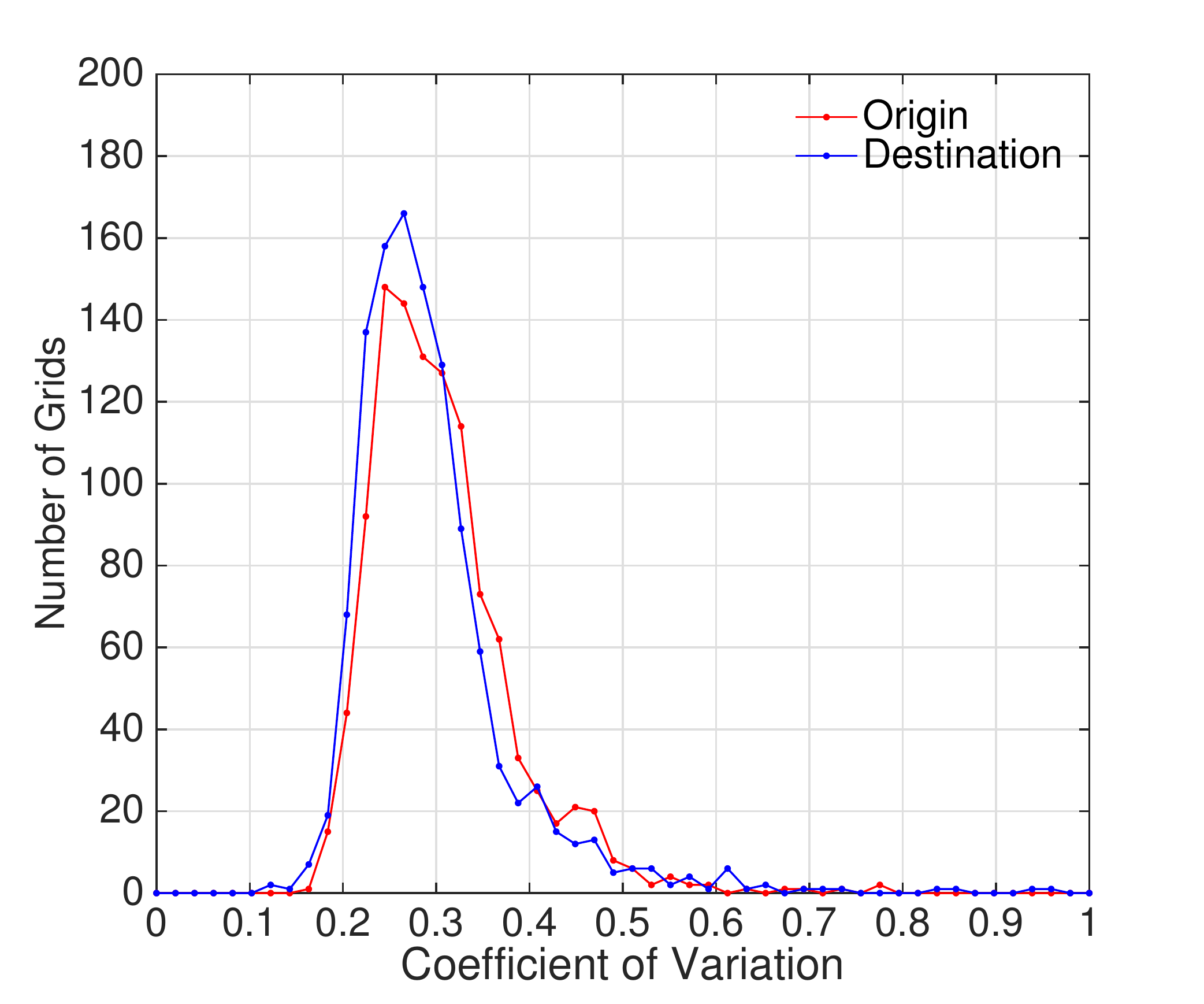}
    \caption{Distributions of coefficient of variation of the origin and destination of trips in grids.}
    \label{fig:Z_D2DInvariantOfGridOD_Day}
\end{figure}

Subsequently, we calculate the following average of $\text{cv}^\Xi_{x,y}(k)$ to measure the multi-day repeatability of a region.
\begin{equation}\label{equ:cv}
	\text{cv}^\Xi_{x,y}= \frac{1}{|K|}\sum_{k\in K} \text{cv}^\Xi_{x,y}(k)
\end{equation}
Figure \ref{fig:Z_D2DInvariantOfGridOD_Day} presents the result and the following can be found. 

\begin{itemize}
\setlength{\itemsep}{0pt}
\setlength{\parsep}{0pt}
\setlength{\parskip}{0pt}	  
	\item[{[8]}] The multi-day repeatability of regional origin and destination numbers is at [0.2 0.4], measured using $\text{cv}^\Xi_{x,y}$ in Equation \ref{equ:cv}. The peak is at 0.25 and few is larger than 0.5.
\end{itemize}

\subsection{Classification of regions}

In this subsection, we classify city regions (i.e., grids here) based on the moment when the peaks of the number of the origin or destination points appear.
First, we identify the peak of a time series as follows. 
Employ a time window with width $\phi$ ($\phi$ should be odd), denoted by $w_1, w_2, ..., w_R$, and move the window from the beginning to the ending of a time series.
A peak, i.e., $\mu^\Xi_{x,y}(w_{\frac{1}{2}\phi})$, is identified when the following two conditions are satisfied.
\begin{equation}\label{equ:PeakCondition}
	\begin{cases}
		 \mu^\Xi_{x,y}(w_1) < \mu^\Xi_{x,y}(w_2) ... < \mu^\Xi_{x,y}(w_{\frac{1}{2}\phi}) >... \mu^\Xi_{x,y}(w_{\phi-1}) > \mu^\Xi_{x,y}(w_{\phi})\\
		 \mu^\Xi_{x,y}(w_{\frac{1}{2}\phi})  > \theta\cdot \overline\mu^\Xi_{x,y}(k) \\
	\end{cases}
\end{equation}
where $\overline\mu^\Xi_{x,y}(k)$ is the mean of $\mu^\Xi_{x,y}(k)$, $k\in{0, 1, ..., 23}$; 
$\theta$ is a coefficient that is larger than 1, indicating that the peak must be $\theta$ times more than the 24-hour average. 
We set $\phi$ and $\theta$ to be 5 and 2 based on our observations of the data, respectively.

Then, we set the following five types for a grid according to the time period when the peak appears.

\begin{itemize}
\setlength{\itemsep}{0pt}
\setlength{\parsep}{0pt}
\setlength{\parskip}{0pt}	  
	\item Type-I: No clear peak.
	\item Type-II.A: A morning peak appears between 6:00 and 11:00
	\item Type-II.B: A noon-and-afternoon peak appears between 11:00 and 16:00
	\item Type-II.C: An evening peak appears between 16:00 and 21:00
	\item Type-II.D: A night peak appears between 21:00 and 5:00
\end{itemize}

\begin{figure}[!htbp]
    \centering
    \subfigure[Origin]{
    \includegraphics[width=3.1in]{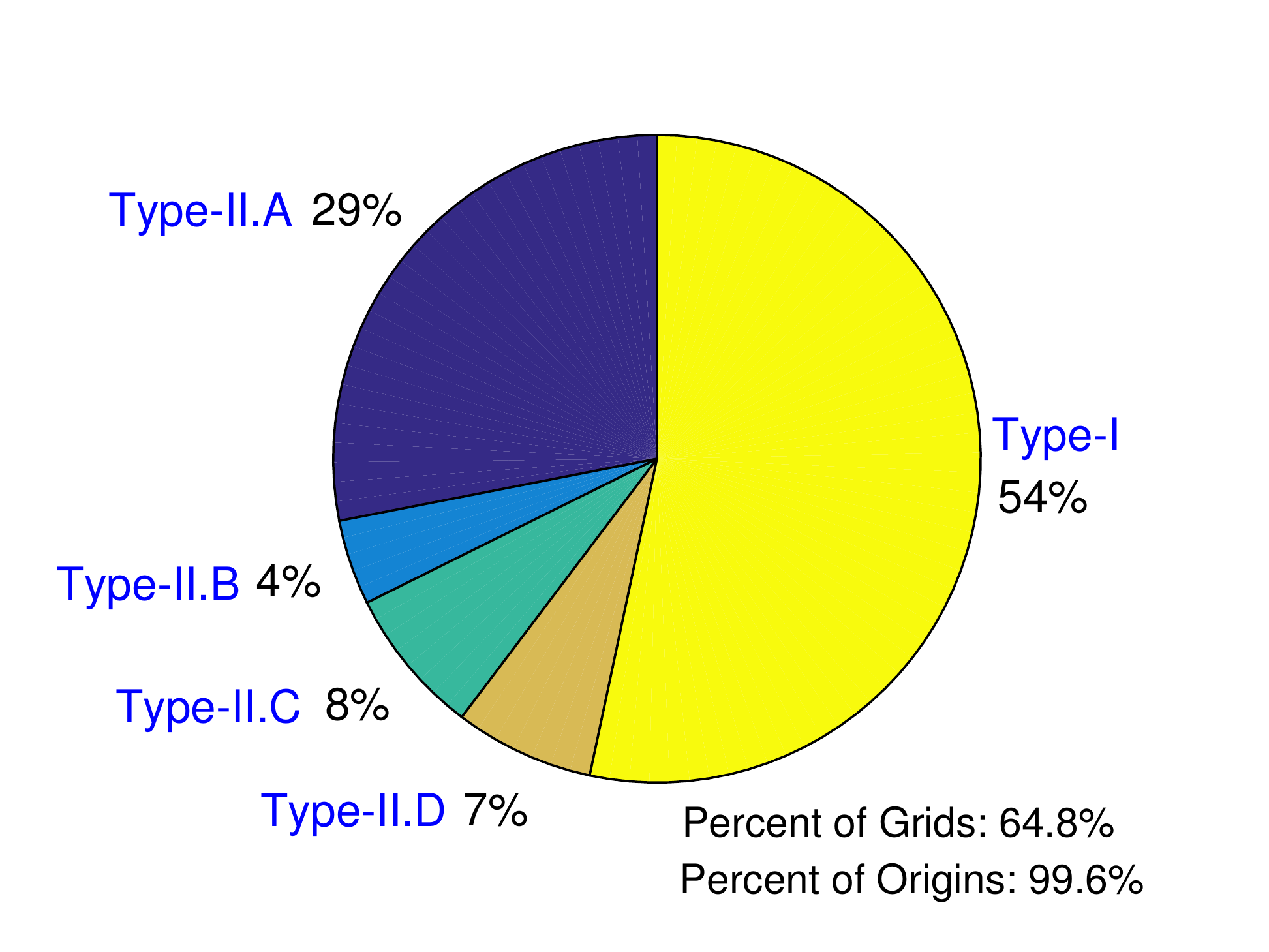}}
    \hspace{-0.4cm}
    \subfigure[Destination]{
    \includegraphics[width=3.1in]{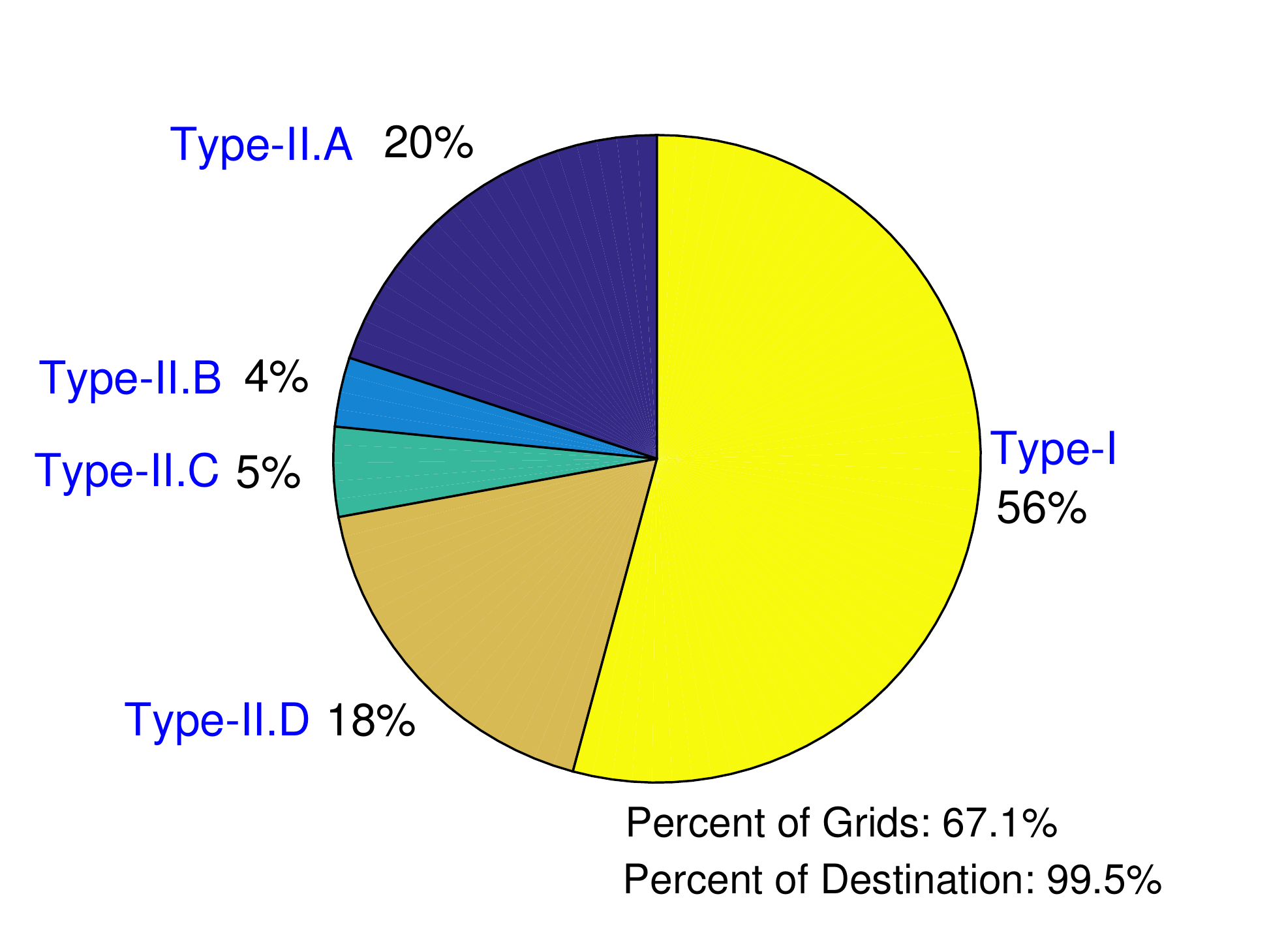}}    
    \caption{Classification results of grids.}
    \label{fig:Z_ClassifyGridTypes}
\end{figure}

As done in Section \ref{sec:repeatability}, 
we remove the grids in which the average number of the origin or destination points during all weekdays is less than 10.
The classification results are plotted in Figure \ref{fig:Z_ClassifyGridTypes} and we have the following observations.

\begin{itemize}
\setlength{\itemsep}{0pt}
\setlength{\parsep}{0pt}
\setlength{\parskip}{0pt}	  
	\item[{[9]}] Most grids have no clear peaks (i.e., Type-I). 
	Considering the fact that we have removed the grids without sufficient origin or destination points, 
	we assert that a large proportion (more than 50\%) of the grids within a weekday have relatively stable demands. 

	\item[{[10]}] Only 4\%$\sim$5\% of the grids show clear noon-and-afternoon and evening peaks (Type-II.B and Type-II.C), 
	unveiling that few grids have suddenly increased-and-dropped trip demand in the noon-and-afternoon and evening.

	\item[{[11]}] The second largest proportion of the grids are in Type-II.A (29\% for origin and 20\% for destination, respectively), i.e., a morning peak. 
	Those grids are active for generating or receiving traffic demand in the morning.

	\item[{[12]}] Type-II.D grids for origins are less than that for destinations, 
	i.e., during the midnight more places become active as travel destinations than as origins.
	It is related to the fact that more activities are back-home at midnight.
\end{itemize}

\section{Driver's perspective: characterization of ride-hailing drivers}\label{sec:Micro}

\begin{figure}[!htbp]
    \centering
    \subfigure[Driver 1]{
    \includegraphics[width=2.2in]{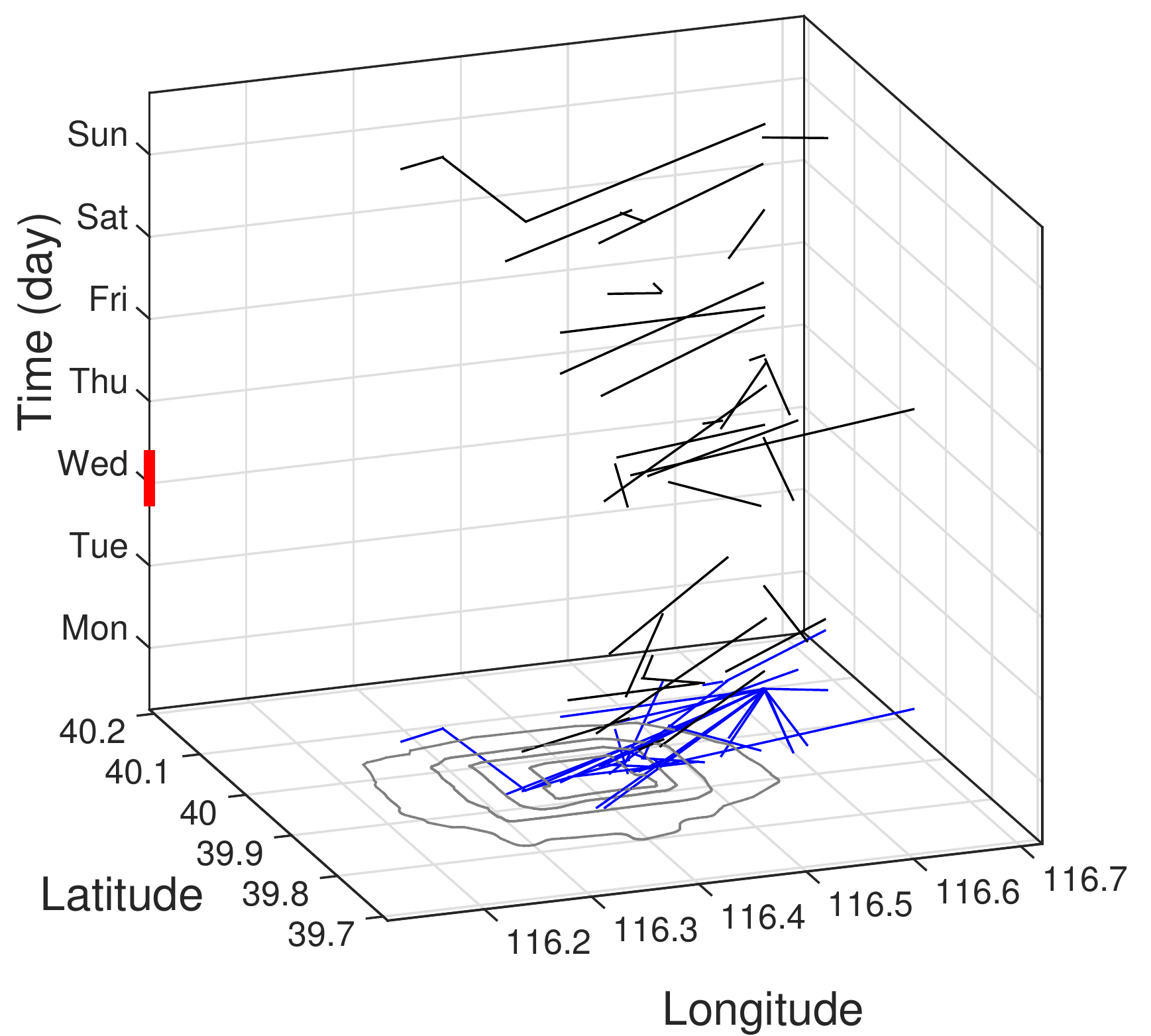}}
    \subfigure[Driver 2]{
    \includegraphics[width=2.2in]{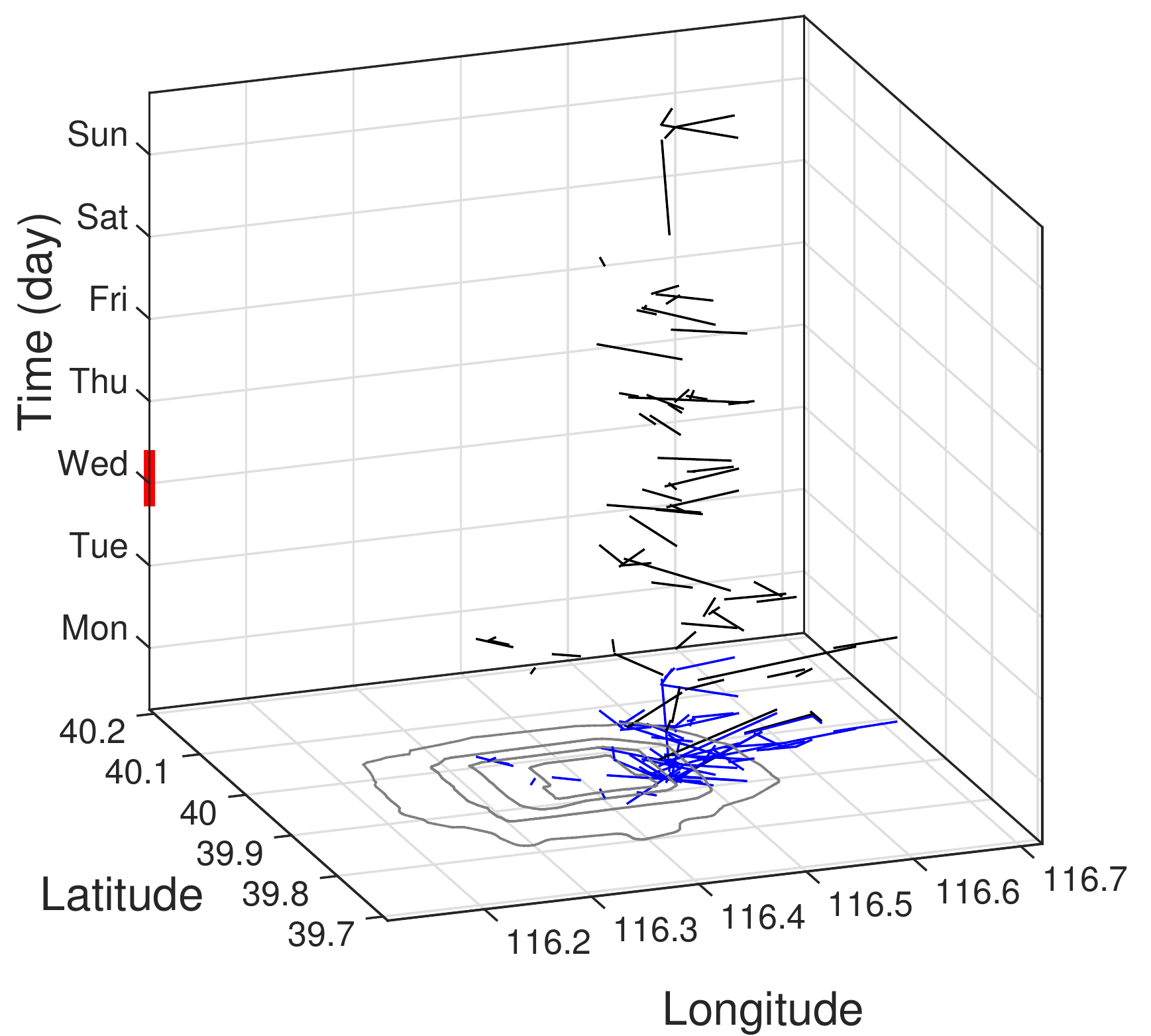}}
    \subfigure[Driver 3]{
    \includegraphics[width=2.2in]{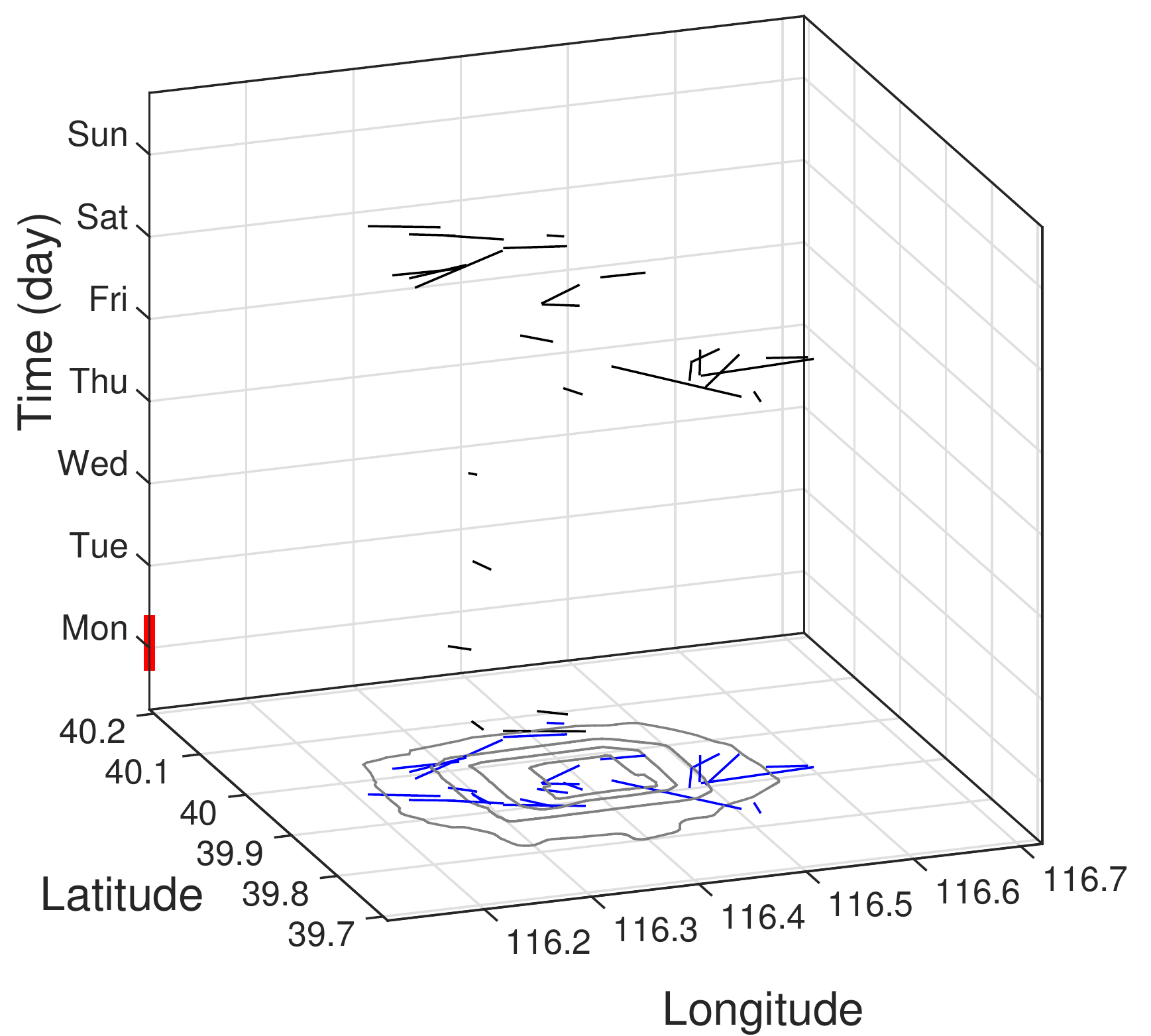}}
    \subfigure[Driver 4]{
    \includegraphics[width=2.2in]{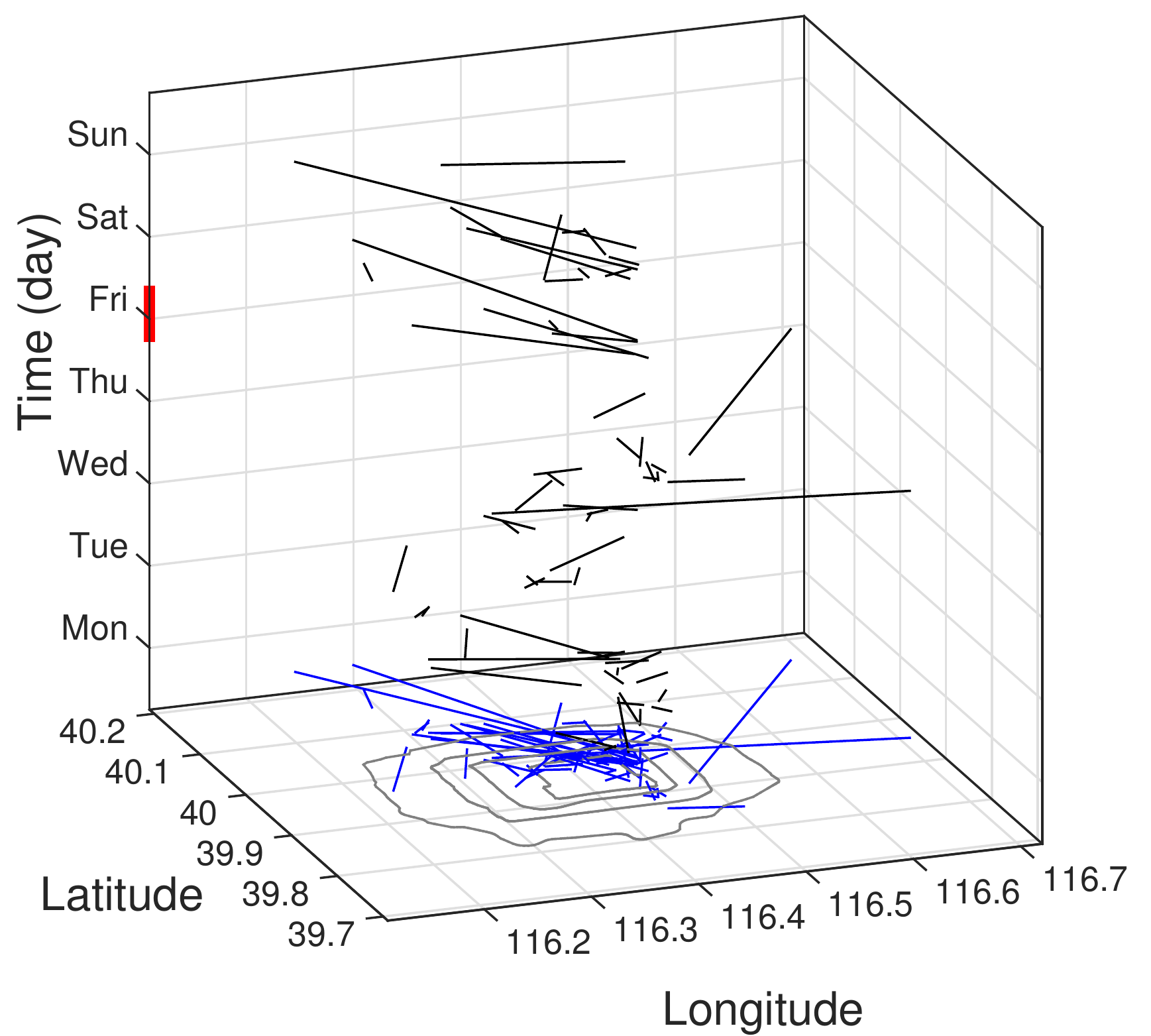}}
    \subfigure[Driver 5]{
    \includegraphics[width=2.2in]{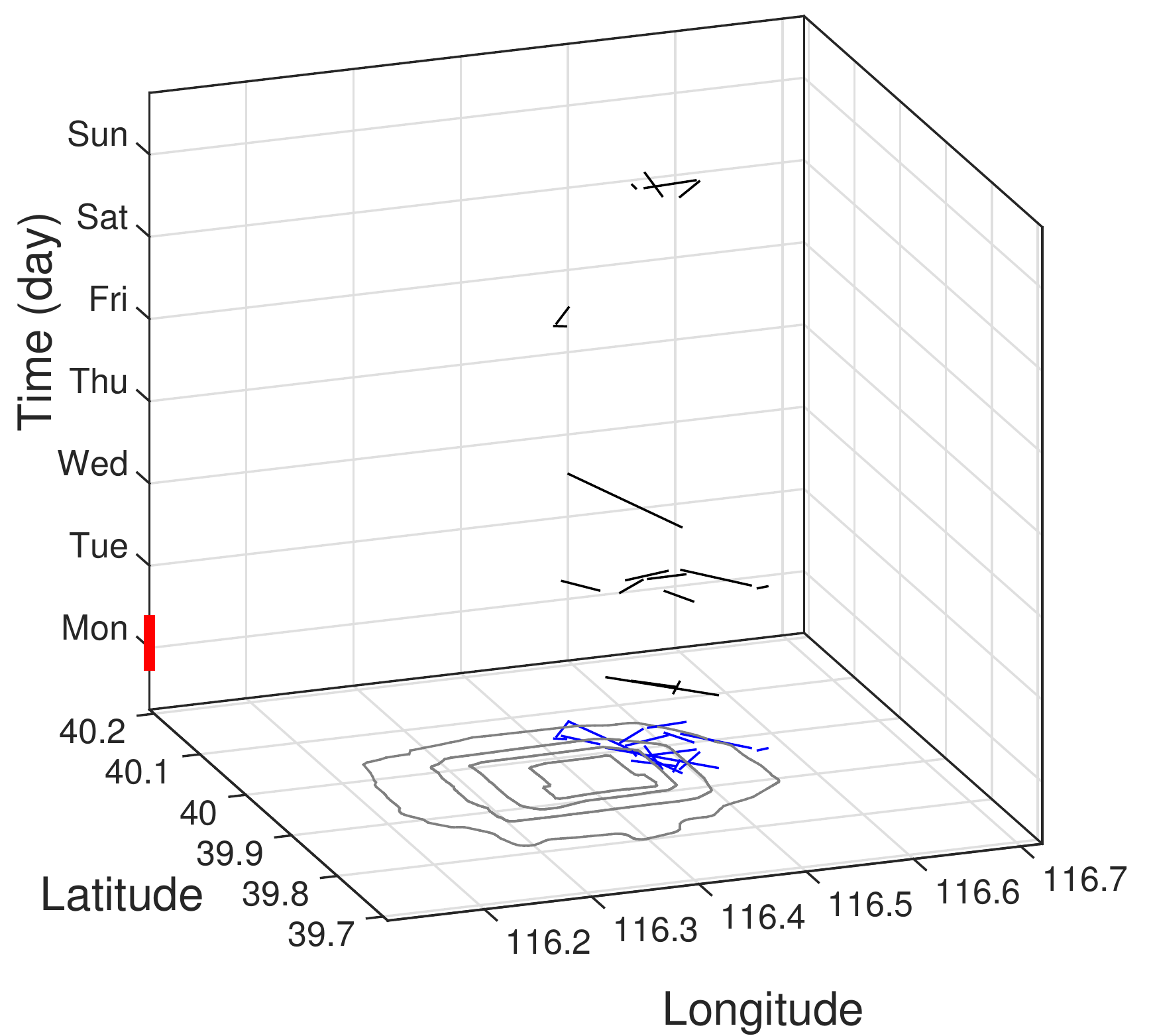}}
    \subfigure[Driver 6]{
    \includegraphics[width=2.2in]{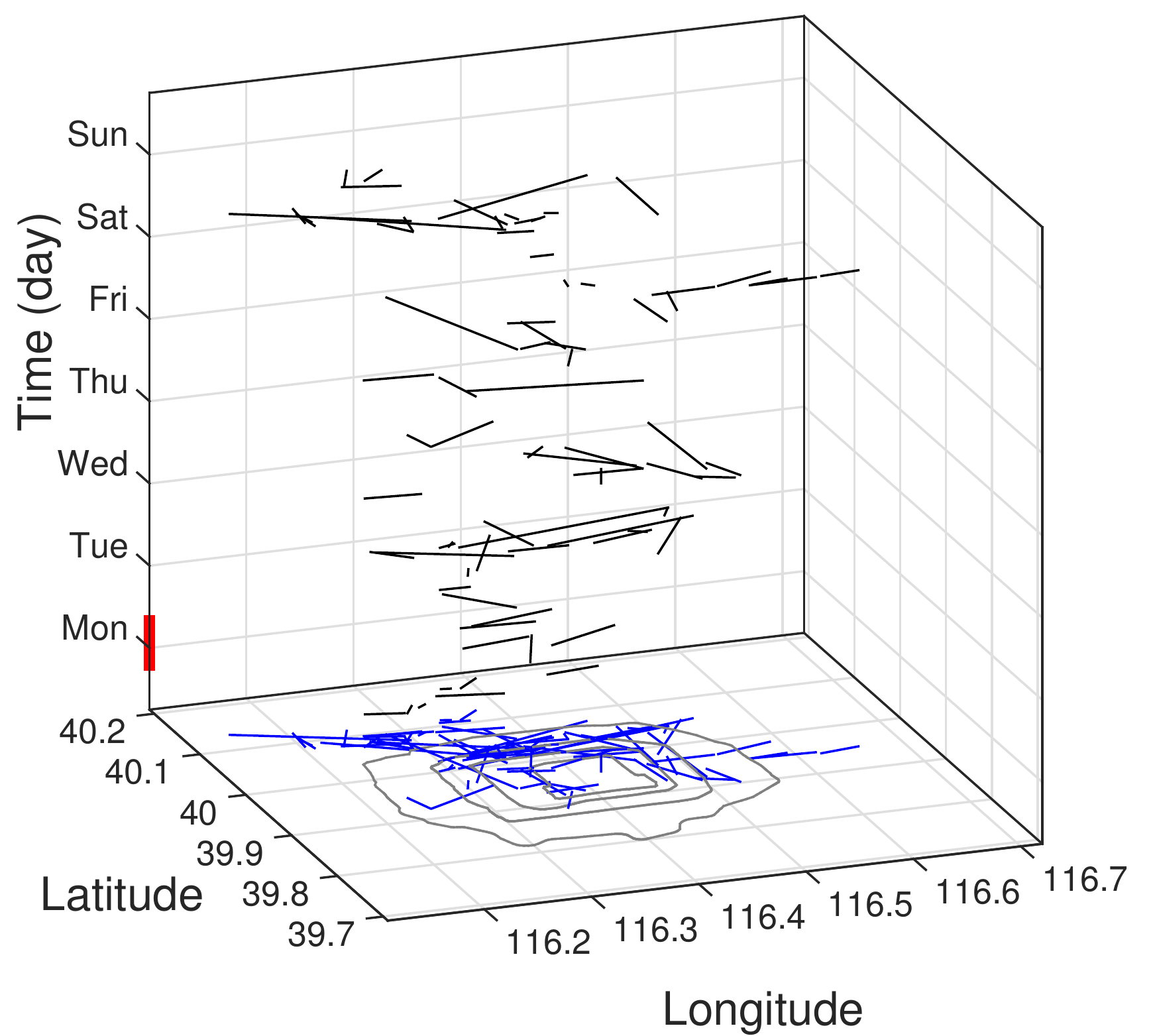}}
    \subfigure[Driver 7]{
    \includegraphics[width=2.2in]{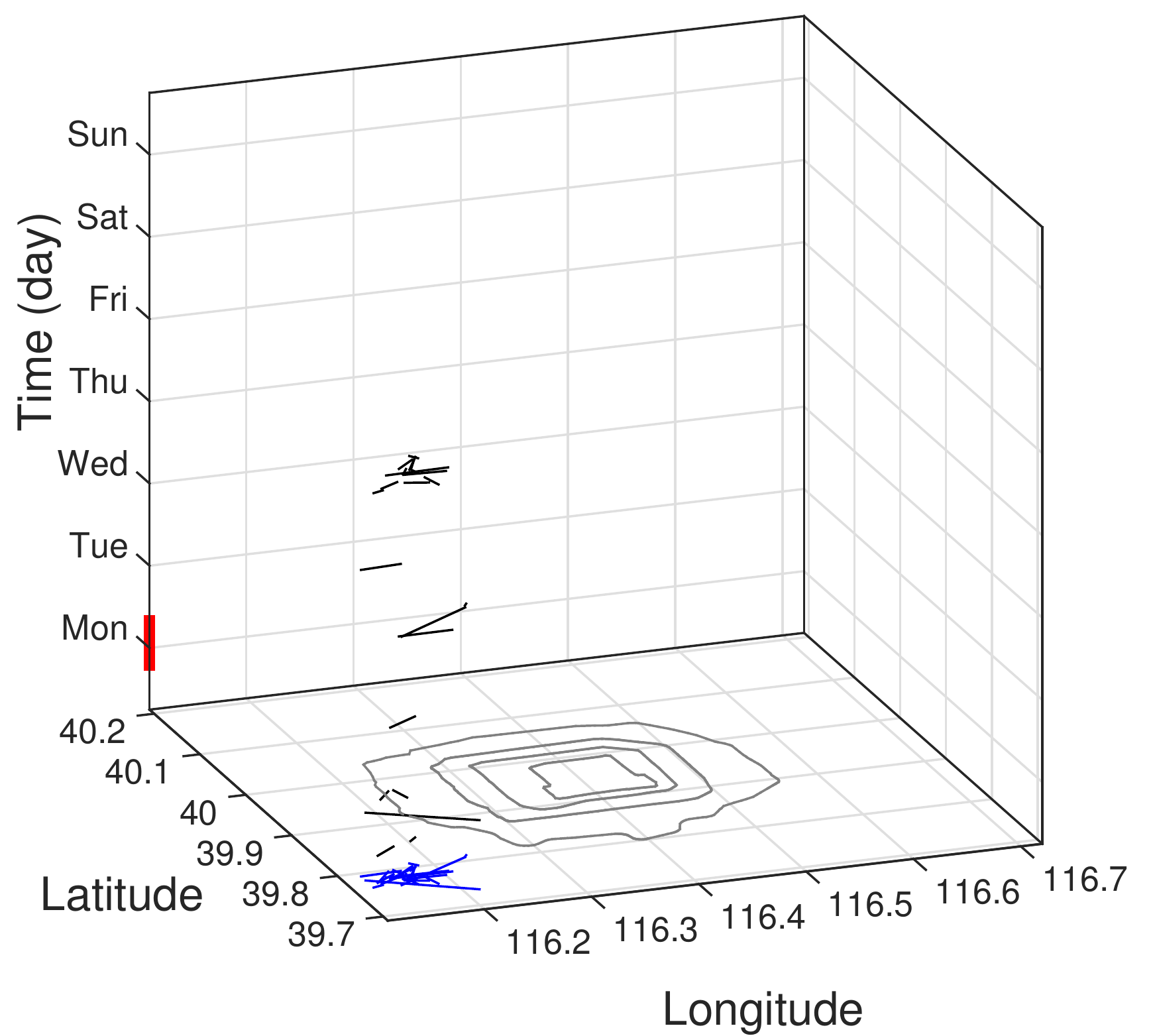}}
    \subfigure[Driver 8]{
    \includegraphics[width=2.2in]{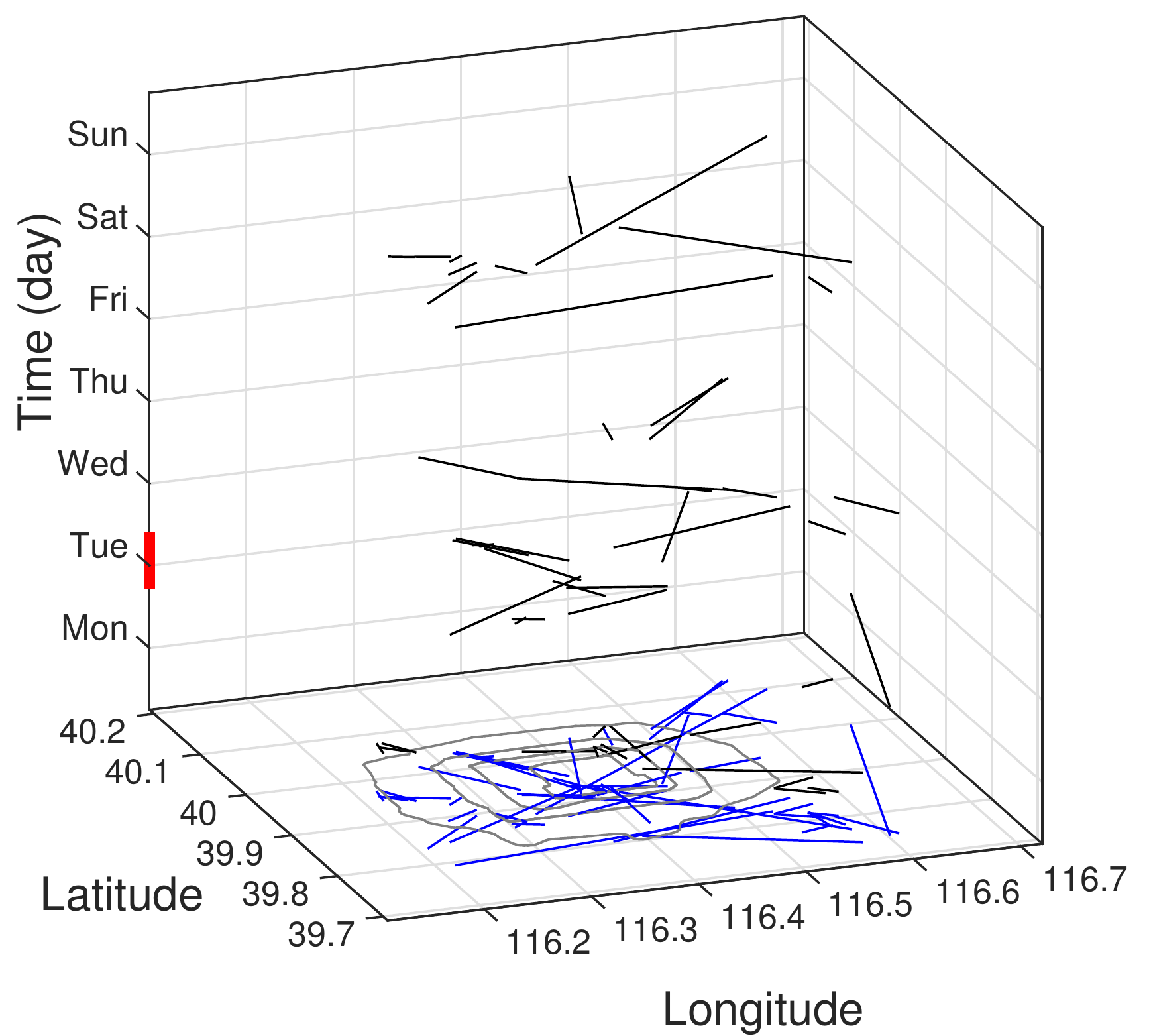}}
    \subfigure[Driver 9]{
    \includegraphics[width=2.2in]{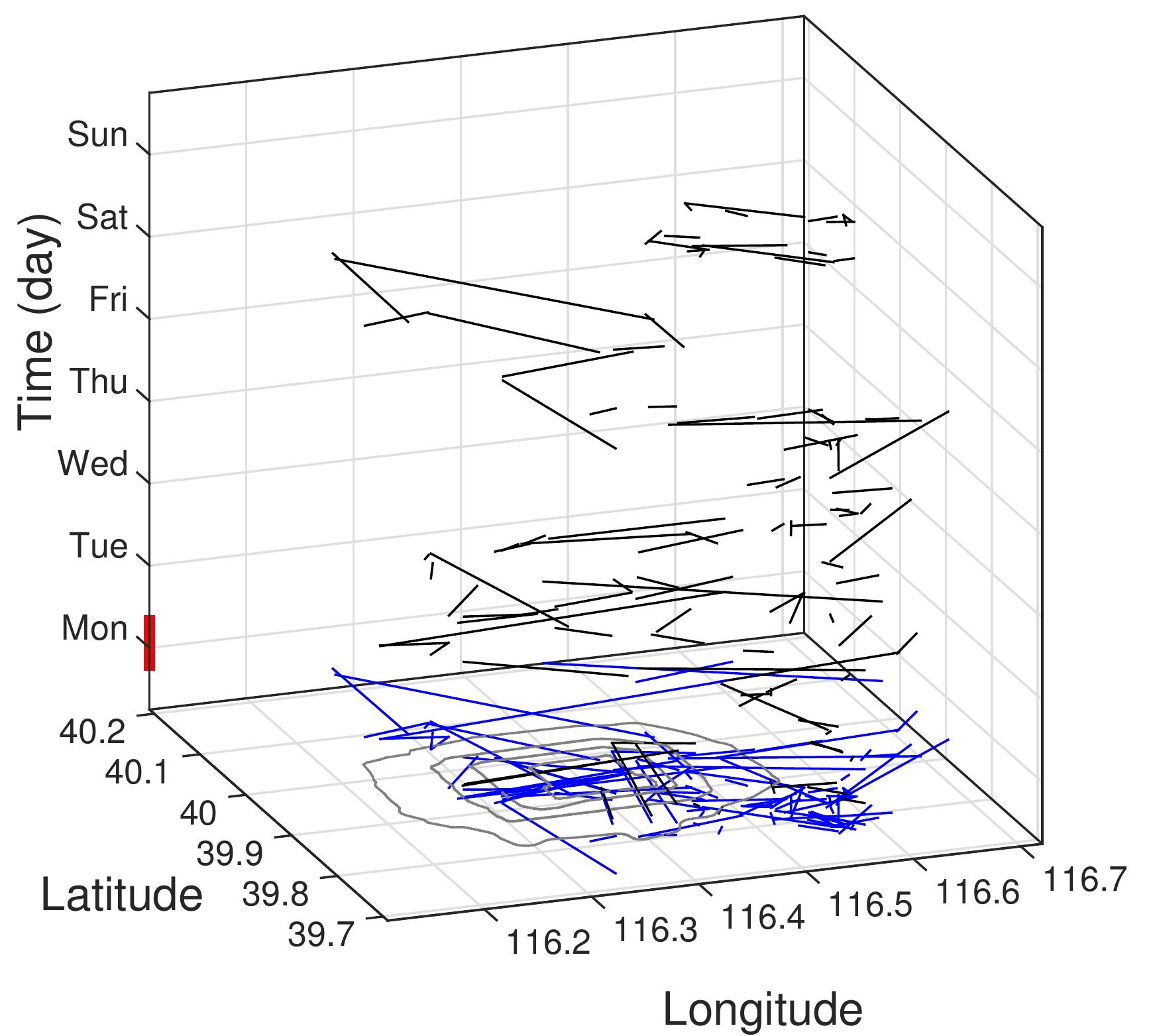}}
    \caption[Individual drivers' trips in a week.]
    {(9 randomly selected) individual drivers' trips in a week.
    Black line: the origin and destination link of a trip;
    Blue line: the spatial projection of the black line;
    Red line on the time-axis: an indication of the day that the vehicle is prohibited to run inside Ring 5 of Beijing due to the license plate restriction policy \citep{Wang2014g,Jia2016a}.}
    \label{fig:Z_VehicleTripsIn3D}
\end{figure}

In this section, we turn our lens to an individual ride-hailing driver and his/her multi-day activities. 
First, to glance, we randomly select 9 drivers and plot their trips in a week in Figure \ref{fig:Z_VehicleTripsIn3D}.
Temporally, it can be seen that some drivers work almost every day in a week (Drivers 1, 2, 4, 6, and 9), while some only on certain days (Drivers 3 (weekends), 5, 7 (weekdays) , and 8).
Spatially, some drivers provide service within large spatial regions (Drivers 1, 3, 4, 6, 8, and 9), while some within small regions (Drivers 2, 5, and 7).
The observations imply the existence of drivers' preferences for providing ride-hailing services.

\subsection{Spatial and temporal distributions of ride-hailing trips}

\begin{figure}[!htbp]
    \centering
    \subfigure[Trip number]{
    \includegraphics[width=3.1in]{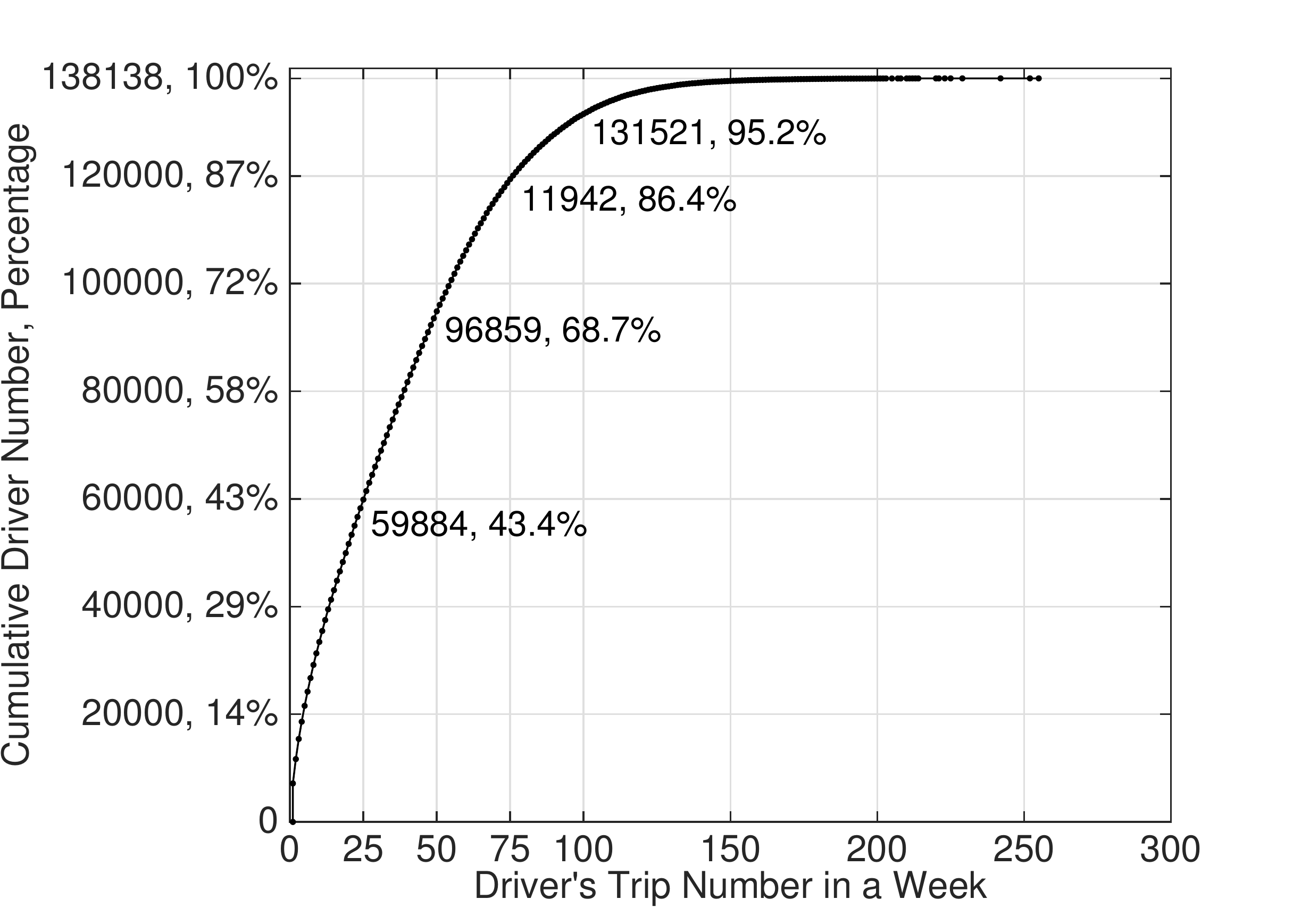}}
    \subfigure[Trip duration]{
    \includegraphics[width=3.1in]{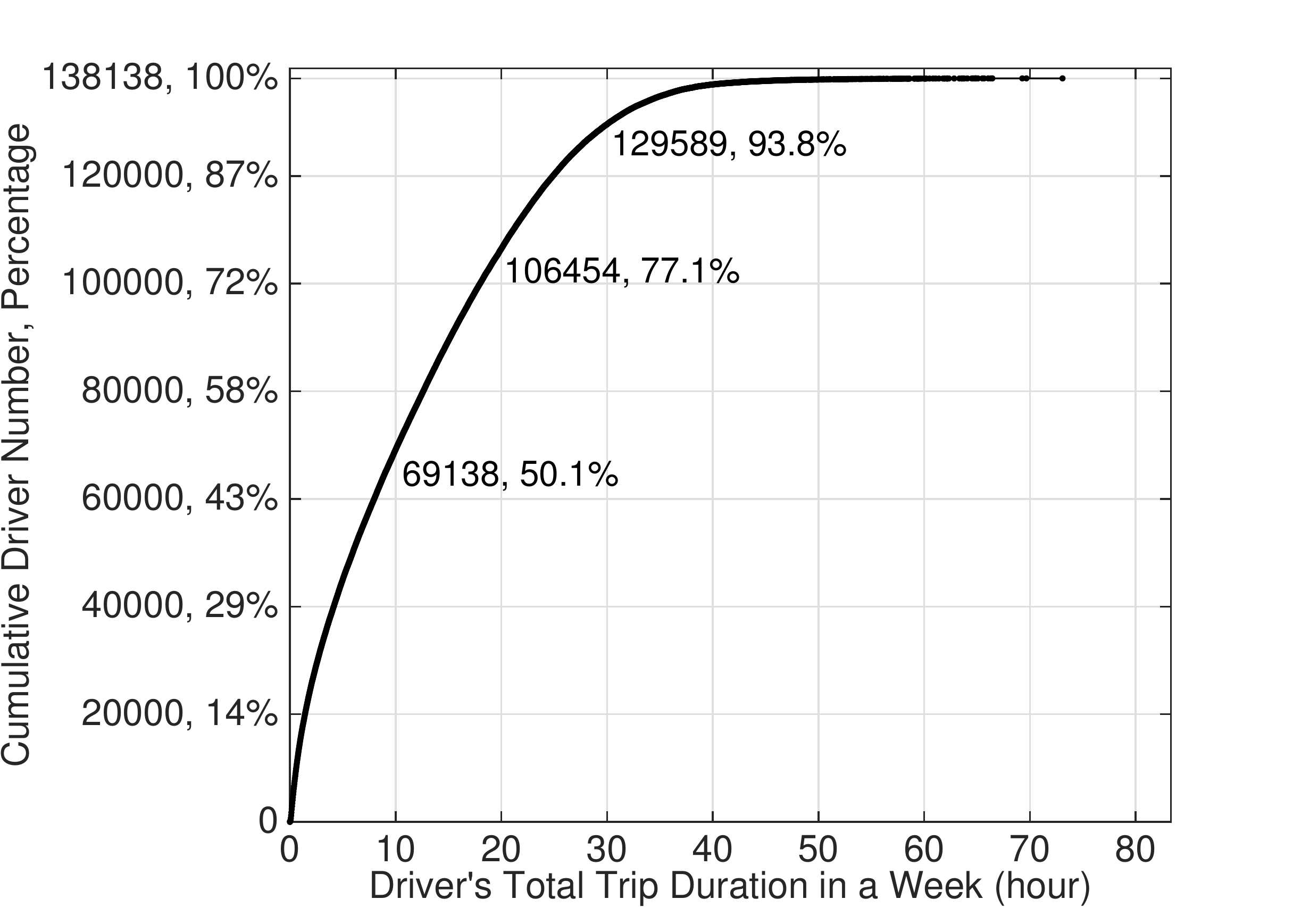}}    
    \caption{Distribution of all drivers' trips in a week.}
    \label{fig:Z_Distribution_TripNumber}
\end{figure}

To acquire the proportions of part- and full-time ride-hailing drivers, we first measure the empirical cumulative distribution of all drivers' trip numbers in a week.
Figure \ref{fig:Z_Distribution_TripNumber} presents the results and we have the following observations. 

\begin{itemize}
\setlength{\itemsep}{0pt}
\setlength{\parsep}{0pt}
\setlength{\parskip}{0pt}	  
	\item[{[13]}] A large part of ride-hailing drivers are part-time drivers. 
	59,884 (43.4\%) drivers take less than 25 trips in a week (Figure \ref{fig:Z_Distribution_TripNumber}(a)), 
	and 69,138 (50.1\%) drivers' total trip durations are less than 10 hours in a week (Figure \ref{fig:Z_Distribution_TripNumber}(b)).
	Although we don't have a universal criterion to distinguish part- and full-time ride-hailing drivers, 
	we can speculate from the result that at least half of the ride-hailing drivers are part-time drivers. 
	\item[{[14]}] Only a small part of the drivers take ride-hailing services as their full-time jobs.
	Specifically, 6617 (=138,138-131,521, or 4.8\%=100\%-95.2\%) drivers provide more than 100 services in a week (Figure \ref{fig:Z_Distribution_TripNumber}(a)), 
	and 8549 (=138,138-129,589, or 6.2\%=100\%-93.8\%) drivers work more than 30 hours (Figure \ref{fig:Z_Distribution_TripNumber}(b)).
	These drivers are surely full-time drivers who intensively accept ride-hailing requests as their jobs.
\end{itemize}
Those observations are consistent with the findings in the existing studies \citep{Chen2017c,Hall2018}.
Moreover, we obtain the exact percentage based on the overall population for the first time.


\begin{figure}[!htbp]
    \centering
    \subfigure[Trip duration]{
    \includegraphics[width=3.1in]{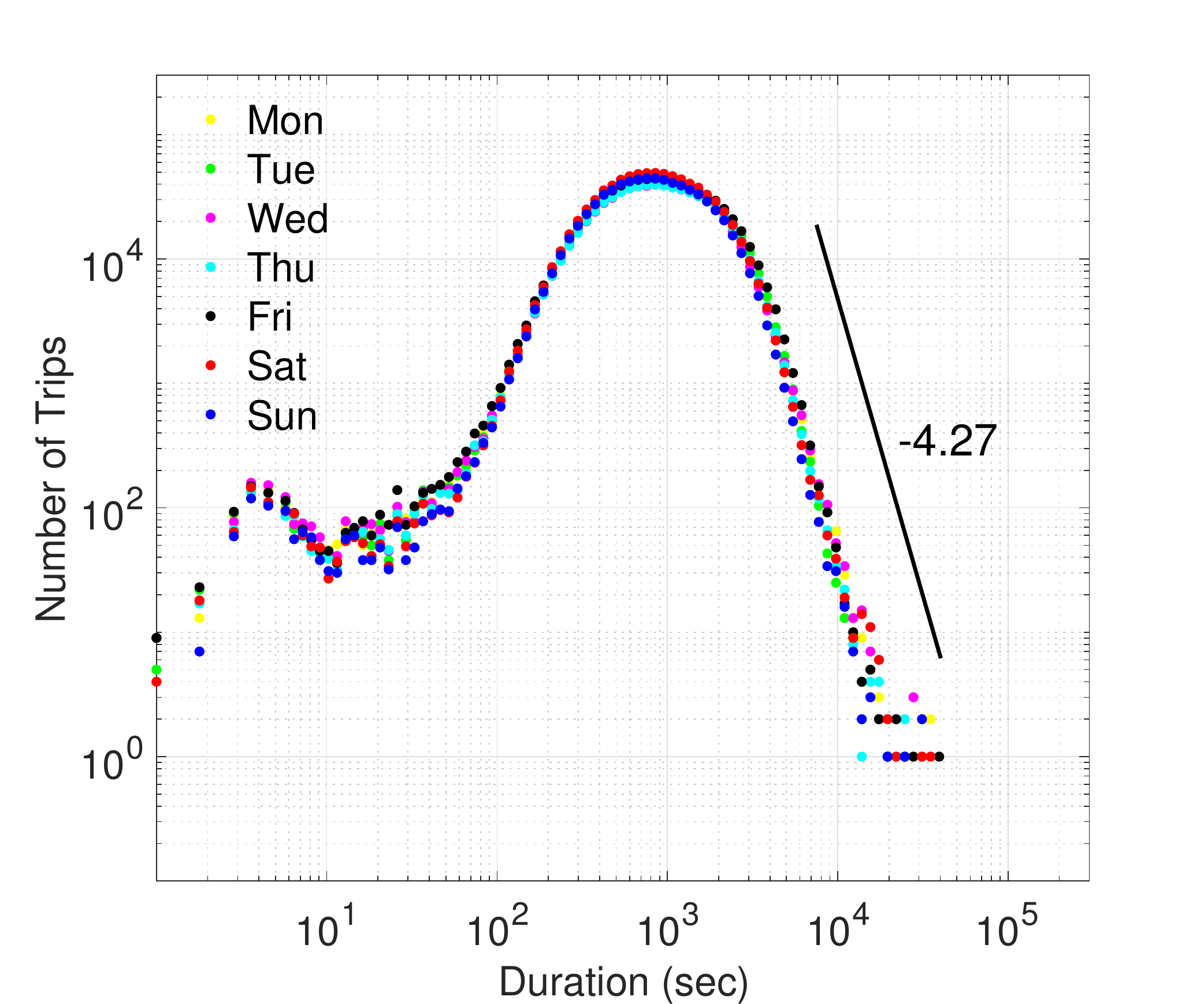}}
    \subfigure[Trip displacement]{
    \includegraphics[width=3.1in]{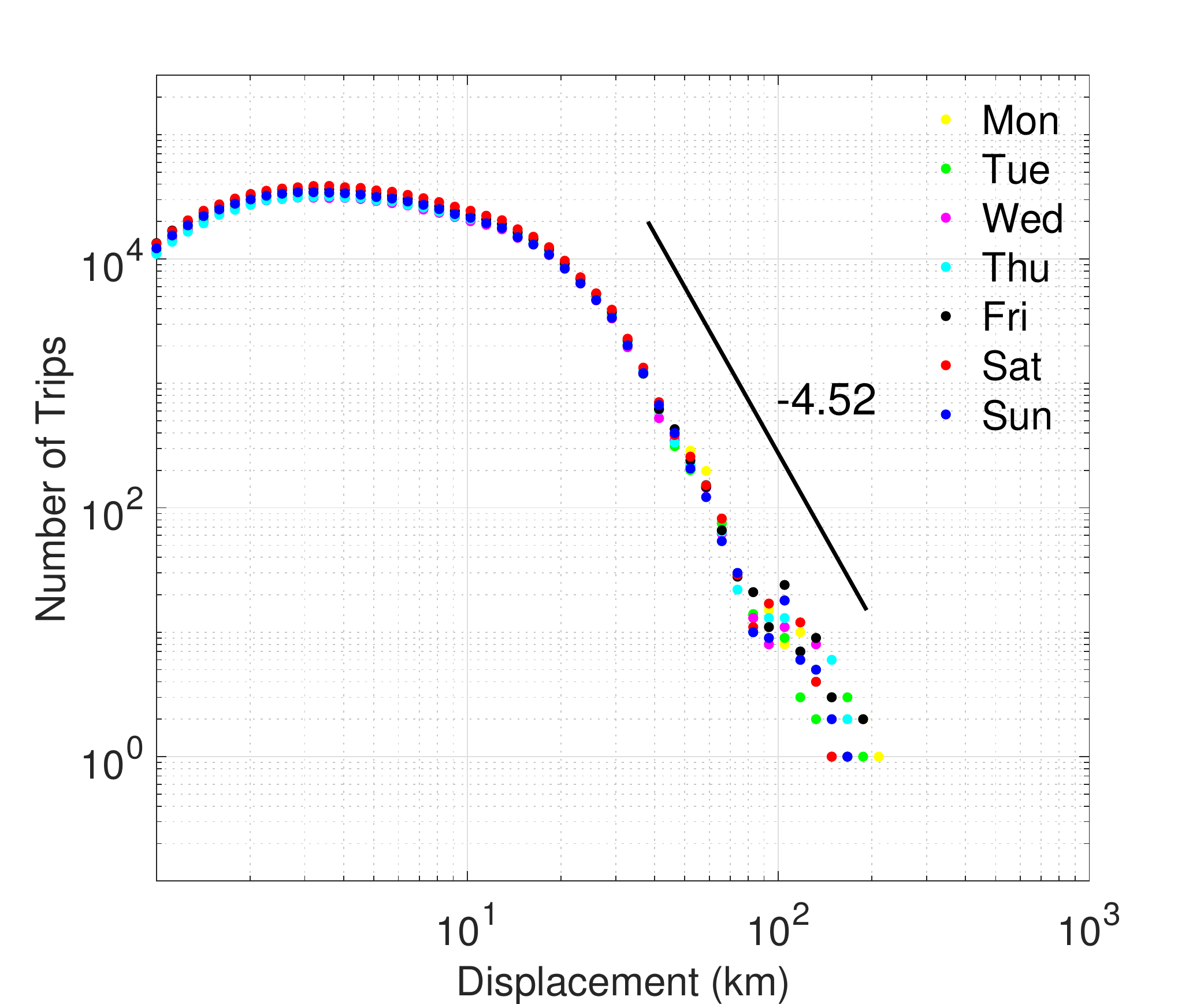}}    
    \caption{Logarithmically binned distributions of all trips on different days.}
    \label{fig:Z_LogLog_DailyTripDuriation}
\end{figure}

In addition to the empirical cumulative distribution, we calculate the durations ($T_{ij}$) and displacements ($L_{ij}$) of all trips on different days following Equation \ref{equ:TL}.
The results are presented in a log-log plot in Figure \ref{fig:Z_LogLog_DailyTripDuriation} and we have the following observations.

\begin{itemize}
\setlength{\itemsep}{0pt}
\setlength{\parsep}{0pt}
\setlength{\parskip}{0pt}	  
	\item[{[15]}] The shapes of the distributions on different days are similar, indicating the high repeatability of the ride-hailing trip duration and displacement distribution on multiple days.
	
	\item[{[16]}] The durations of most trips are between 300 sec and 3000 sec (i.e., between 5 min and 50 min), and the displacements of most trips are less than 20 km. 
	Those values reflect (or depend on) the size of the Beijing city.
	
	\item[{[17]}] The distribution tails of both the trip duration and displacement follow the power law with the similar exponent values of -4.27 and -4.52 \textcolor{black}{(Table \ref{tab:Hypothesis})}, respectively, indicating the existence of occasional long-distance trips.
	The power-law observation is largely different from the exponential distribution exhibited by taxis \citep{Liang2012}.
	The observation that the exponent values of two distributions are approximately equal implies the positive correlation between the trip duration and displacement. 
\end{itemize}

\begin{table}[htbp]\centering\footnotesize
\textcolor{black}{\setlength{\tabcolsep}{1.5mm}{
    \centering \caption{Kolmogorov-Smirnov test of the distribution tails of trip duration and displacement.}\label{tab:Hypothesis}
    \begin{tabular}{lcccc}
    \toprule
	& NULL HYPOTHESIS:  & power exponent &  $p$-value &   confidence coefficient\\
	& Follow power distribution & ($b$) &  & \\	
	& ($y=a x^b$) &  &  & \\		
    \midrule
	Trip duration ($>$6000 sec) & Accept & -4.27  & 0.7710 &  0.05 \\
	Trip displacement ($>$40 km) & Accept & -4.52  & 0.6725 &  0.05  \\	
    \bottomrule
    \end{tabular}}}
\end{table}

\subsection{Temporal characterization}

This subsection focuses on the characterization of a ride-hailing driver from the temporal perspective, i.e., which time period is preferred by a driver to provide ride-hailing services.

First, we define ``{\it work}" for a ride-hailing driver as follows: 
{\it a work consist of a series of successive trips and the time interval between two successive trips is short (qualified by using a threshold).}
When a driver is working, the driver intensively provides service within a time period. 
Let $\varepsilon$ be the threshold and set it as $\varepsilon=3600$ sec with the consideration that a driver will take another ride-hailing order within one hour if the driver is working.
Mathematically, driver $i$'s trips during work $r$ is written as follows.
\begin{equation}
	W_{ir}=\{ TR_{ij}\ | \ t^O_{i(j+1)} - t^D_{ij}< \varepsilon \}
\end{equation}


To temporally characterize ride-hailing drivers' working time preferences, we divide a day into four periods:
$P^\kappa_1$=[0:00, 6:00], 
$P^\kappa_2$=[6:00, 12:00], 
$P^\kappa_3$=[12:00, 18:00], and 
$P^\kappa_4$=[18:00, 24:00], where $\kappa=1$ indicates that the day is a weekday and $\kappa=2$ a weekend.
Then, calculate the percentage of the working time in each period as follows.
\begin{equation}
	\delta_{il} = \frac{|W^T_{ir} \bigcap P^\kappa_l |}{|P^\kappa_l|}	
\end{equation}
where $W^T_{ir}$ is the time span of driver $i$'s work $r$; 
$l=1,2,3,4$ indicates the four periods.
The total length of the periods on the five weekdays is $|P^1_l|= 6\times 5= 30$ h and that on the two weekends is $|P^2_l|= 6\times 2= 12$ h.

\begin{figure}[!htbp]
    \centering
    \includegraphics[width=3.8in]{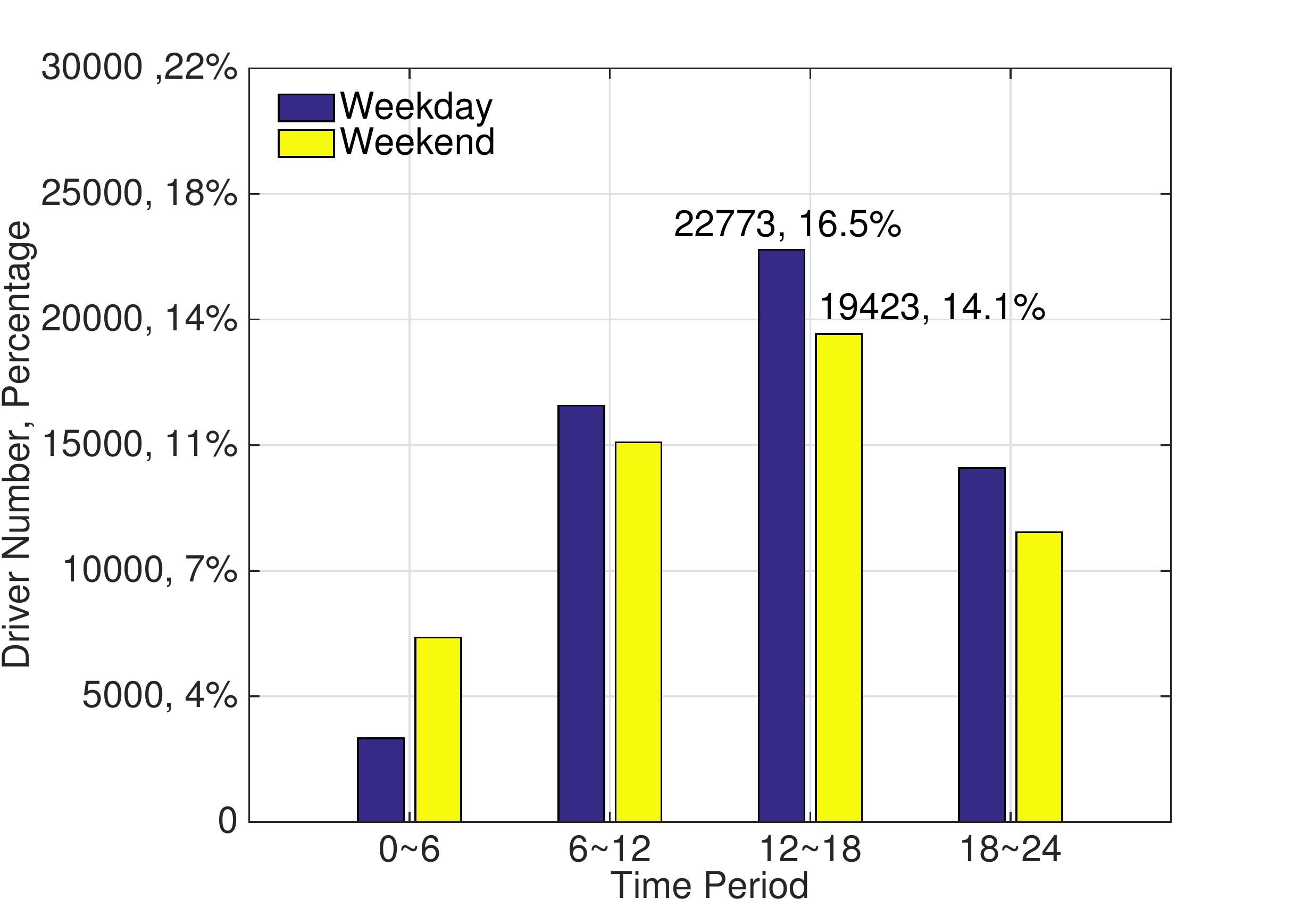}
    \caption[The number of the drivers contained in time periods.]
    {The number of the drivers contained in each time period $P_{l}^\kappa$ ($\delta^*=$ 40\%).}
    \label{fig:Z_Cluster_WorkTime}
\end{figure}

Then, we say that driver $i$ usually works in period $P_{l}^\kappa$ if $\delta_{il}>\delta^*$ where $\delta^*$ is a threshold.
Figure \ref{fig:Z_Cluster_WorkTime} presents the temporal characterization resulting from the perspective of working periods after setting $\delta^*=$ 40\% and we have the following findings.

\begin{itemize}
\setlength{\itemsep}{0pt}
\setlength{\parsep}{0pt}
\setlength{\parskip}{0pt}	  
	\item[{[18]}] $P^\kappa_3$=[12:00, 18:00] is the period that contains the largest number of frequently working drivers.
	The numbers are approximately 22,773 (16.5\% of total drivers) and 19,423 (14.1\%) on weekdays and weekends, respectively.
	
	\item[{[19]}] The second and third preferred working periods are $P^\kappa_2$=[6:00, 12:00] and $P^\kappa_4$=[18:00, 24:00], respectively.
	
	\item[{[20]}] Only for $P^\kappa_1$=[0:00, 6:00], the working drivers on weekends are more than those on weekdays, indicating that more drivers select to continuously work at the weekend night to serve people's demand of night-life traveling. 
	
\end{itemize}

Moreover, we characterize working time preferences from the individual perspective, i.e., drivers' preferred working periods. 
Table \ref{tab:Z_Ranking_WorkTime} presents the results and we have the following observations.

\begin{itemize}
\setlength{\itemsep}{0pt}
\setlength{\parsep}{0pt}
\setlength{\parskip}{0pt}	  
	\item[{[21]}] Most drivers who frequently work during a day only work in a time period (i.e., 6 hours; see Ranks 1 to 7, 25,884 drivers in total, i.e., 18.7\%), implying that the ride-hailing drivers usually don't work as long as those for other fixed-time jobs (e.g. working in an office), although they also frequently work in a fixed time period. 
	
	\item[{[22]}] The second most of the drivers are the ones working in two periods (Ranks 8, 10, 11, 13, and 14; 7,729 drivers in total, i.e., 6\%) and they usually work during 6:00 and 18:00 or during 12:00 and 24:00. 
	
	\item[{[23]}] Like [20], few drivers select to work during 0:00 and 6:00 on weekdays, while more (2529) drivers work in the same period on weekends.
\end{itemize}

\begin{table}[!htbp]\centering\footnotesize
\setlength{\tabcolsep}{3mm}{
\caption{The number of the drivers who work in each time period ($\delta^*=$ 40\%).}\label{tab:Z_Ranking_WorkTime}
\begin{tabular}{ccccccccccc}
\toprule
 \multicolumn{4}{c}{Weekday (Time Period)} & & \multicolumn{4}{c}{Weekend (Time Period)} &  Driver Number & Rank\\
\cline{1-4}\cline{6-9}
0$\sim$6 & 6$\sim$12 & 12$\sim$18 & 18$\sim$24 &  & 0$\sim$6 & 6$\sim$12 & 12$\sim$18 & 18$\sim$24 & (Percentage) &  \\
\midrule
 &  & \checkmark &  & &  &  &  &  & 5789 (4.2\%) & 1 \\ \rowcolor{light-gray} 
 &  &  &  & &  &  & \checkmark &  & 4485 (3.2\%) & 2 \\  
 & \checkmark &  &  & &  &  &  &  & 3900 (2.8\%) & 3 \\ \rowcolor{light-gray} 
 &  &  & \checkmark & &  &  &  &  & 3364 (2.4\%) & 4 \\  
 &  &  &  & &  & \checkmark &  &  & 3312 (2.4\%) & 5 \\ \rowcolor{light-gray} 
 &  &  &  & & \checkmark &  &  &  & 2529 (1.8\%) & 6 \\  
 &  &  &  & &  &  &  & \checkmark & 2505 (1.8\%) & 7 \\ \rowcolor{light-gray} 
 & \checkmark & \checkmark &  & &  &  &  &  & 2372 (1.7\%) & 8 \\  
 &  & \checkmark &  & &  &  & \checkmark &  & 2105 (1.5\%) & 9 \\ \rowcolor{light-gray} 
 & \checkmark & \checkmark &  & &  & \checkmark & \checkmark &  & 1427 (1.0\%) & 10 \\  
 &  &  &  & &  & \checkmark & \checkmark &  & 1423 (1.0\%) & 11 \\ \rowcolor{light-gray} 
 & \checkmark &  &  & &  & \checkmark &  &  & 1361 (1.0\%) & 12 \\  
 &  & \checkmark & \checkmark & &  &  &  &  & 1350 (1.0\%) & 13 \\ \rowcolor{light-gray} 
 &  &  &  & &  &  & \checkmark & \checkmark & 1157 (0.8\%) & 14 \\  
\bottomrule
\end{tabular}}  
\end{table}

\subsection{Spatial characterization}

This subsection characterizes a driver from the perspective of his/her activity space.  
To that end, we employ the hierarchical clustering to cluster all origin and destination points of a driver's all trips according to their spatial positions.

The hierarchical clustering is a cluster analysis method, which works 
by first building a cluster tree (a dendrogram) to represent data 
and then cutting the tree to group the data.
When building the cluster tree here, \textcolor{black}{the input is all origin and destination points of a driver in the week,} and the metric of measuring the distance between two points is the distance on the earth surface.
The linkage criteria of point sets (i.e., cluster distance) is the shortest distance between the points in the two sets.
In cutting the tree, a threshold $\beta$ is employed, meaning that the distance between any two final clusters is larger than $\beta$.
Then, we enclose the maximum cluster and calculate its area by using the Delaunay triangulation.

\begin{figure}[!htbp]
    \centering
    \subfigure[Driver 10]{
    \includegraphics[width=2.2in]{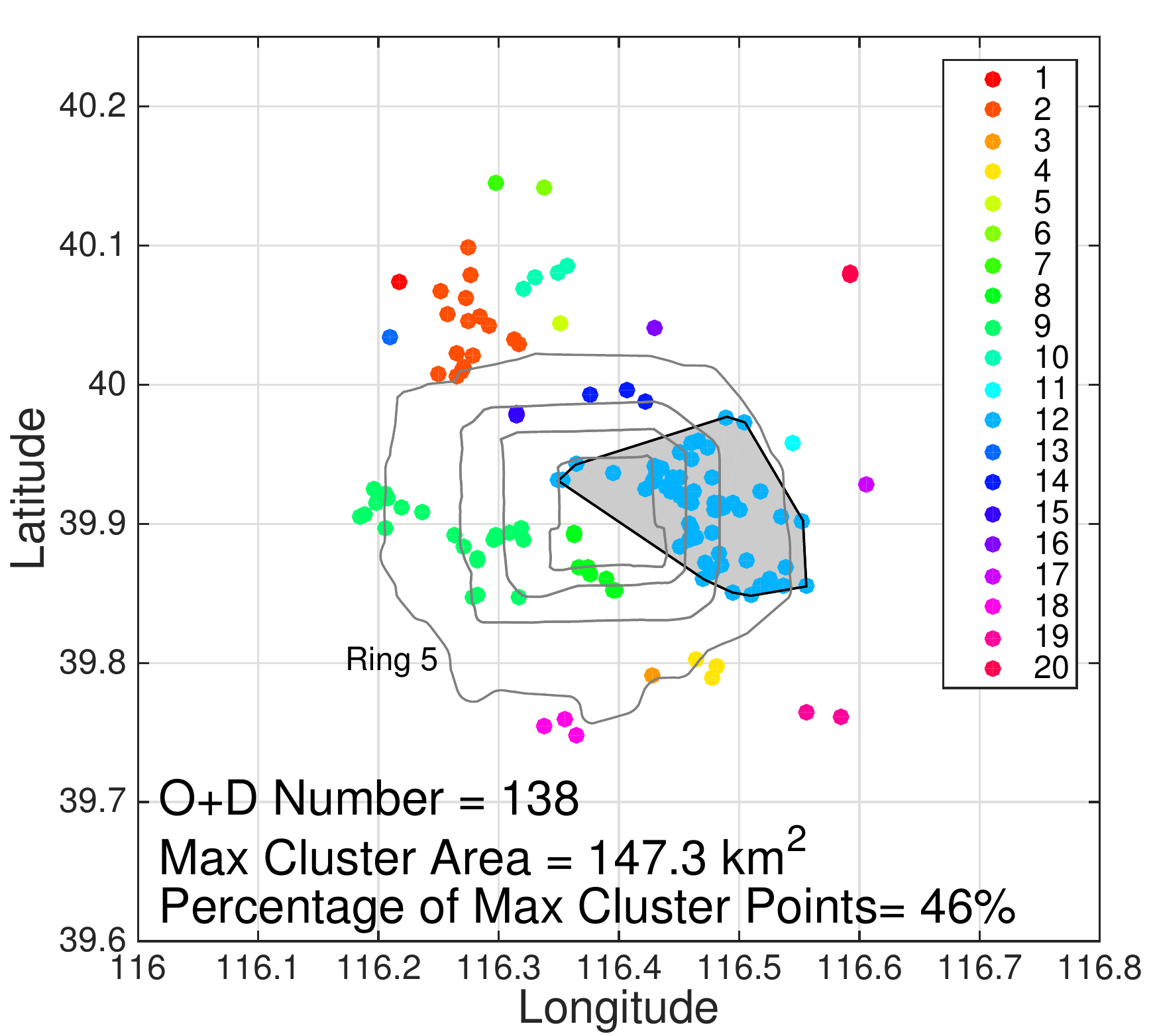}}
    \subfigure[Driver 11]{
    \includegraphics[width=2.2in]{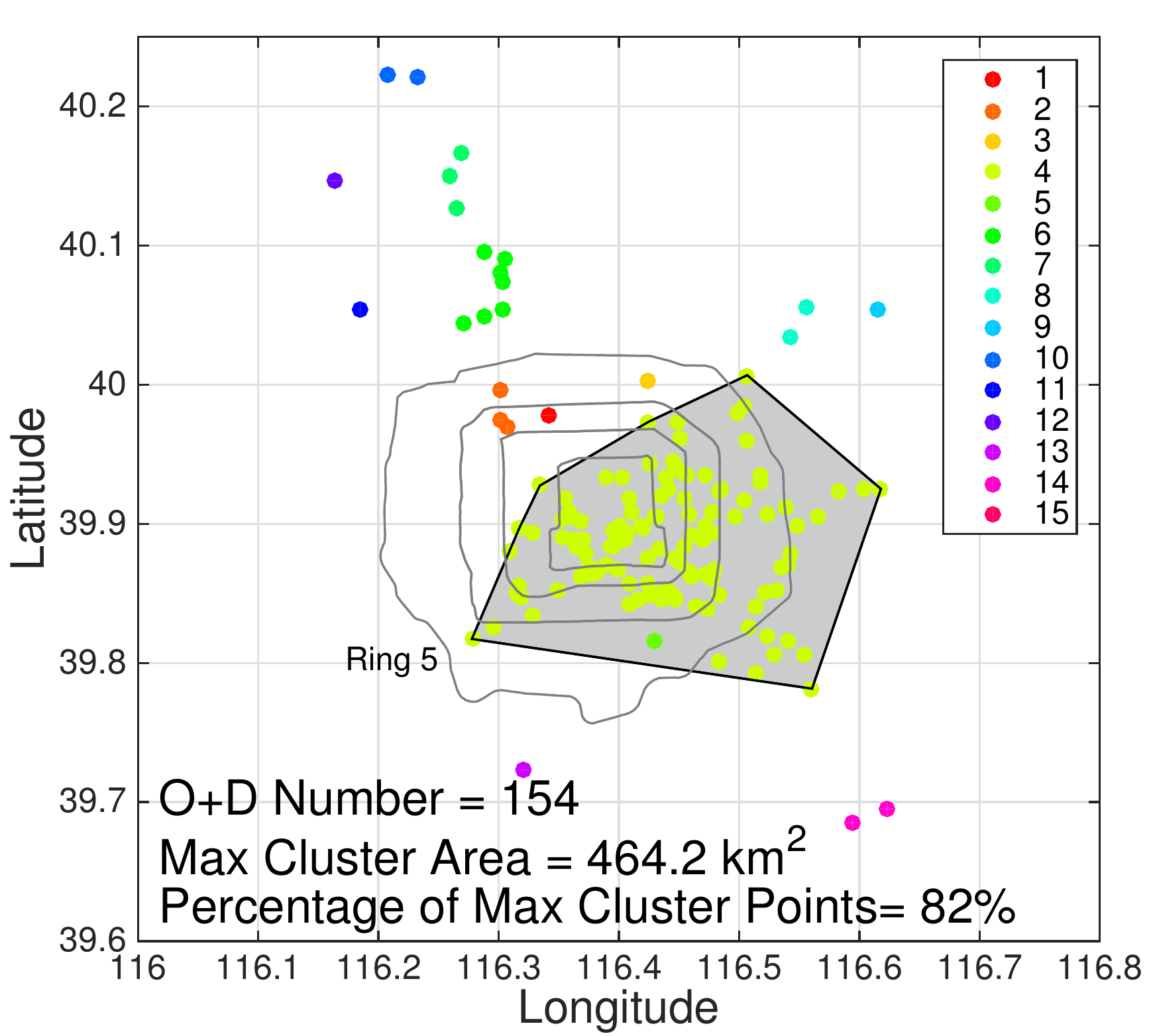}}
    \subfigure[Driver 12]{
    \includegraphics[width=2.2in]{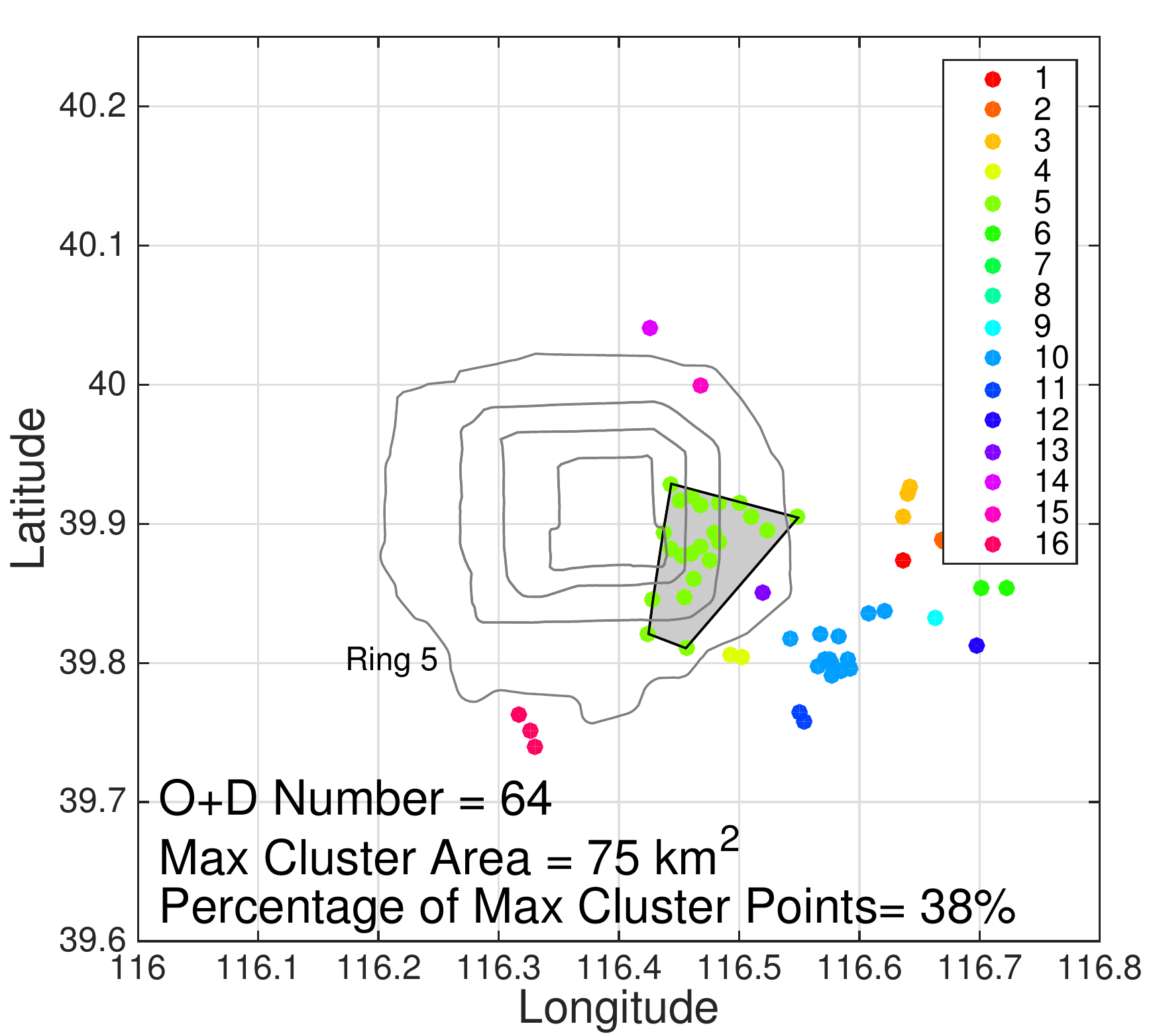}}
    \subfigure[Driver 13]{
    \includegraphics[width=2.2in]{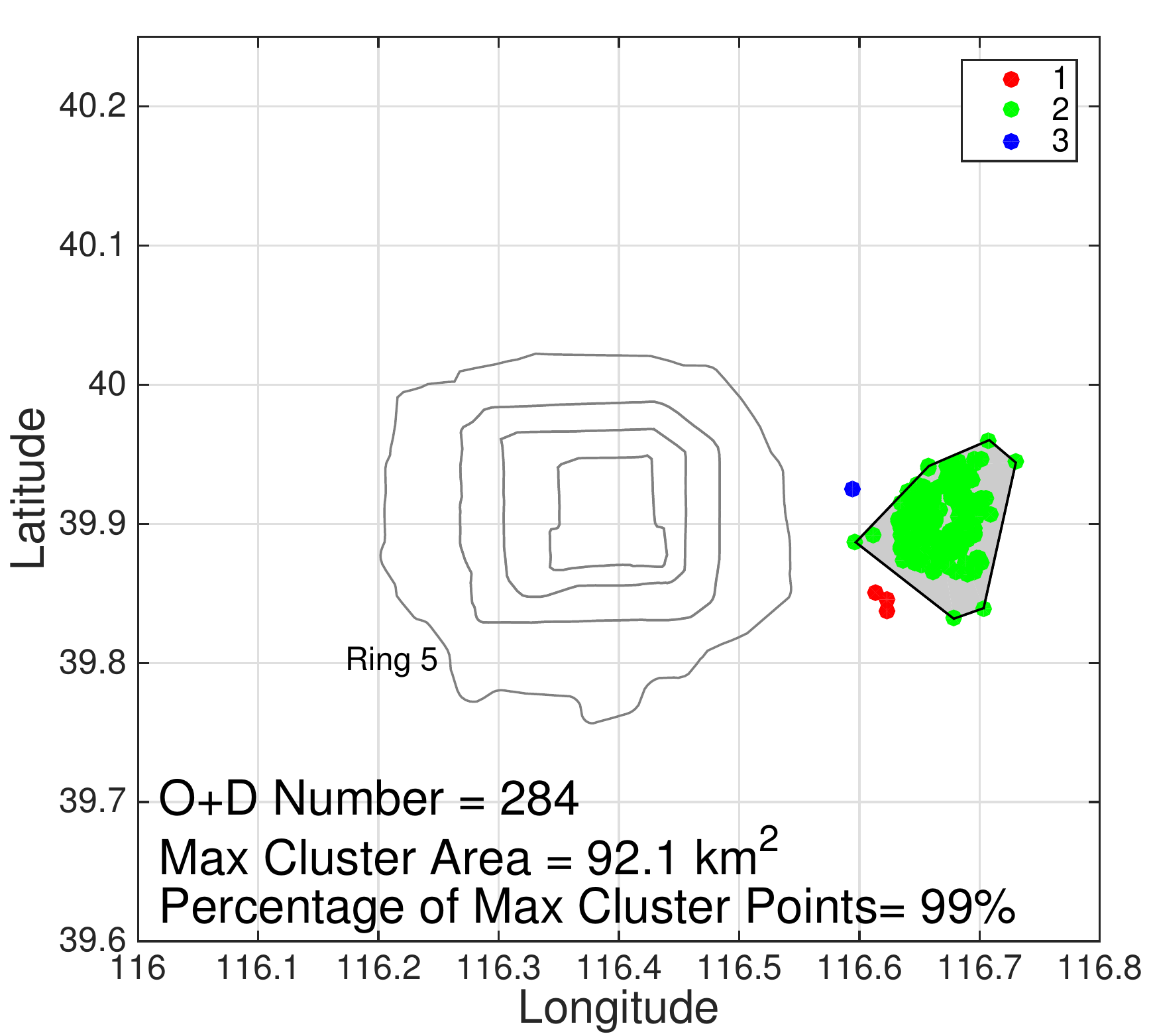}}
    \subfigure[Driver 14]{
    \includegraphics[width=2.2in]{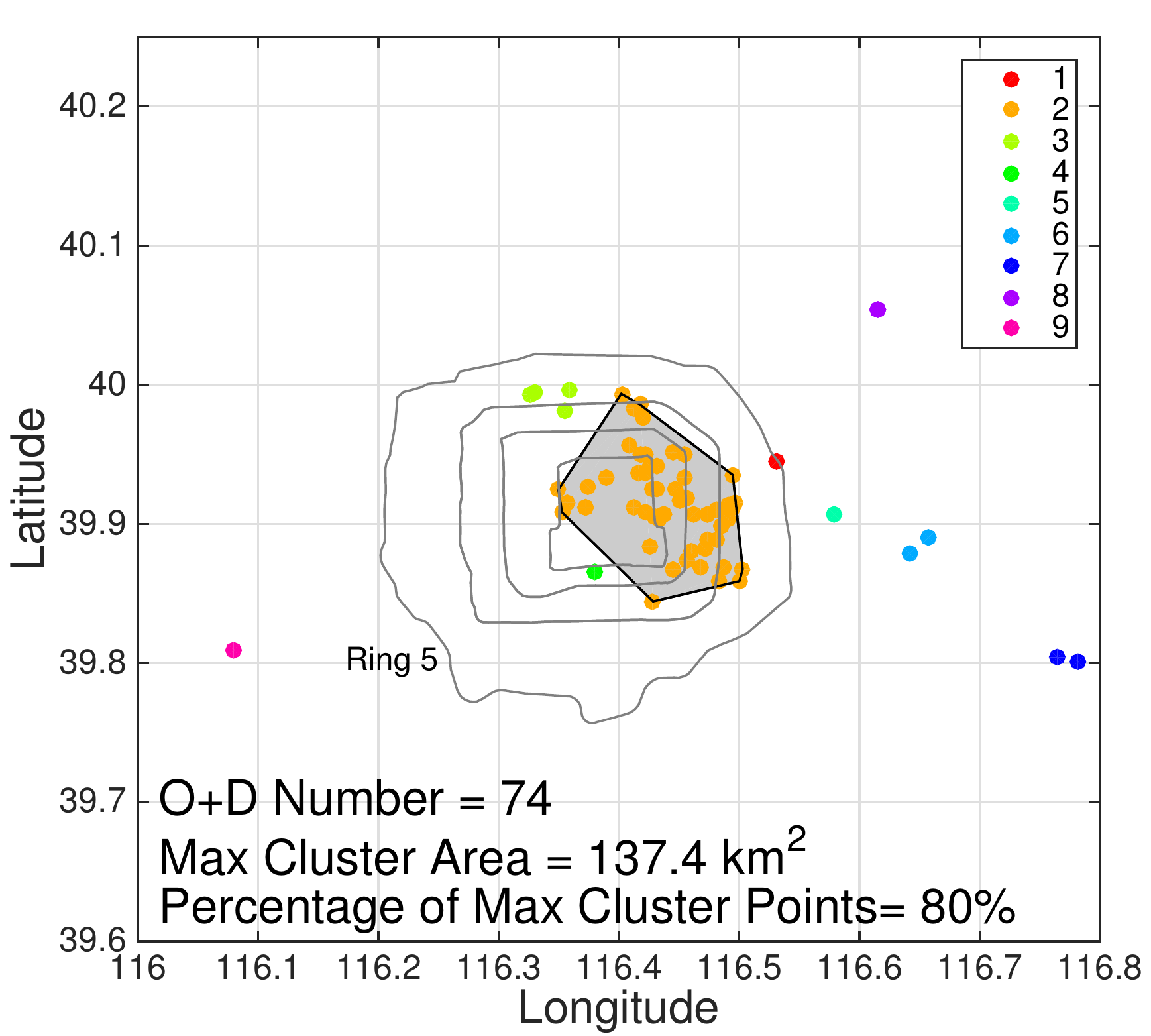}}
    \subfigure[Driver 15]{
    \includegraphics[width=2.2in]{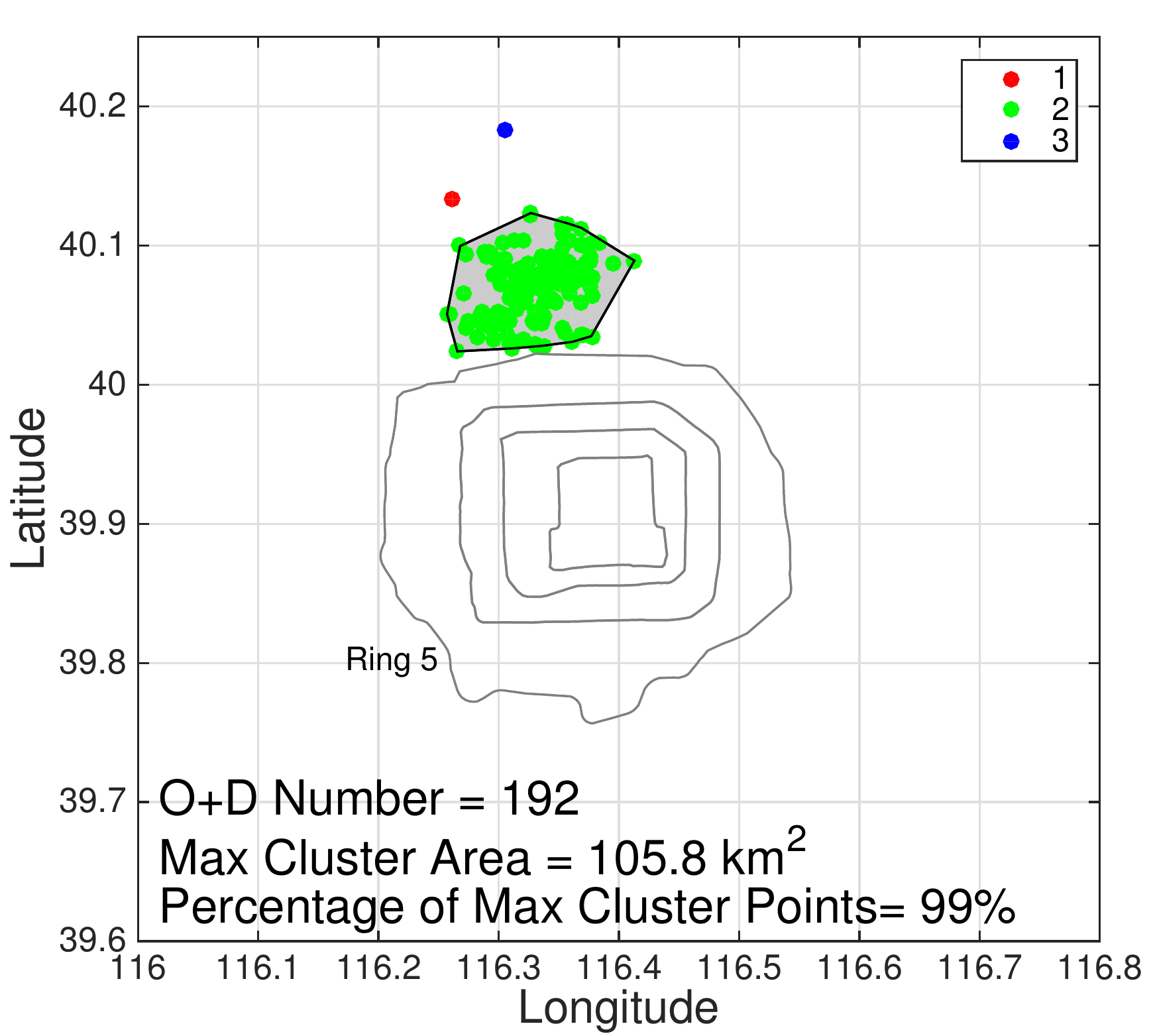}}
    \subfigure[Driver 16]{
    \includegraphics[width=2.2in]{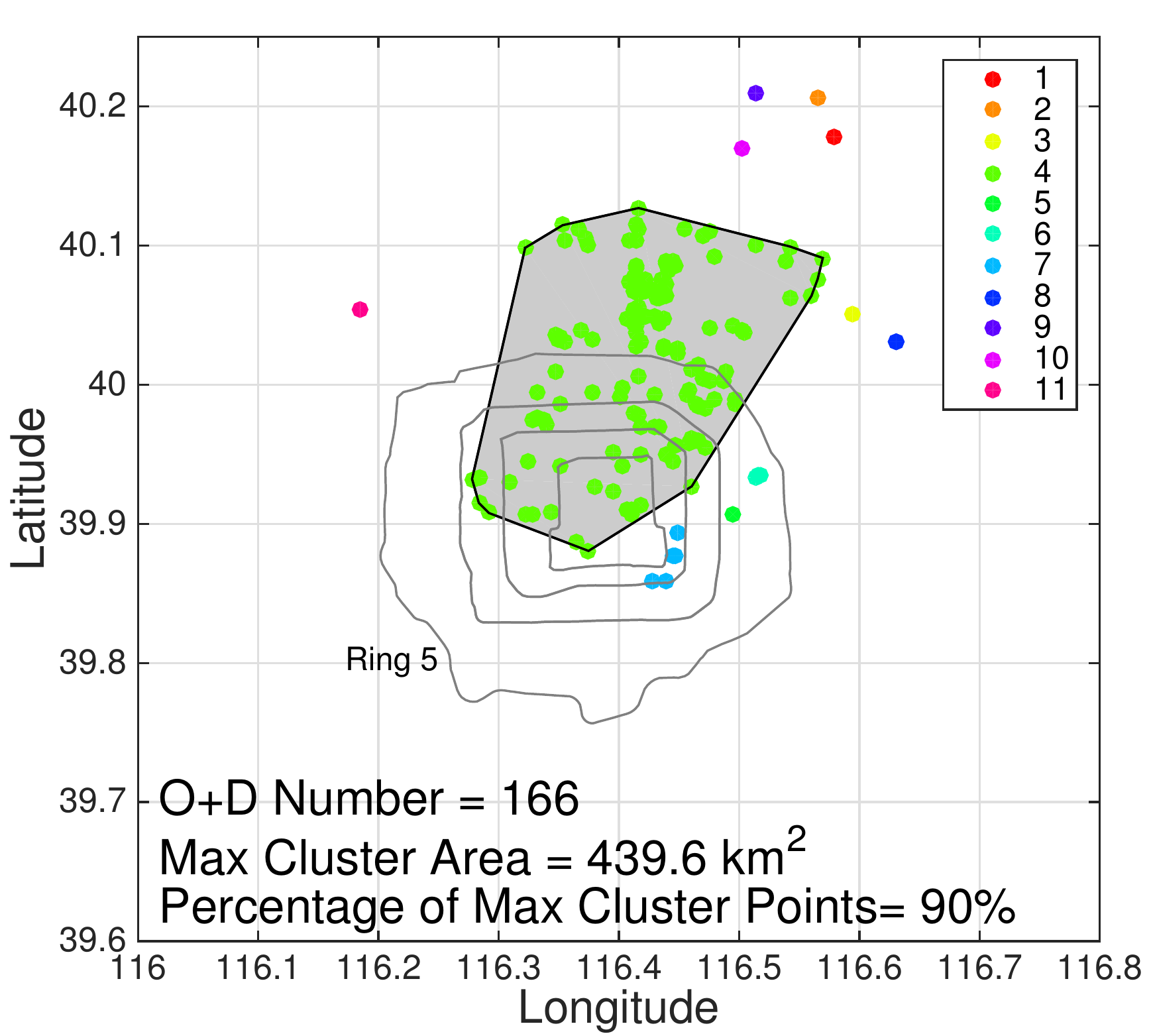}}
    \subfigure[Driver 17]{
    \includegraphics[width=2.2in]{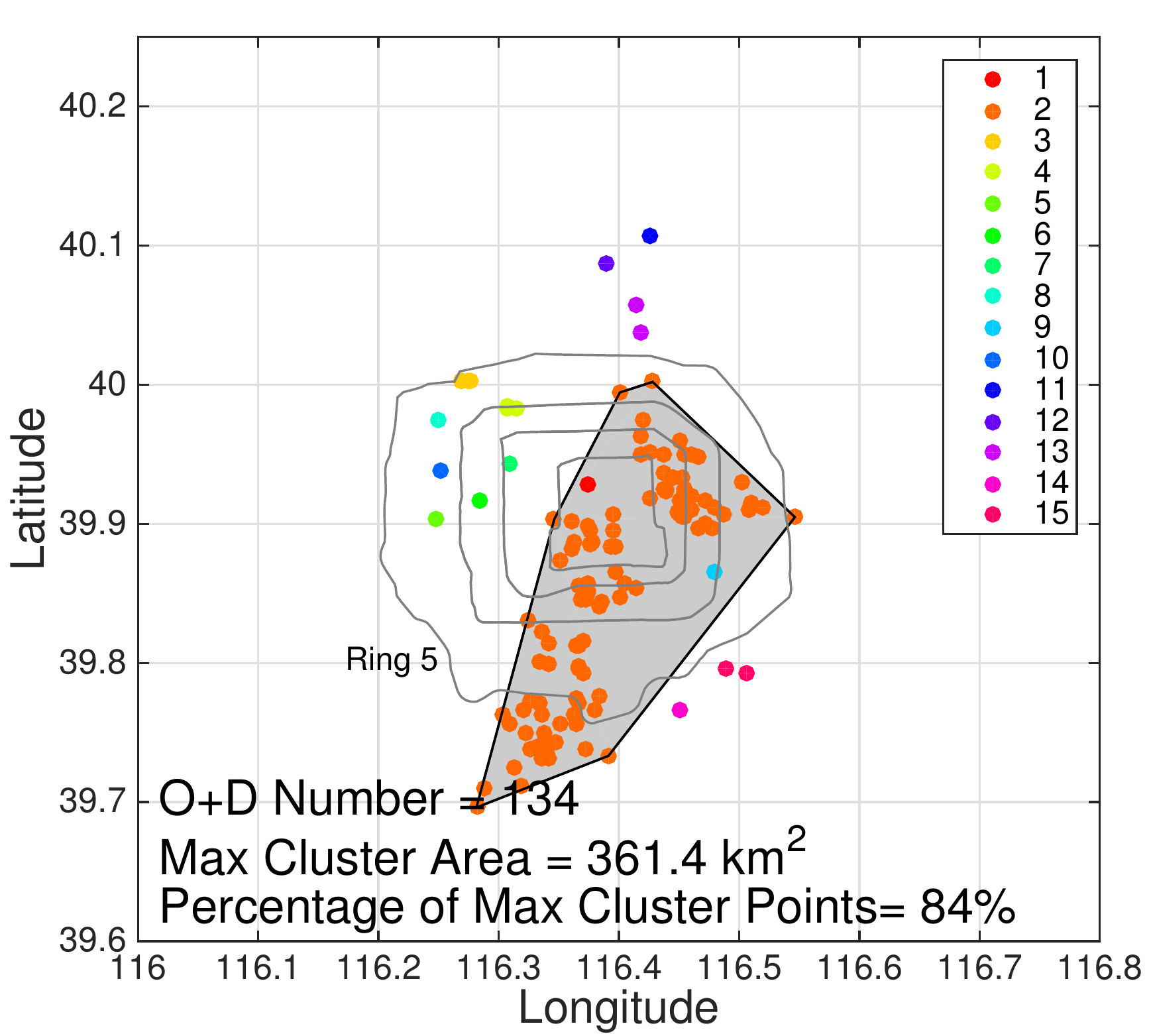}}
    \subfigure[Driver 18]{
    \includegraphics[width=2.2in]{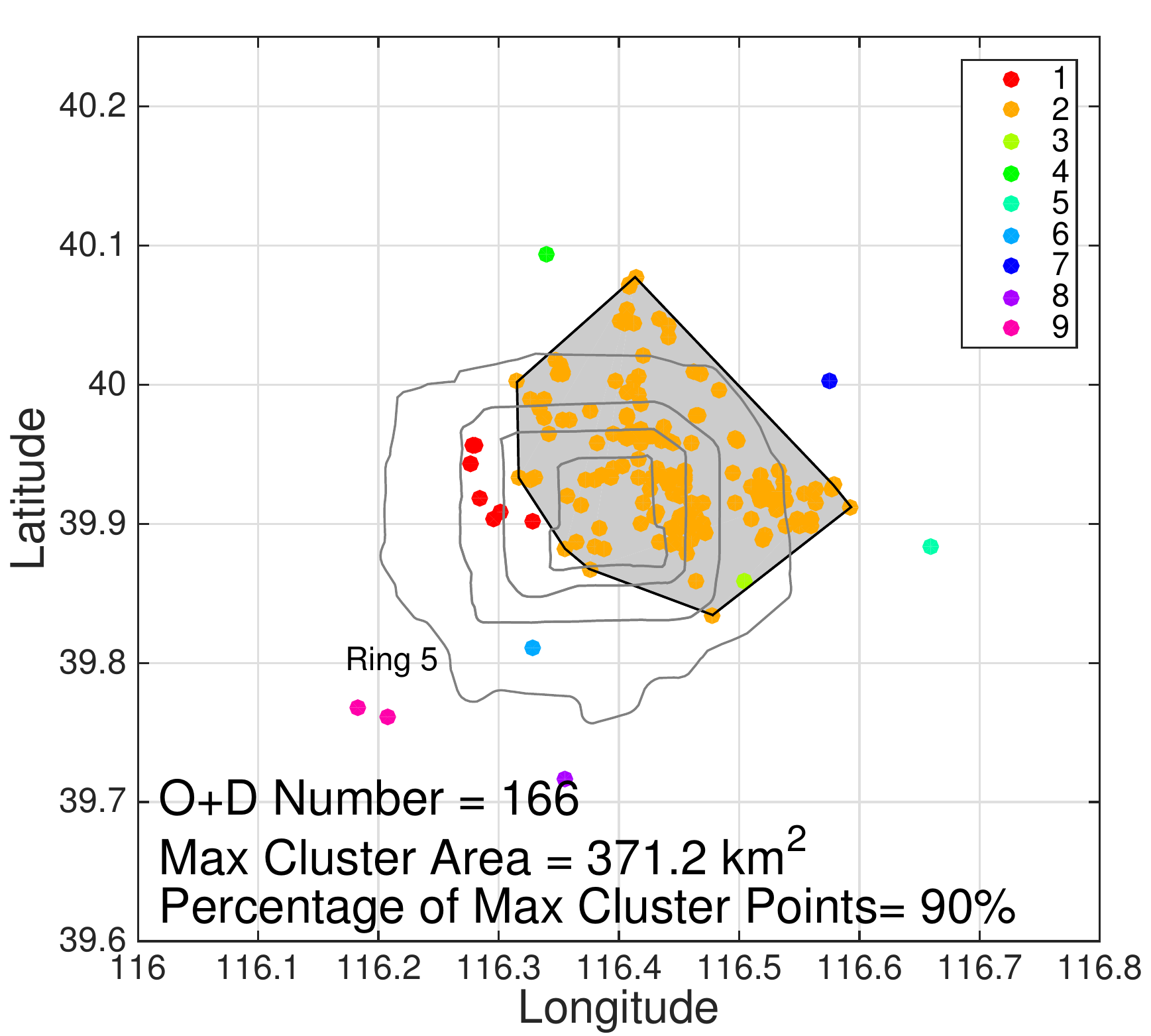}}
    \caption[Clustering of driver's origin and destination points of trips in a week.]
    {Clustering of spatial distribution of (9 randomly selected) driver's origin and destination points of the trips in a week ($\beta=5$ km).
    Grey polygon: the maximum cluster.}
    \label{fig:Z_Cluster_Spatial}
\end{figure}

To give a direct visualization of ride-hailing drivers' activity space, we primarily plot 9 randomly selected drivers' working space in Figure \ref{fig:Z_Cluster_Spatial}.
It is found that some drivers provide service in a large spatial range even covering the whole central area of the city (such as Drivers 10, 11, 12, 16, 17, and 18), while some only work in a small area (such as Drivers 13, 14, and 15).

\begin{figure}[!htbp]
    \centering
    \subfigure[Percentage]{    
    \includegraphics[width=3in]{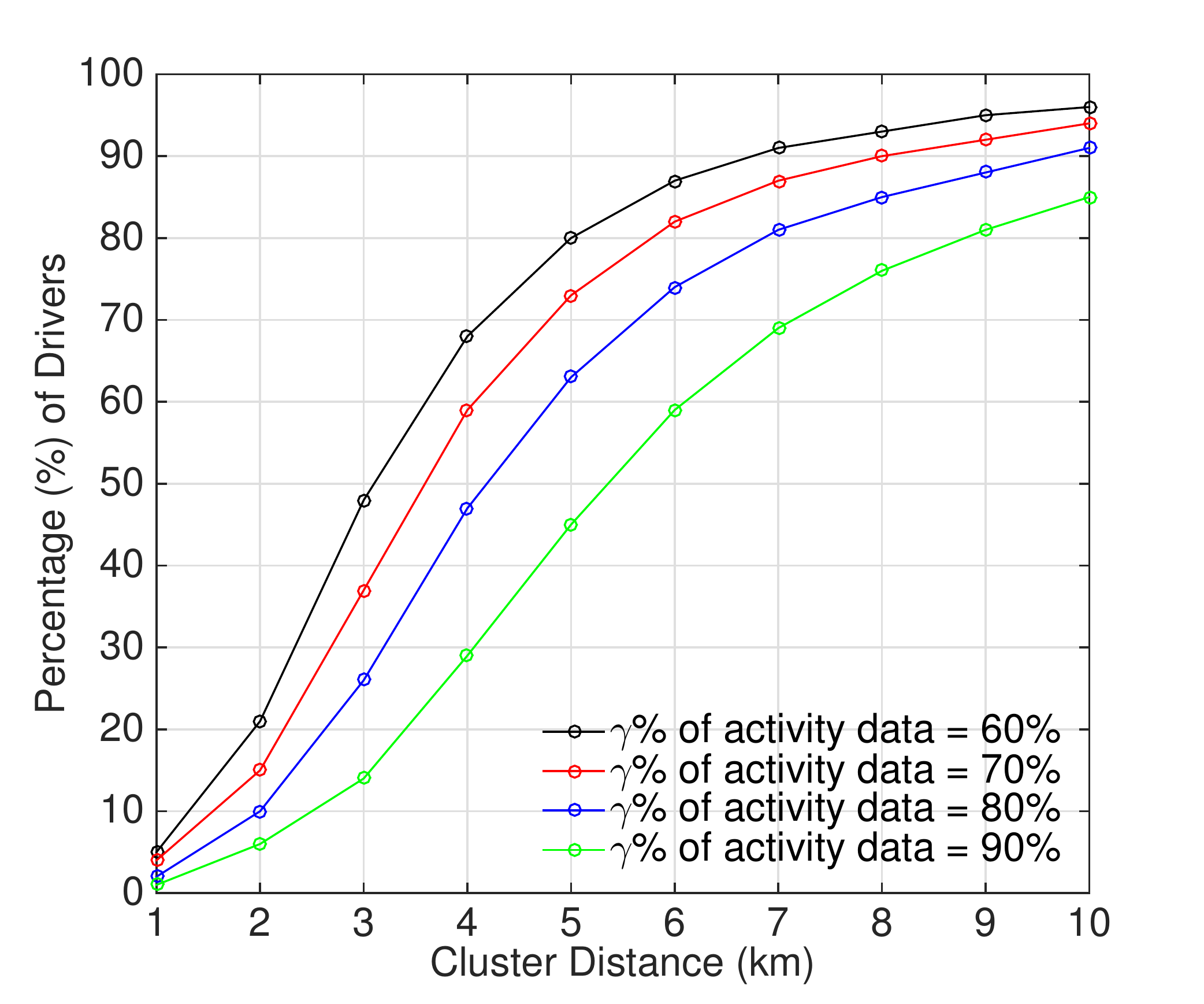}}
    \subfigure[Marginal increase of percentage]{
    \includegraphics[width=3in]{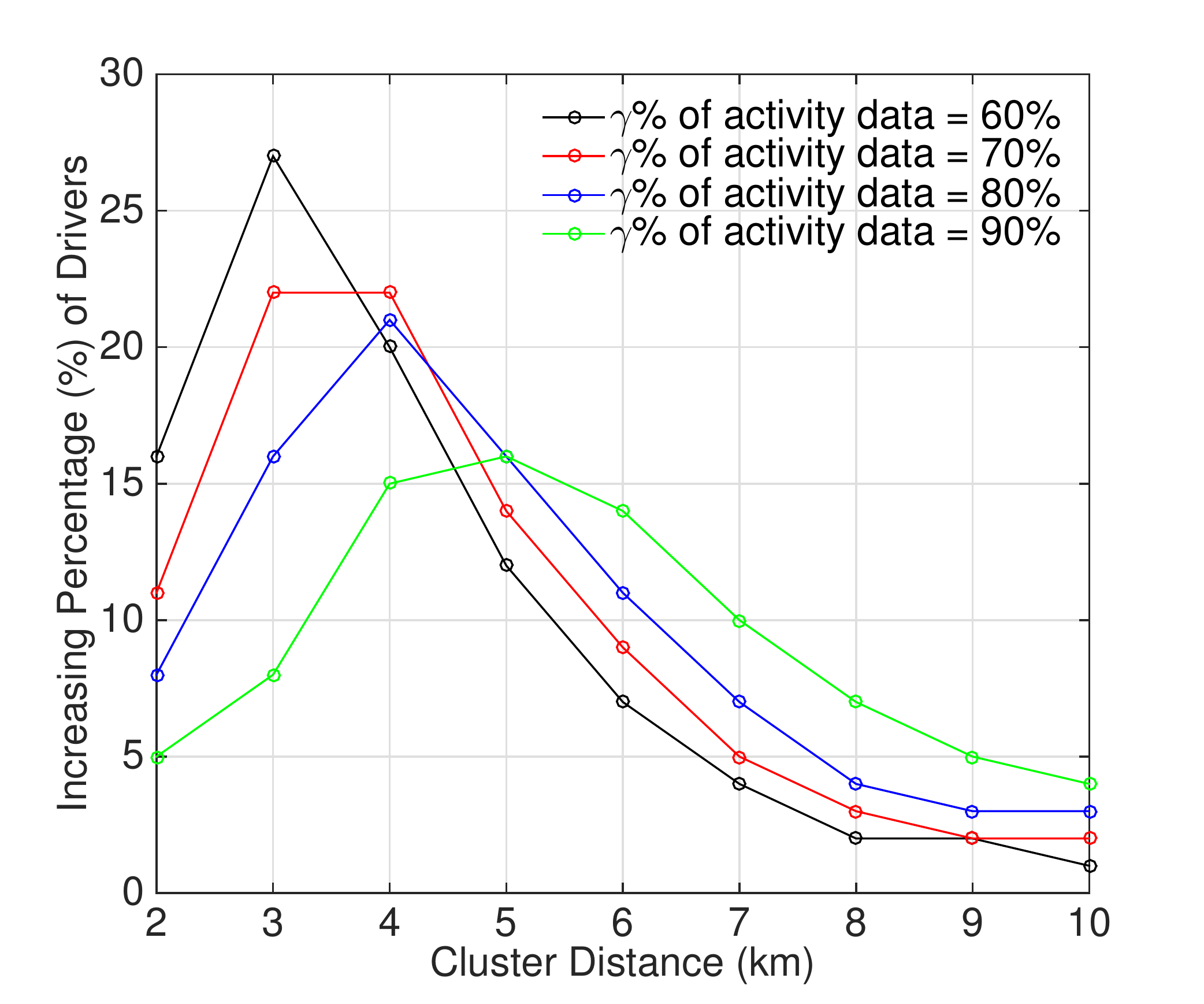}}        
    \caption[Percentage of the drivers whose $\gamma$ (\%) of all activity data (origin and destination points) are in the maximum cluster.]
    {Percentage of the drivers whose $\gamma$ \% of all activity data (origin and destination points) are in the maximum cluster.}
    \label{fig:Z_DifferentClusterDistance}
\end{figure}

Naturally, we are interested in understanding the impact of the cluster distance on the clustering.
According to the hierarchical clustering, the distance between any two data points in a cluster is less than the cluster distance, and thus the cluster distance roughly indicates the density of a driver's activities in a cluster, i.e., the preference of accepting long- or short-distance trips. 
The larger the cluster distance is, the lower the activity density in a cluster will be, i.e., more preferring to the short-distance trip demand. 
Focusing on the maximum cluster, we calculate the relationship between the cluster distance and the percentage of drivers whose $\gamma$\% of all origin and destination points are within the maximum cluster. 
The results are presented in Figure \ref{fig:Z_DifferentClusterDistance} and we have the following findings.

\begin{itemize}
\setlength{\itemsep}{0pt}
\setlength{\parsep}{0pt}
\setlength{\parskip}{0pt}	  
	\item[{[24]}] Cluster distance of 10 km could make the maximum cluster contain most driver's activities 
	(In Figure \ref{fig:Z_DifferentClusterDistance}(a), the percentage is 85\% when $\gamma\%=90$\%, while the percentage is 96\% when $\gamma\%=60$\%),	 
	implying that most drivers are not willing to go to a place where is 10 km away from his/her regularly-cruising region (i.e., the maximum cluster).
	
	\item[{[25]}] 
	To involve more drivers working in their maximum cluster (i.e., increasing along the y-axis of Figure \ref{fig:Z_DifferentClusterDistance}(a)), increasing the cluster distance (i.e., increasing along the x-axis of Figure \ref{fig:Z_DifferentClusterDistance}(a)) apparently takes effect, while the effect first increases and then decreases (Figure \ref{fig:Z_DifferentClusterDistance}(b)), implying a part of drivers work spatially intensely (the drivers corresponded by the left of the peak), while the other part of drivers work relatively loosely; after quickly involving those spatially intensely-working drivers, the pace becomes slow in terms of involving those spatially loosely-working drivers. 
	In particular, the peak when $\gamma\%=90$\% occurs at the cluster distance of 5 km that is greater than the cluster distance at the peak of smaller $\gamma$. 
	
\end{itemize}


\begin{figure}[!htbp]
    \centering
    \includegraphics[width=3.8in]{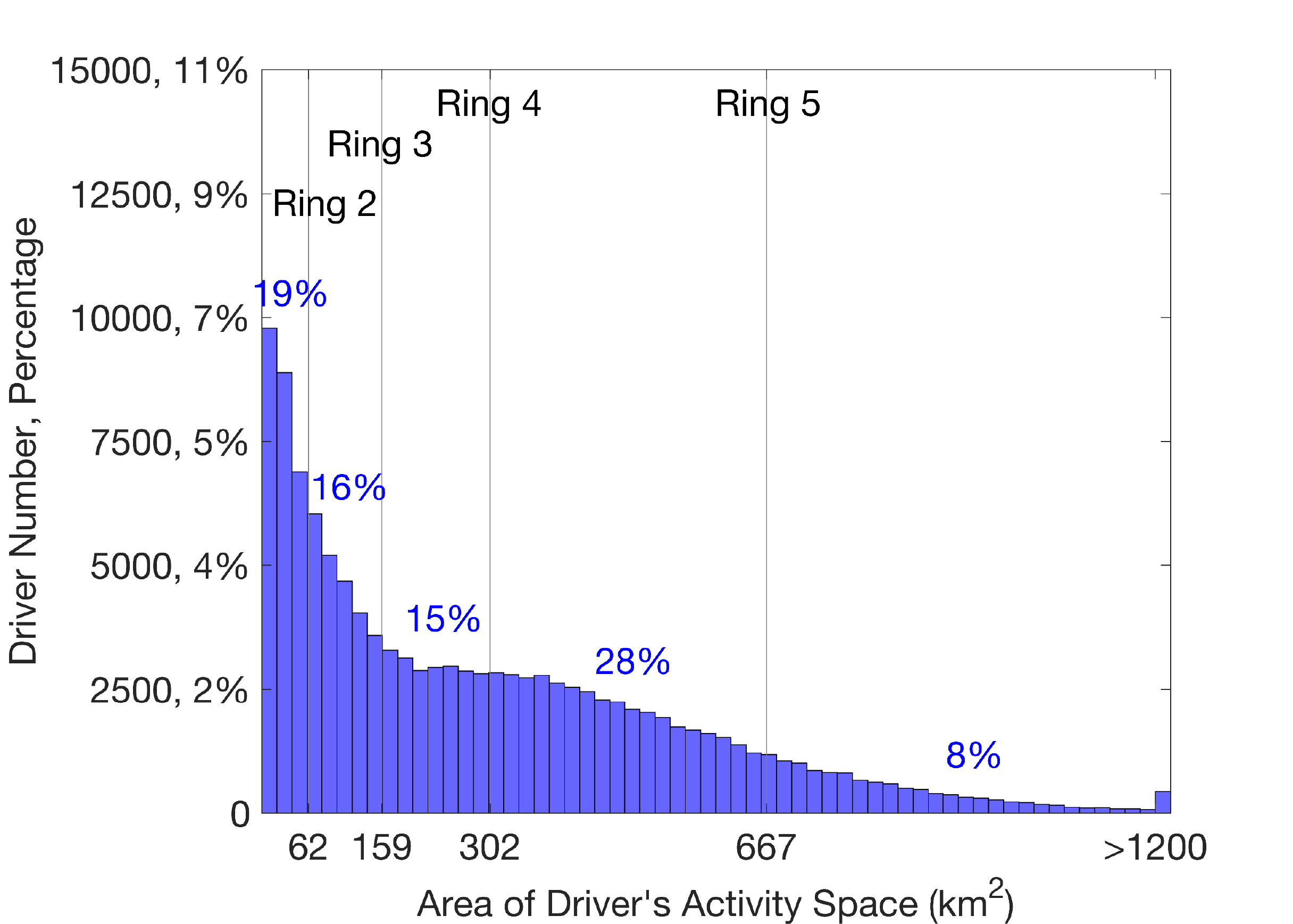}
    \caption[Distribution of ride-hailing drivers' largest activity-space areas]
    {Distribution of ride-hailing drivers' largest activity-space areas (cluster distance is set to 5 km).
    The regions that are smaller 5 km$^2$ are removed from the statistics, since very small area may be resulted by few ride-hailing activities.}
    \label{fig:Z_Distribution_MaxClusterArea}
\end{figure}

Then, we look at the distribution of the areas of ride-hailing drivers' daily activity spaces, which is presented in Figure \ref{fig:Z_Distribution_MaxClusterArea}.
\textcolor{black}{
Let A$_2$, A$_3$, A$_4$, and A$_5$ be the areas inside Rings 2$\sim$5, and A$_2$= 62 km$^2$, A$_3$= 159 km$^2$, A$_4$= 302 km$^2$, and A$_5$= 667 km$^2$.
We have the following observations by taking the areas inside Rings 2$\sim$5 in Beijing as references.}

\begin{itemize}
\setlength{\itemsep}{0pt}
\setlength{\parsep}{0pt}
\setlength{\parskip}{0pt}	  

	\item[{\textcolor{black}{[26]}}] \textcolor{black}{\textcolor{black}{
	The percentages of the drivers, the areas of whose activity spaces are smaller than A$_2$, between A$_2$ and A$_3$, between A$_3$ and A$_4$, between A$_4$ and A$_5$, and larger than A$_5$, are 19\%, 16\%, 15\%, 28\%, and 8\%, respectively.}}
	
	\item[{\textcolor{black}{[27]}}] \textcolor{black}{Amongst those groups of drivers, the largest group (i.e., 28\%) is the drivers who prefer to work in the spaces whose areas are between A$_4$ and A$_5$. 
	The areas of drivers' activity spaces are approximately equivalent to the area of the central area of Beijing, since it commonly treats inside Ring 5 as the central area of Beijing.}
	
	\item[{\textcolor{black}{[28]}}] \textcolor{black}{The second largest group (i.e., 19\%) is the drivers who prefer to work in relatively small spaces, whose areas are smaller than A$_2$.
	The smallest group (i.e., 8\%) is the drivers who usually work in rather large spaces, whose areas are larger than A$_5$. 
	}

\end{itemize}

\subsection{Correlation between temporal and spatial characterizations}

\begin{figure}[!htbp]
    \centering
    \includegraphics[width=7in]{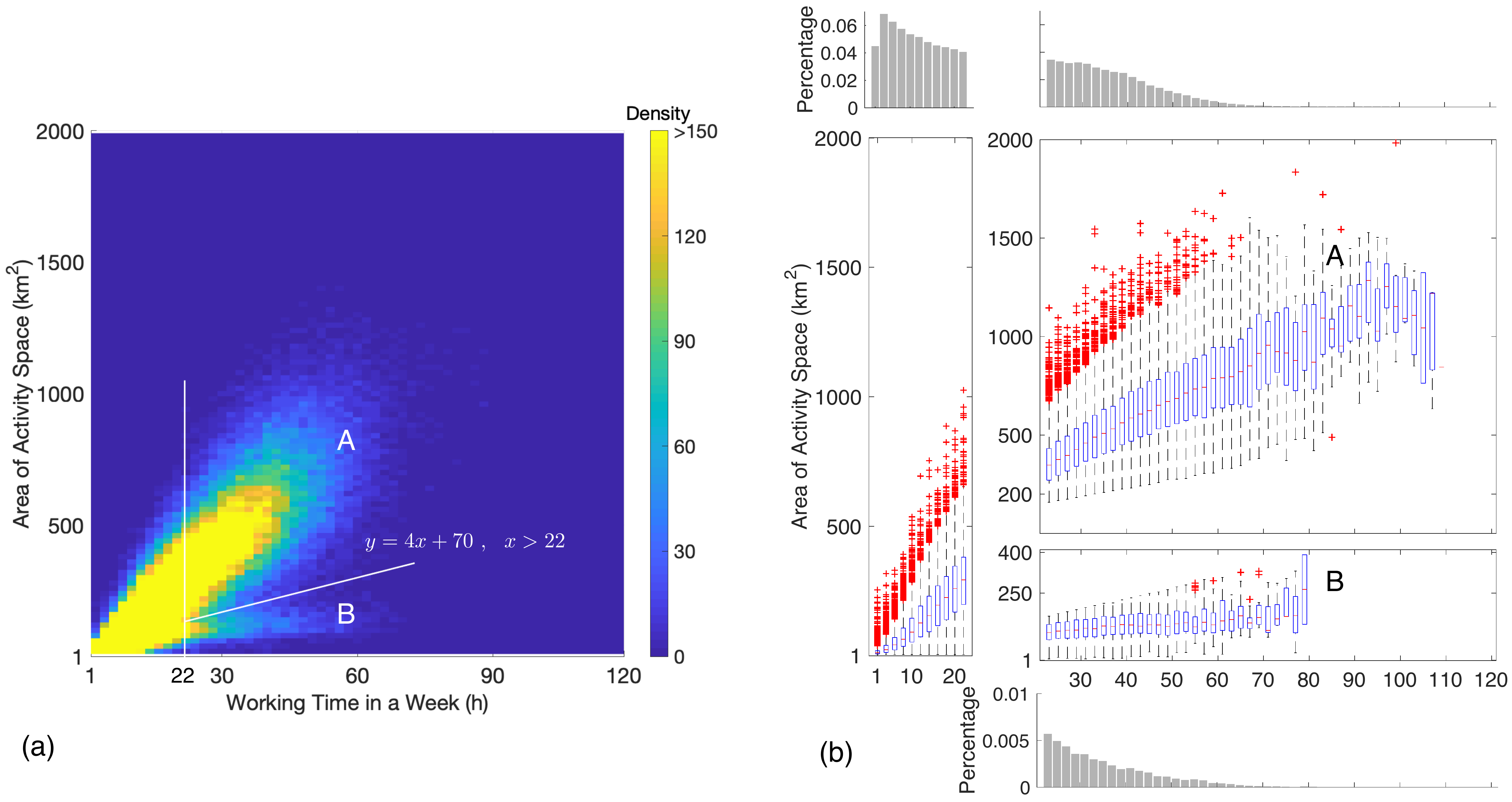}
    \caption{\textcolor{black}{Correlation between driver's working time and activity space.
    (a) Heat map; (b) Box plot (The histograms show the percentage of data points (drivers) contained by the corresponding box plot in the total points (drivers).)}}
    \label{fig:Z_TimeSpaceCorrelation}
\end{figure}

This subsection combines the temporal and spatial characterizations to see if some correlations exist.
To that end, we jointly plot drivers' total working time in a week and the areas of their activity spaces
in Figure \ref{fig:Z_TimeSpaceCorrelation}.
Interesting observations are found as follows.

\begin{itemize}
\setlength{\itemsep}{0pt}
\setlength{\parsep}{0pt}
\setlength{\parskip}{0pt}	  
	\item[{[29]}] The correlation plot shows two branches when the working time is larger than 22 h (A and B in Figure \ref{fig:Z_TimeSpaceCorrelation}), 
	indicating two categories of ride-hailing drivers.
	The majority of the drivers are in Category A and the activity space is enlarged with the growth of the working time.
	A minority of the drivers (i.e., Category B), whose activity spaces are not increased with the growth of the working time, are observed for the first time. 
	Unlike those in Category A, the drivers in Category B only prefer to work in a small space.	
	
	\item[{[30]}] The correlation between the working time and the activity space is approximately linear positive, i.e., the increase of working time results in the expansion of the driver's activity space. 
	It is particularly obvious for Category A.

\end{itemize}

This observation confirms the existence of ride-hailing drivers' selection and preferences in providing service as discussed in Section \ref{sec:perspective}.
This is a major difference between the ride-hailing and taxi drivers\footnote{It is known that taxi drivers cannot select passengers, i.e., they have to go anywhere that their passengers would like to go.}.

\section{Discussion and conclusion}\label{sec:Conclusion}

Using the multi-day ride-hailing driver activity data in an entire city, this paper characterizes the ride-hailing activities from regional and driver's perspectives, respectively.
A series of findings that are labeled as from [1] to [30] in Sections \ref{sec:Macro} and \ref{sec:Micro} are obtained, which are summarized as follows.

\noindent {\it Spatiotemporal flowing of ride-hailing trips: A regional perspective}

\begin{itemize}
\setlength{\itemsep}{0pt}
\setlength{\parsep}{0pt}
\setlength{\parskip}{0pt}	
	\item{} {\it Observation [1]-[2]}.
	On Friday and Saturday, the ride-hailing demands are slightly larger than those on the other days.
	For the ride-hailing demand, there is no clear off-peak at noon and the high-demand peak lasts to midnight.
	Those are different from the observations from daily traffic and taxi demand that usually exhibit clear morning and evening peaks.
	
	\item{} {\it Observations [3]-[5]}.
	There are two peaks (at 15:00 and midnight) for the appearance of the regions with the dense origins, while only one peak (at noon) for the regions with dense destinations. The shrinking and expanding processes, which reflect the spatiotemporal rhythm of a city in terms of using the ride-hailing services, are also observed.
	
	\item{} {\it Observations [6]-[8]}.
	For the same time interval on each weekday, the number of the origin or destination points that are associated with the same (1 km $\times$ 1 km) regions do not change largely, indicated by the coefficients of variation that are between 0.15 and 0.25. 
	Only for several intervals (usually at morning and evening peaks), the repeatability is low, indicated by the coefficients of variation are between 0.3 and 0.4.
	On average, the coefficients of variation are between 0.2 and 0.4.
	The results imply that, due to the existence of the high repeatability of the ride-hailing demands, a newly-proposed advanced prediction method could be deemed as really taking effect only when its prediction accuracy is higher than that resulted from the repeatability at the natural condition. 
	
	\item{} {\it Observations [9]-[12]}.
	A large proportion (more than 50\%) of the regions have relatively stable ride-hailing demands during a whole weekday, and the second large proportion is the regions showing a morning demand peak.
	
\end{itemize}

\noindent {\it Characterization of ride-hailing drivers: A driver's perspective}

\begin{itemize}
\setlength{\itemsep}{0pt}
\setlength{\parsep}{0pt}
\setlength{\parskip}{0pt}	
	\item{} {\it Observation [13]-[14]}.
	More ride-hailing drivers are part-time drivers.
	Specifically, 43.4\% of the drivers take less than 25 trips in a week and the total trip duration of 50.1\% of the drivers is less than 10 hours.
	Only a small part of the drivers take the ride-hailing as their full-time jobs.
	4.8\% of the drivers provide more than 100 ride-hailing services in a week, and 6.2\% of the drivers work more than 30 hours. 
	
	\item{} {\it Observation [15]-[17]}.
	The trip-duration distributions in different days show an algebraic power-law tail with the exponent of -4.27. 
	Likewise, it is -4.52 for all trip-displacement distributions.
	The distributions are quite similar in different days, indicating high repeatability of ride-hailing trips in terms of distributions.
	In addition, the duration and displacement of most trips are between 5 min and 50 min and less than 20 km, respectively. 
	Meanwhile, the power law tails indicate the fact that long trips may appear occasionally, which is different from those shown by taxis.
	
	\item{} {\it Observation [18]-[20]}.
	[12:00, 18:00] is the period that contains the largest number of frequently working drivers. 
	The percentages are 16.5\% and 14.1\% on weekdays and weekends, respectively.
	[6:00, 12:00] and [18:00, 24:00] are the second and third preferred working periods.
	Only for [0:00, 6:00], the working drivers on weekends are more than those on weekdays, which is consistent with the common sense that more night-life activities occur at weekend midnight. 
	
	\item{} {\it Observation [21]-[23]}.
	Most frequently-working drivers prefer to work approximately 6 hours per day (18.7\% of all drivers), implying that the ride-hailing drivers usually don't work as long as those for other jobs (e.g. working in an office).  
	Besides, 6\% of all drivers select to work for about 12 hours.

	\item{} {\it Observation [24]-[28]}.
	Through spatially clustering drivers' origin and destination points, it is found that most drivers in Beijing are not willing to go to a place where is 10 km away from his/her regularly-cruising region.
	28\% of ride-hailing drivers provide the ride-hailing services in a city-wide space; 19\% of all drivers are active in a relatively small space; Only 8\% of all drivers would like to provide service in a space that is larger than the central area of the city.
	
	\item{} {\it Observation [29]-[30]}.
	Two categories of ride-hailing drivers are found after combining the temporal and spatial characterizations of ride-hailing drivers.
	One (the majority) is the drivers whose activity spaces are linear positively correlated to the working time, while the other is the drivers who only prefer to work within a limited space.
	This observation confirms the existence of ride-hailing drivers' selection and preferences in providing service, which is a major difference between ride-hailing and taxi drivers.

\end{itemize}

From those findings, we can see that ride-hailing mobility has its own characteristics, such as the shrinking and expanding processes, the power distributions of trip duration and displacement, and the two categories of drivers. 
Many of them are quite different from our knowledge of other travel modes such as taxis \citep{Liang2012,Cai2016,He2017,Dong2018,He2019b,He2020}. 
Understanding these characteristics not only benefits TNCs but also helps traffic managers.
For example, the TNC could design a more-targeted dynamic incentive system to adjust the supply of ride-hailing vehicles. 
Traffic managers could consider to treat those full-time ride-hailing drivers as professional drivers and regulate them as done to taxi drivers. 
The debate if TNCs increase traffic congestion could be answered from a more detailed time-space dimension with high precision.
For the insights that are more related to scientific researchers, the region-dependent repeatability provides the related ride-hailing demand prediction with a baseline, indicating a capable method should result in the prediction errors that are smaller than the fluctuation shown by the demand itself \citep{Ke2017a,Zhang2020b}. 
The ride-hailing drivers' limited selection behavior within a small spatiotemporal scope also deserves comprehensive analysis and modeling.

As the most existing studies\footnote{Limited by the data, most existing studies are case studies, since they usually only focus on the ride-hailing in one city, 
such as San Francisco \citep{Rayle2016,Shaheen2016}, Las Vegas \citep{Alemi2018}, Beijing \citep{Dong2018}, Santiago de Chile \citep{Tirachini2019a}, and Austin \citep{Yu2019b}
} did, this paper only focuses on the ride-hailing activities in Beijing, China.
Nevertheless, we attempt to avoid mentioning specific locations, and we believe the above findings not only directly benefit transportation researchers and managers in Beijing, but also contribute to the general understanding of ride-hailing activities.

\textcolor{black}{The grid size used for the regional analysis in Section \ref{sec:Macro} is set to be 1 km.
Obviously, using different sizes will change the results with specific numbers, while we don't think it will radically differentiate the patterns that we observed due to the fact that the analysis is conducted from the macroscopic perspective \citep{He2020b}.
The selection of the grid size is relevant to the modifiable areal unit problem, and it always causes information loss and result bias no matter what size is chosen \citep{Fotheringham1991,Clark2014,Nelson2017,Zhou2020}.
Therefore, it is interesting to start an investigation to evaluate the sensitivity of the analysis result to the grid size.}

Thanks to the uniqueness of the data used here, we obtained many new observations by only using simple and direct analysis methods. 
In the future, it is meaningful to mine the data with more advanced (e.g., machine learning) methods to understand the latent factors dominating ride-hailing mobility.
In particular, theoretically modeling ride-hailing mobility and drivers' choice are of importance. 
Combining more data sources, such as point-of-interest data, traffic flow data, built-environment data, will also enrich our understanding of ride-hailing and the factors associated. 
\textcolor{black}{
Other research direction is to compare with the ride-hailing activities in other cities or with other travel modes.
Although the requirement for data is absolutely higher, the comparative studies on different regions and modes are significant for both practice and theory.}

\section*{Acknowledgement}
The author is grateful to Prof. Ning Jia, Prof. Xiaoyong Yan, Dr. Geqi Qi, Dr. Long Cheng, Mr. Shouzheng Pan, and Ms. Lin Wang for their insightful suggestions.
The research is funded by 
National Natural Science Foundation of China (71871010).

\bibliographystyle{model2-names}
\bibliography{library}

\end{spacing}
\end{document}